\documentclass[sigconf]{acmart}

\pdfoutput=1

\usepackage{tikz}
\usepackage{amsmath}

\usepackage{algorithm}
\usepackage{algpseudocode}
\usepackage{graphicx}
\usepackage{subfigure}
\usepackage{url}
\usepackage{color}
\usepackage{multirow}
\usepackage{amssymb}
\usepackage{amsmath}
\usepackage{gensymb}
\usepackage{longtable}
\usepackage{tikz}
\usepackage{comment}

\usepackage{etoolbox}
\usepackage{courier}

\usepackage{xspace}

\newcommand{\ouralg}{$\mathsf{NoDE}$\xspace}
\newcommand{\OurAlg}{$\mathsf{NoDE}$\xspace}

\newcommand{\ouralgSection}{{NoDE}\xspace}
\newcommand{\rNoise}{$\mathsf{Random Noise}$\xspace}
\newcommand{\rPower}{$\mathsf{Random Power}$\xspace}
\newcommand{\rNoiseBf}{$\mathbf{Random Noise}$\xspace}
\newcommand{\rPowerBf}{$\mathbf{Random Power}$\xspace}

\setcopyright{none}

\renewcommand\footnotetextcopyrightpermission[1]{} 
\acmConference{ACM SIGMETRICS}{June 08--12, 2020}{Boston, MA}
\begin{document}

\title{Your Noise, My Signal: Exploiting Switching Noise for Stealthy Data Exfiltration
	from Desktop Computers}

\author{Zhihui Shao}
\authornote{Both authors contribute equally to this work.
}
\affiliation{%
  \institution{University of California, Riverside}
}
  \email{zshao006@ucr.edu}

\author{Mohammad A. Islam}
\authornotemark[1]
\authornote{This work was partially done while
M. A. Islam was at UC Riverside.}
\affiliation{%
  \institution{University of Texas at Arlington}
 }
\email{mislam@uta.edu}

\author{Shaolei Ren}
\affiliation{%
	\institution{University of California, Riverside}
}
\email{shaolei@ucr.edu}

\begin{abstract}
Attacks based on power analysis have been long existing and
studied, with some recent works focused on data exfiltration from victim systems without using conventional communications (e.g., WiFi). Nonetheless, prior works typically rely on intrusive direct power measurement, either by implanting meters in the power outlet or tapping into the power cable, thus jeopardizing the stealthiness of  attacks. 
In this paper, we propose \ouralg (Noise for Data Exfiltration),
a new system
for stealthy data exfiltration from
enterprise desktop computers. 
Specifically, \ouralg achieves data exfiltration
over a building's power network by exploiting high-frequency voltage
ripples (i.e.,
switching noises) generated by power factor correction circuits built into today's computers.
Located at a distance and even from a different room, the receiver
can non-intrusively measure the voltage of a power outlet to capture the high-frequency
switching noises for online information decoding without supervised training/learning.
To evaluate \OurAlg, we run experiments on seven different computers
from  top vendors and using top-brand power supply units.
Our results
show that for a single transmitter, \ouralg achieves a rate of up to 28.48 bits/second
with a distance of 90 feet (27.4 meters) without the line of sight,
demonstrating a practically stealthy threat. Based on
the orthogonality of switching noise frequencies of different computers, we also demonstrate
simultaneous data exfiltration from four computers using only one receiver.
Finally, we present a few possible defenses, such as
installing noise filters, and discuss their limitations.

\end{abstract}



\maketitle
\thispagestyle{empty}

\section{Introduction}\label{sec:introduction}

The total number of registered malware samples has grown by 36\% in the past year
and reached
an all-time high of 690 million,  let alone the huge
number of undiscovered malware \cite{Security_Malware_ThreatReport_McAfee_March_2018}.
Importantly, more than 70\% of the malware threats are in the form of phishing, spywares,
and Trojans that aim at stealing sensitive information, especially
from end users in companies, universities, among
others (which we collectively refer to as \emph{enterprise}) \cite{Security_Threat_Survey_SANS_2017}.

In the wake of growing risks of data theft,
a proactive defense is to keep sensitive data
{within} an enterprise network at all times.
Nonetheless, this approach is vulnerable to various types of {covert} channel attacks, through which sensitive data is {stealthily} transferred to a program that can access
external networks and eventually send information to the outside \cite{Security_Covert_Channels_SANS_2010,Security_Covert_Channel_Mobile_ZhiqiangLin_SecureComm_2014_Chandra2014TowardsAS,Security_Lampson_Note_on_Confinement_Problem_1973_Lampson:1973:NCP:362375.362389}.
For example, one program's usage pattern of CPU resources, if detected by another program,
can be modulated for information transfer between the two \cite{Security_Lampson_Note_on_Confinement_Problem_1973_Lampson:1973:NCP:362375.362389,Security_CovertChannel_Mobile_Survey_2012_Marforio:2012:ACC:2420950.2420958}.
Consequently, to mitigate data theft risks,
enterprise users
commonly have restricted access to outside networks ---
all data transfer
from and to the outside is tightly scrutinized.

Nevertheless, such systems may still suffer from data exfiltration attacks that bypass the conventional communications protocols (e.g., WiFi) by transforming the affected computer into a transmitter and establishing a covert channel.
For example, the transmitting computer can modulate the intensity of the generated acoustic noise by varying its cooling fan or hard disk spinning speed to carry 1/0 bit information (e.g., a high fan noise represents ``1'' and ``0'' otherwise), while a nearby receiver with a microphone can hear the noise and decode the carried bits \cite{Security_AirGap_Acoustic_DiskFiltration_DiskNoise_ESORICS_2017,Security_AirGap_Fansmitter_Acoustic_arXiv_DBLP:journals/corr/GuriSDE16,Security_AirGap_Acoustic_Ultrasound_FPS_Canada_2014,Security_AirGap_Acoustic_Ultrasound_MeshNetworks_NotIsraeli_Journal_2013}.
Likewise, the power consumption \cite{Security_AirGap_Mobile_USB_Charging_ACNS_Mobile_CovertChannel_Spolaor2017NoFC,Security_AirGap_Electric_PowerHammer}, the amount of generated heat \cite{Security_AirGap_Thermal_BitWhisper_CSF_2015_Guri:2015:BCS:2859845.2859982},
the electromagnetic interference (EMI) \cite{Security_AirGap_EMI_GSMem_Security_2015,Security_AirGap_EMI_AirHopper_TIST_Journal_2017_Guri:2017:BAG:3055535.2870641},
the system status LEDs \cite{Security_AirGap_Optimal_Keyboard_LED_Loughry:2002:ILO:545186.545189,Security_AirGap_Optical_HardDrive_LED_Guri2017LEDitGOL},
and magnetic signal (to escape a Faraday cage) \cite{Security_AirGap_Magnetic_Mobile_ASP_DAC_2016_Matyunin2016CovertCU,Security_AirGap_Magnetic_ODINI_FaradyCage_Bypass}
can all be modulated in a similar manner for data exfiltration.

\textbf{Our contribution.}
We contribute to the existing body of research by designing a new data exfiltration system,
called \ouralg (Noise for Data Exfiltration), where a malware modulates the victim computer's power consumption to send data over a building's power network to the attacker's receiver.
The key novelty is that \ouralg uniquely exploits  high-frequency voltage ripples (i.e.,
electronic switching noises)
generated by power factor correction (PFC) circuits built into today's
power supply units for power-consuming devices like computers.
%
Like the existing data exfiltration attacks (Table~\ref{table:exfiltration_summary}), \ouralg exhibits several desirable properties:
a reasonable achievable bit
rate (28.48 bits/s), good effective distance (27.4 meters), and no line-of-sight requirement.
Additionally, based on an in-depth investigation of
how PFC-induced switching noises relate to a computer's power consumption,
\ouralg adds the following two distinguishing features to the literature.

$\bullet$ \emph{Indirect power measurement}. \ouralg uses {indirect} power measurement that does not require any tampering of the building's power network and hence is more stealthy. Here, being {indirect}
means that the target computer's
current does not directly flow through the attacker's sensing device at the receiver; instead,
the receiver only measures voltage signals containing PFC-induced switching noises which we find are correlated with the target computer's power/current. Nonetheless, the existing power-based data exfiltration attacks rely on direct power measurement and hence are less stealthy \cite{Security_AirGap_Mobile_USB_Charging_ACNS_Mobile_CovertChannel_Spolaor2017NoFC,Security_AirGap_Electric_PowerHammer,Genkin:2016:EKE:2976749.2978353}: a power meter is directly connected to the outlet or  a sensing apparatus is placed along the cable directly powering the target device.

$\bullet$ \emph{Simultaneous data exfiltration.} We identify the (approximate) orthogonality nature of PFC-induced
switching noises in practice, thus allowing simultaneous data exfiltration from multiple
computers to a single receiver without much interference.
Thus,
if multiple computers within the same power network are infected, only a single receiver is needed
to exfiltrate data from these computers in parallel. This results in a higher overall exfiltration rate due to the multiple parallel data streams.

More concretely, we focus on an enterprise environment, with the goal
of stealthy data exfiltration from a desktop computer.
Note that, we do not target military-grade systems that have sophisticated and expensive defense against information leakages (e.g., TEMPEST \cite{tempestGoodman2001}).
We first observe that
the amplitude information of a computer's electric current, and hence the power consumption,
is contained in the voltage at any other power outlet connected
to the same building's power network.

In practice, however, it is very challenging to directly extract the current amplitude of the target computer from the voltage measurement,
which
consists of a blend of current amplitude information of all the devices within the same power network (Section~\ref{sec:covert_channel_overview_power_network}). This is further compounded by the power grid's random voltage fluctuations which can be several order-of-magnitude larger than the voltage variation caused by current amplitude variations.

We find that all desktop computers today are mandated to have built-in power
PFC circuits in their power supply
units to reduce harmonics \cite{pfcHandbook,EnergyStar_Computer_Requirement_PFC_V7_2018,iecRegulation}.
Importantly, these PFC circuits result in prominent high-frequency current ripples
between 40kHz and 150kHz \cite{pfcHandbook}, whose amplitude changes with the computer's power consumption --- the higher power consumption, the taller ripples, and vice-versa. These high-frequency current ripples also produce high-frequency voltage ripples at other power outlets, which are referred to switching noises. Thus, by properly filtering the received voltage signals at a power outlet, switching noises can be retained and the receiver is able to successfully extract information about
the transmitter's modulated current amplitude.
Further, the switching noise frequencies are typically different for different computers
and hence are not subject to much interference.
This orthogonality of switching noises allows
simultaneous data exfiltration attacks from multiple computers
by a single receiver.

We present an end-to-end design of \ouralg.
Like
any normal programs, \ouralg only uses the transmitter's CPU resource without
any special privilege (e.g., Kernel access). Any device with data storage, such
as a laptop and a cellphone, as well as an added analog-to-digital
converter (ADC, to digitize the voltage signals)
plugged into a power outlet can be used as a receiver.
Moreover, \ouralg does not require any offline training or calibration using supervised classification algorithms.

To demonstrate the practical applicability of \ouralg,
we run experiments on seven computers with different configurations and vendors in four different labs/offices in two separate buildings.
We also achieve successful exfiltration even when four computers send data simultaneously to a single receiver.
We show that \ouralg achieves successful data exfiltration
with an effective rate of up to 28.48 bits/second, which
is reasonably high compared to many existing covert
channels  \cite{Security_AirGap_Bridgeware_ACM_Communications_2018_Guri:2018:BAM:3200906.3177230,Security_AirGap_EMI_GSMem_Security_2015,Security_AirGap_Mobile_USB_Charging_ACNS_Mobile_CovertChannel_Spolaor2017NoFC}.
More importantly, the receiver can be located in a different room
approximately 90 feet (27.4 meters) away from the transmitter.
Finally, we also present a set of possible defense
mechanisms, such as installing noise filters, and discuss
their limitations.

\section{Features of NoDE and Preliminaries}

In this section, we first discuss \ouralg's distinguishing features and advantages under its own subclass of power analysis-based side/covert channel attacks. Then, we provide a note
on the current PFC design in power electronics.

\subsection{Power Analysis-based Attacks}\label{sec:power_analysis_based_attacks}

\ouralg falls under the power analysis-based attacks, and we identify the key differences of \ouralg from the existing works.
With power usage information
of the victim, prior studies have achieved secret key extraction from smart cards and mobile devices \cite{mangard2008power, Genkin:2016:EKE:2976749.2978353, Camurati:2018:SCE:3243734.3243802}, anomaly detection in embedded systems \cite{clark2013wattsupdoc,Liu:2016:CET:2976749.2978299, park2018power}, tracking websites \cite{clark2013current,Security_AirGap_Mobile_Charging_WebsiteFingerprinting_Powe_Side_Channel_Yang2017OnIB}, among others. Besides
the orthogonal context and objective,
our work stands apart from the prior studies in the following key aspects.

\begin{figure}[!t]
	\centering
	\subfigure[Direct vs indirect  power measurement]{\label{fig:limitation_of_direct_power_new}\includegraphics[width=0.2\textwidth,page=2]{figures/limitation_of_direct_power}}\hspace{0.1cm}
	\subfigure[Direct sensing device locations]{\label{fig:limitation_of_direct_power}\includegraphics[width=0.26\textwidth,page=1]{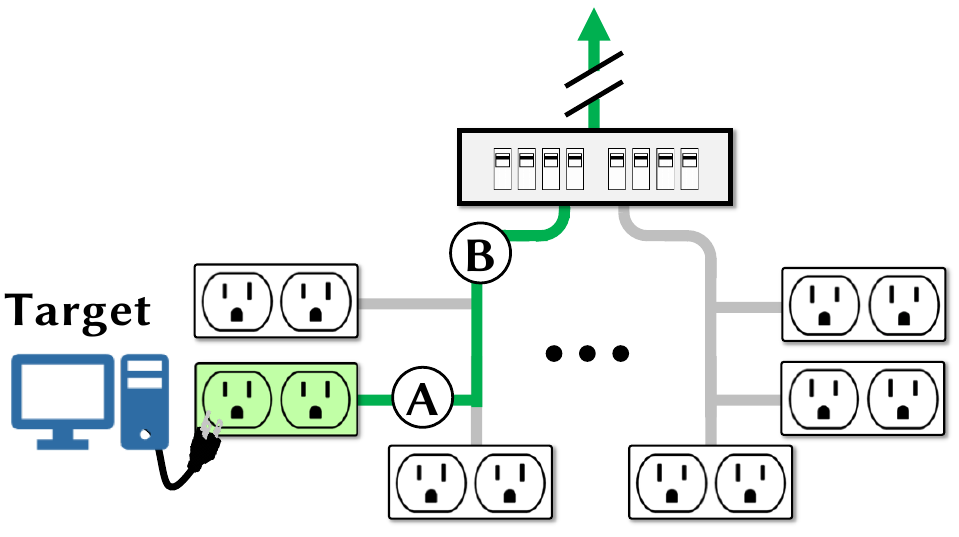}}
	\caption{(a) \textcircled{\raisebox{-.9pt} {1}} is direct measurement using a sensing coil or register placed
		along the direct power path . \textcircled{\raisebox{-.9pt} {2}} is indirect voltage sensing. (b) Possible direct sensing device locations, such as \textcircled{\raisebox{-.9pt} {A}}
		and \textcircled{\raisebox{-.9pt} {B}}. However, location \textcircled{\raisebox{-.9pt} {B}} measures superimposed power consumption of multiple outlets.}
\end{figure}

\textbf{Power measurement.}
A prominent assumption made by many existing
attacks is that the target system's power consumption (or current) can be directly
measured \cite{clark2013current,Security_AirGap_Mobile_Charging_WebsiteFingerprinting_Powe_Side_Channel_Yang2017OnIB}. Nonetheless,
this can only
be accomplished if the sensing apparatus is placed directly
inside the target system or at the nearest power outlet (indicated
by Fig.~\ref{fig:limitation_of_direct_power_new} and \textcircled{\raisebox{-.9pt} {A}}  in Fig.~\ref{fig:limitation_of_direct_power}).
Moreover,
the way that  the transmitter's current is measured
(e.g., measuring current at \textcircled{\raisebox{-.9pt} {B}} in Fig.~\ref{fig:limitation_of_direct_power}) \cite{Security_AirGap_Electric_PowerHammer}
results in a very low bit rate because of strong interference
from other computers and/or devices.

By sharp contrast, \ouralg  can collect
voltage signals from \emph{any} outlet within the same power network
as the transmitter, and identifies and focuses on a unique frequency
band 
for each transmitter, thus achieving
stealthy data exfiltration (even simultaneously from multiple transmitters).

\textbf{Offline training.}
The studies on side channel attacks using power analysis are prolific, e.g., recognizing TV content out of voltage signal measurement
from a nearby outlet \cite{UWashington_TV_CCS_2011_Enev:2011:TVP:2046707.2046770}
and identifying human gestures using body-induced electric signals \cite{UWashington_PowerNoiseBodyAntenna_CHI_2011_Cohn:2011:YNM:1978942.1979058}.
Nonetheless, these studies typically apply supervised machine learning models and need to extract a set of features from the collected power signals and compare them against a set of pre-recorded patterns  for recognition.
Thus,
offline model training on the target system under a controlled environment is required. Clearly,
this is not feasible in our context for stealthy reasons.

Our design is fundamentally different and does not need to match received
signals against pre-trained patterns, which can significantly differ
from runtime conditions. Instead,
to receive new information from the target computer
on the fly, \ouralg adapts its 1/0 bit decision threshold
and filter's passband based on a small pre-determined pilot sequence on a frame-by-frame basis.

\textbf{Simultaneous transmission using orthogonal signals.}
In addition to the intrusive nature, direct power measurement requires dedicated power meters (i.e., receivers) for each target computer. Placing a power mater at a higher level in the power network (e.g., at \textcircled{\raisebox{-.9pt} {B}} in Fig.~\ref{fig:limitation_of_direct_power}) to capture multiple transmitters does not work as the current signals from the transmitters as well as other equipment are superimposed in the readings and hard to distinguish. While \ouralg also collects a mix of signals from all equipment, 
it can still separate each transmitted signal in the high frequency (10kHz$\sim$150kHz) using the insight that PFC circuits generate orthogonal high-frequency switching noises (Fig.~\ref{fig:spectrum_voltage_multi}).

While some side
channel attacks directly look at the current amplitude
\cite{clark2013current,Security_AirGap_Mobile_Charging_WebsiteFingerprinting_Powe_Side_Channel_Yang2017OnIB},
the study \cite{Enev:2011:TVP:2046707.2046770} considers a wide
frequency band (e.g., 1$\sim$60kHz) of collected voltage signals.
Thus, when other appliances or multiple victims are present, strong interference can be produced.
For example,
the attack is demonstrated on only one TV in a home or small lab 
with a relatively clean power network \cite{Enev:2011:TVP:2046707.2046770}.
By contrast,
for each transmitter, we precisely extract the frequency feature over a proper narrow band (e.g., 60Hz) without
strong interference from other devices, thus achieving simultaneous
data exfiltration from multiple transmitters in a large lab with about 30 active computers (Section~\ref{sec:multi_TX}).

\OurAlg also differs from \cite{Shaolei_Colo_Voltage_CCS_2018} which
utilizes a wider frequency band (i.e., 1kHz or more) and estimates a data center-wide \emph{aggregated} power consumption over
a much lower time resolution (i.e., once every minute) for load injection attacks. Whereas, \ouralg is specifically designed to detect individual power consumption at a time resolution of milliseconds.



%

\subsection{NoDE and TEMPEST} 


Electromagnetic emission has long been known as a major source of information leakage \cite{highland1986random}. Notably, the eavesdropping on electromagnetic radiation (EMR) of communication equipment in the early-mid 1900s leads to the development of, to date partially classified, defense technologies code-named TEMPEST \cite{tempestGoodman2001}. The recently declassified TEMPEST defense imposes stringent restrictions on electromagnetic radiation from computer systems both over the air and through power lines \cite{tempestGoodman2001}. However, TEMPEST defense, shielding equipment and/or Faraday caging, is very expensive and mainly used in military application. For example, NATO countries spend billions of dollar of their defense budget each year for TEMPEST shielding \cite{Anderson99softtempest}. While the defense requirement is outlined, the technology for TEMPEST attacks still remains classified. There is skepticism on the feasibility of TEMPEST attacks \cite{tempestForbes2000,tempestGoodman2001}, not to mention the sophisticated sensing equipment necessary to carry out such an attack. Consequently, typical enterprise environments (e.g., companies), which are our target systems, may not necessarily adhere to the costly TEMPEST defense practices.

A relatively inexpensive alternative to TEMPEST using software, named Soft-TEMPEST is proposed in \cite{kuhn1998soft}. Soft-TEMPEST targets obsolete CRT monitors and mainly reduces the electromagnetic emission distance, whereas \OurAlg utilizes power networks for data exfiltration at a distance. 

\subsection{Power Electronics}

It is well known in
the field of power electronics that the high-frequency switching operation
in PFC circuits produces voltage noises
and is fundamental for improving the power factor of appliances \cite{pfcHandbook,PFC_frequency_analysis_VT_zhang2001evaluation}.
The existing research on  PFC designs is primarily
from the \emph{efficiency} perspective, e.g., how
to select the conduction mode and switching frequency to achieve
the best energy efficiency of power supply units
and meet regulation compliances \cite{pfcHandbook}.
In sharp contrast, there is much less understanding
from the \emph{adversarial} perspective:
\emph{how can the PFC-induced switching noises be exploited as
	useful signals for stealthy data exfiltration attacks?}
\ouralg fills
the gap by performing an in-depth study of how the
switching noise amplitude relates to  a computer's power consumption
and uniquely transforming switching noises of a computer's power
supply unit into data-carrying signals for data exfiltration.

\section{Threat Model}\label{sec:threat_model}

We consider a broadly-interpreted enterprise environment (such as company and
university) which is a primary target for data theft \cite{Security_Data_Theft_Wiki,Security_Data_Theft_Enterprise_Survey_cheng2017enterprise}. We focus on a desktop computer,
because it is the predominant type of computer (especially
for storing important data) 
and its built-in PFC circuit is suitable for information
transfer
(Section~\ref{sec:covert_channel_extracting_pfc}). 
For brevity, we also use ``transmitter'' or
``computer'' to denote the transmitting desktop computer.

\begin{figure}[!t]
	\centering
	\includegraphics[trim=0cm 14cm 11.5cm 2.8cm,clip,  width=0.48\textwidth,page=1]{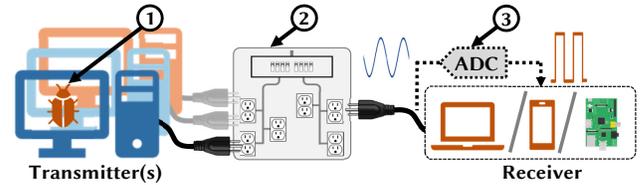}
	\caption{Threat model. \raisebox{.5pt}{\textcircled{\raisebox{-.9pt} {1}}} Modulation
		program in the transmitter.
		\raisebox{.5pt}{\textcircled{\raisebox{-.9pt} {2}}} Building's power network. \raisebox{.5pt}{\textcircled{\raisebox{-.9pt} {3}}} An analog-to-digital converter inside an innocuous-looking device.}\label{fig:threat_model}
\end{figure}

Our threat model is illustrated in Fig.~\ref{fig:threat_model},
including one or more transmitters and a receiver.
Both the transmitters and receiver are connected to the
same building's power network.
Note that, because typically there are filters 
between different buildings' power networks, even dedicated PLC adapters cannot reliably communicate across different buildings \cite{PowerLineCommunications_Survey_JSAC_2016_Cano:2016:SAP:2963140.2963846}.

\textbf{Transmitter.} A transmitter is a desktop computer infected
by malware that intends to send sensitive information (e.g.,
password and financial information) to the outside without using any network or removable storage. 
Like in the existing literature on covert channels,  
our threat model builds upon the malware's capability of obtaining sensitive
information, and is not intended for sending large files due to rate limits.
The malware can use the transmitter's CPU like any normal programs, but no special privileges
are assumed by \ouralg.
Importantly,
while it leverages the building's power network,
\ouralg does not need to intrusively install a new or use an existing dedicated PLC adapter in the target transmitter.

\textbf{Receiver.} The receiver can be any innocuous-looking device that is plugged into an outlet
in the same building's power network as the transmitter. The receiver
needs an ADC for digitizing its received voltage, which is universally used by any
signal-collecting digital systems (e.g., digital temperature sensor) and can be easily hidden
inside a laptop/cellphone charger. Moreover, the receiver
can be located in a distant room different than the transmitter. 
Thus, there is no prohibitive requirement for a receiver.
Even though the building is solely occupied
by a single enterprise, guests are typically still allowed to plug
their 
laptops or cellphones into an outlet for charging.

\textbf{Practicality of malware injection.}
While there are various ways that malware can get into a computer, we classify them into two categories --- easy and hard --- based on the level of difficulty and effort required for malware infection.

$\bullet$ \emph{Easy}:
We consider the malware injection ``easy'' when the target computer can take input from the outside through external media and/or networks. For example, the target computer can get infected with malware when it visits malicious webpages or is connected to an effected USB drive, during which malware can be implanted without being noticed.
It is the predominant approach of malware injection
for today's enterprise environments \cite{Security_Threat_Survey_SANS_2017}.
Not to mention those malware injection incidents affecting
millions of Internet users every day \cite{Security_Malware_ThreatReport_McAfee_March_2018,Security_Threat_Survey_SANS_2017},
a  striking example of infecting a mission-critical system is that the malware Stuxnet infected
Iran's nuclear program through an infected USB drive \cite{langner2011stuxnet,stuxnetRealStoty}.

$\bullet$ \emph{Hard}:  We consider that the malware injection is ``hard'' when
there is no easy approach to malware injection.
For example, the target computer can be almost completely isolated from all external networks, which is also known as ``air-gaping'' \cite{Security_AirGap_Bridgeware_ACM_Communications_2018_Guri:2018:BAM:3200906.3177230}.
In such a scenario, malware can still be injected by exploiting hardware/software backdoors
throughout the supply chain. For example, it has been recently reported that
some microchips were added to servers' motherboards during the manufacturing process without the knowledge of a major server vendor \cite{superMicroImplant_bllomberg}. In other instances,
ShadowPad was implanted into a software program developed by a third-party vendor, affecting hundreds of large businesses \cite{Security_Backdoor_Discovery_Kaspersky_2017}, while malicious
batteries \cite{Security_AirGap_Mobile_Malicious_Battery_Technion_UTAustin_2018} and other hardware Trojans \cite{Security_HardwareTrogan_Survey_2010_Tehranipoor:2010:SHT:1724965.1725002} are all known threats to data security. An even more striking example is that over 100,000 computers
that had \emph{never} been connected to any network were also implanted with information-stealing hardware Trojans by using a classified technology \cite{Security_AirGap_NSA_100000_Implant_Hardware_NYTimes_2014}.

A complete data exfiltration attack involves two major steps. The first step is to inject a malware program into the target computer and collect sensitive information.
The second step is to establish a covert communication channel to send sensitive data to the receiver.
Like the existing covert channel literature \cite{Security_AirGap_Acoustic_MOSQUITO_ArXiv_2018,Security_AirGap_Bridgeware_ACM_Communications_2018_Guri:2018:BAM:3200906.3177230,Security_AirGap_Electric_PowerHammer,Security_AirGap_Mobile_Malicious_Battery_Technion_UTAustin_2018}, our work focuses on the step of establishing a covert communication channel.
Thus,
we can embed \ouralg into an existing information-collecting
malware.
Concretely, we describe the procedure of integrating \ouralg with malware created by the recently released malware toolkit L0rdix \cite{malwareToolL0rix_1,malwareToolL0rix_2} as follows. The L0rdix toolkit comes with a variety of pre-built and configurable functionalities. It can steal information from the victim system by collecting login information, browser cookies, and files matching pre-configured extensions. It can also monitor clipboard content to steal data matching predefined strings. L0rdix toolkit offers botnet capabilities such as opening a specific URL, execute commands, kill processes, upload and download files and run executable files. L0rdix also comes with USB infection, mining, detection prevention,  anti-analysis, and anti-VM capabilities as well.  In our context, we leverage the malware's existing capabilities of system infiltration and information collection, and utilize its botnet functionality with our covert channel communication. We can either attach our code with L0rdix malware or configure L0rdix to download our program as malicious code. Other malware generators such as Senna Spy FTP, which spread as a Trojan bundled with free software, can also be used to secretly download the code of \ouralg into an infected system \cite{toolSennaSpy,toolSennaSpy2}.

\section{Power Network as a Covert Channel}\label{sec:covert_channel}

In this section,
we present our covert channel residing in
a building's power network.

\subsection{Overview of a Building's Power Network}\label{sec:covert_channel_overview_power_network}

\begin{figure}[!t]
	\centering
	\includegraphics[trim=0cm 8cm 2.3cm 0cm,clip,  width=0.48\textwidth,page=2]{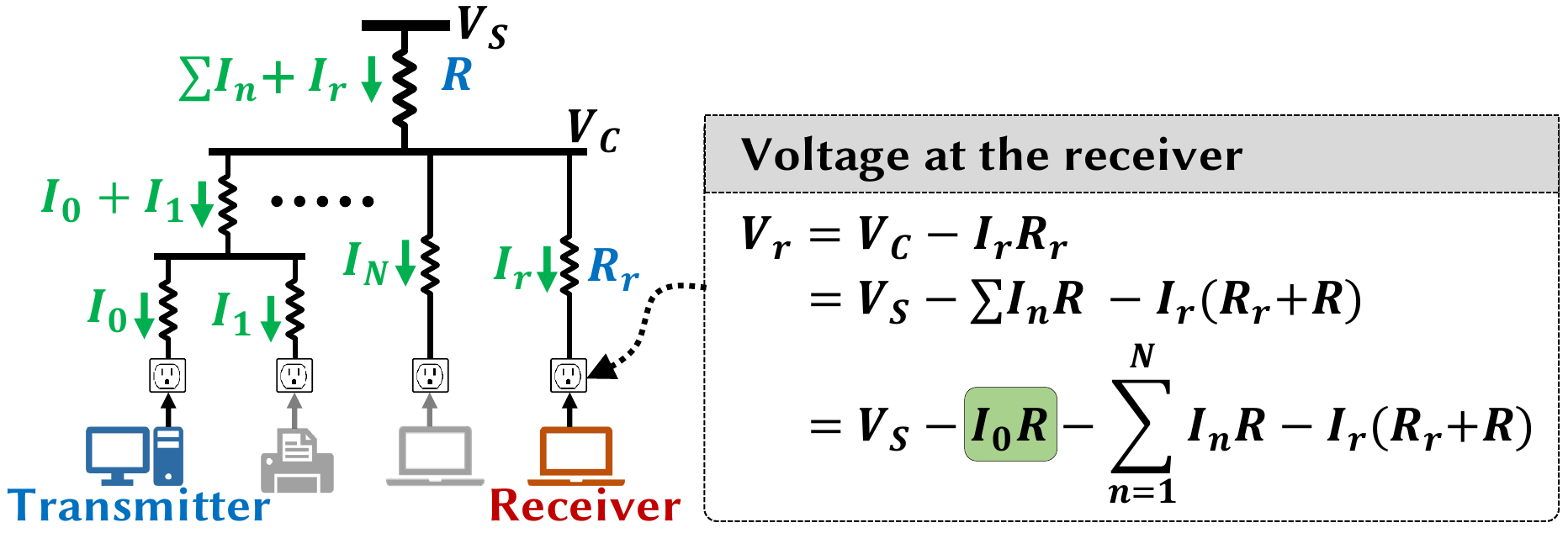}
	\caption{Illustration of a building's power network topology showing possible locations of transmitter and receiver. The transmitter current $I_0$ is embedded in receiver's voltage $V_r$. }\label{fig:power_network_circuit}
\end{figure}

The utility power typically enters a building through a single point.
Then, through a distribution box,
power is split to different floors/rooms in parallel and finally to different power strips/outlets.
For large buildings, a multi-level distribution hierarchy may be used.
Further, the utility may provide three-phase power to buildings, and the three phases
can be divided 
depending on functions (e.g., offices
on one phase, and the central air conditioner on
another) or physical topologies. 
For illustration, Fig.~\ref{fig:power_network_circuit} provides a simplified example  of a single-phase power network of a building, highlighting
the parallel connection of different power outlets.
The common source voltage $V_S$ enters a distribution/panel box and
reduces to $V_C$ due to a voltage drop caused by the line resistance. Then, the voltage
$V_C$ is supplied to different
rooms/outlets.
We can write the receiver's voltage signal $V_r $ as
\begin{equation}\label{eq:voltage_equation_overall}
V_r = V_C - I_r R_r = V_S - I_0R - \sum_{n=1}^N I_n R - I_r (R+R_r),
\end{equation}
where $R$ is the resistance of the common line from which all currents flow,
and $R_r$ is the resistance of the line directly supplying power to the receiver.
Clearly, the receiver's voltage signal $V_r$ contains the transmitter's current denoted
by $I_0$ in Fig.~\ref{fig:power_network_circuit}, whose amplitude
can be modulated by varying the CPU load to carry information (Section~\ref{sec:design_transmitter}).
Thus, if the receiver is able to  extract $I_0R$ out of its signal $V_r$, it can exfiltrate
information from a computer through the power network.

\subsection{Computer's Power Supply Unit}

We now look at the anatomy of a computer's power supply unit
and identify an important component --- PFC circuit ---
which generates
high-frequency current ripples that can be detected by the receiver.

\subsubsection{\textbf{A closer look at computer's power supply unit}}

\begin{figure}[!t]
	\centering
	\includegraphics[trim=0cm 11.5cm 7cm 2cm,clip,  width=0.48\textwidth,page=1]{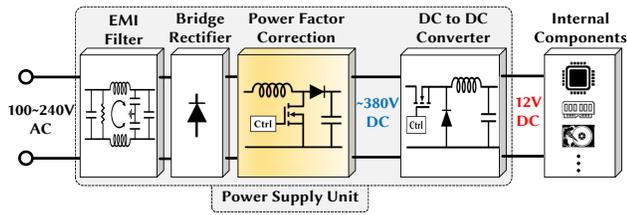}
	\caption{Components of a desktop computer's power supply unit.}\label{fig:computer_PSU_block}
\end{figure}

As shown in Fig.~\ref{fig:computer_PSU_block}, a power supply unit based
on the standard SMPS design draws 100$\sim$240V AC voltage from an outlet and then, after multiple
stages, provides regulated 12V DC voltage to internal components such as CPU.
Specifically, at the front end, there is
an EMI filter to limit frequency components of greater than 150kHz both
coming from and conducted back to the power network, in compliance
with  international regulations
\cite{PFC_EMI_Understanding_TI,PFC_Federal_EMI_ElectronicCode_Website}.
Then, a rectifier  converts the incoming AC voltage to a pulsating DC voltage (unipolar half-sine waves), followed by a PFC circuit which improves the power
factor by regulating
the input current waveform and making it resemble the entering voltage's
sine wave. Fig.~\ref{fig:pfc_server_current_dell} shows a snapshot of
the current waveform drawn by our Dell computer with PFC.
The PFC elevates the voltage to around 380V, which is
stepped down 
and becomes 12V DC voltage
for internal components.

Harmonic distortion is undesirable since it reduces the power factor
and causes unwanted power losses in the power system \cite{pfcHandbook}.
Low-power devices with SMPS (switch mode power supply) are allowed
to have a low power factor without PFC (see Fig.~\ref{fig:pfc_impact_harmonics} in
Appendix~\ref{sec:appendix_current_microsoft} for the current waveform).
Nonetheless, regardless of the actual power consumption,
all devices with a power rating of 75W (applicable for desktop computers)
must have PFC circuits for
mitigating harmonics  as mandated
by international regulations \cite{EnergyStar_Computer_Requirement_PFC_V7_2018,iecRegulation,pfcHandbook}.
Thus, a crucial point we highlight is that
\emph{the PFC requirement for mitigating
	harmonics universally applies to {all} of today's desktop computers}.

\subsubsection{\textbf{Frequency spikes generated by PFC}}\label{sec:covert_channel_pfc_spikes}

\begin{figure}[!t]
	\centering
	\subfigure[Current waveform]{\label{fig:pfc_server_current_dell}\includegraphics[trim=0cm 0cm 0cm 0cm,clip,  width=0.23\textwidth,page=1]{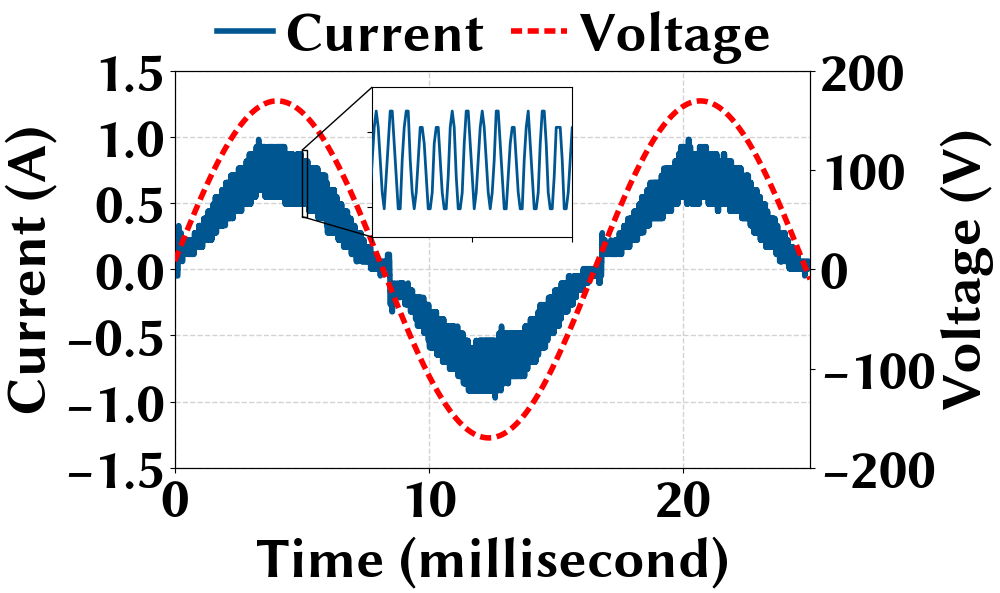}}\hspace{0.2cm}
	\subfigure[Frequency analysis of current]{\label{fig:pfc_with_psd_dell}\includegraphics[trim=0cm 0cm 0cm 0cm,clip,  width=0.23\textwidth,page=1]{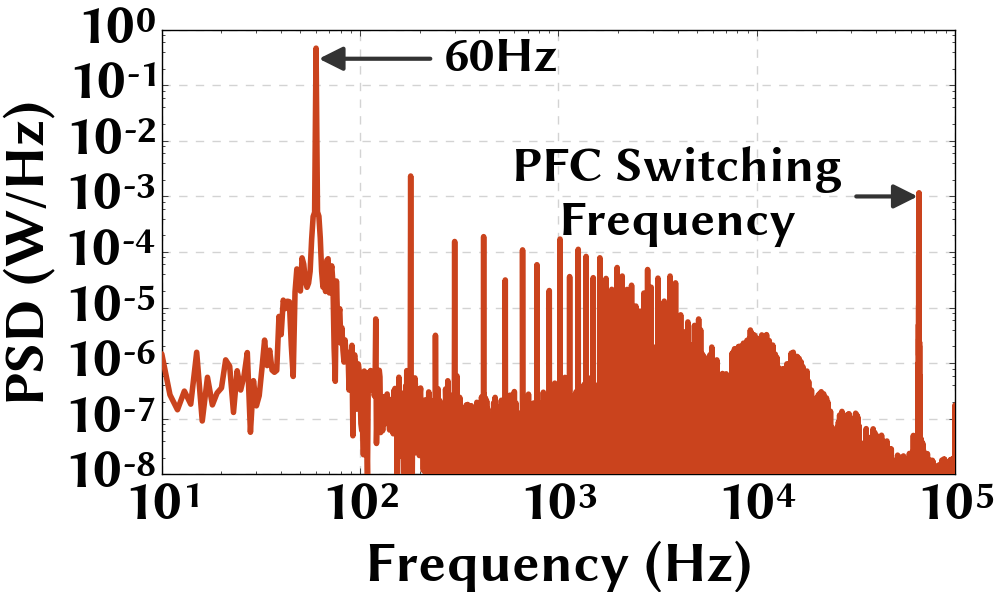}}
	\caption{(a) Current drawn by our Dell PowerEdge computer. (b) The current has a high-frequency
		spike due to PFC operation under CCM.}\label{fig:pfc_with_dell_summary_label}
\end{figure}

As shown in Fig.~\ref{fig:pfc_server_current_dell}, while improving power factor,
the PFC circuit also produces
high-frequency current ripples due to its working principle \cite{pfcHandbook}.
Specifically, the core of a PFC circuit is rapidly switching the incoming current between two modes --- a rising mode where the current  increases, and a falling mode where the current decreases.
Through switching, the current drawn from the power
outlet resembles a sine waveform following the voltage signal.
The switching frequency is determined by a controller as well as the PFC
components (e.g., inductor), and typically falls into the range of $<$40kHz, 150kHz$>$,
which is not subject to EMI regulations that set limits on frequencies greater than 150kHz \cite{pfcHandbook,PFC_Federal_EMI_ElectronicCode_Website}.

There are three basic modes for PFC switching (a.k.a., conduction mode) as summarized
in Table~\ref{table:pfc_designs} in Appendix~\ref{sec:appendix_conduction_mode}.
In practice, vendors may adopt  proprietary designs using variants
of the basic modes \cite{pfcHandbook}.
Naturally, the PFC's switching operation
results in high-frequency current ripples, thus generating a PSD (power spectrum density) spike
around the switching frequency.
Illustrative current waveforms
and frequency analysis results are shown in Table~\ref{table:pfc_designs}.
Note that when PFC is not used, there exist no such PSD spikes
within $<$40kHz, 150kHz$>$
(see Fig.~\ref{fig:pfc_without_psd_log}).

Desktop computers' power supply units almost all
have a rated capacity of over 300W to accommodate
extensibility. Thus,
the continuous conduction mode
(CCM) is most widely-used due to its low peak current.
We show the current and its frequency analysis
for our Dell computer in Fig.~\ref{fig:pfc_with_dell_summary_label}.
It can be
observed that with PFC, the harmonics are an order of magnitude smaller
than the 60Hz component.
Furthermore, the current waveform and frequency analysis match
with those illustrative figures for CCM in Table~\ref{table:pfc_designs} in Appendix~\ref{sec:appendix_conduction_mode},
clearly  showing a prominent high-frequency PSD spike generated by the PFC circuit.

\begin{figure*}[!t]
	\centering
	\subfigure[Transmitter's CPU load and current amplitude]{\label{fig:powerVoltage_systemLoad}\includegraphics[trim=0cm 0cm 0cm 0cm,clip,  width=0.23\textwidth,page=1]{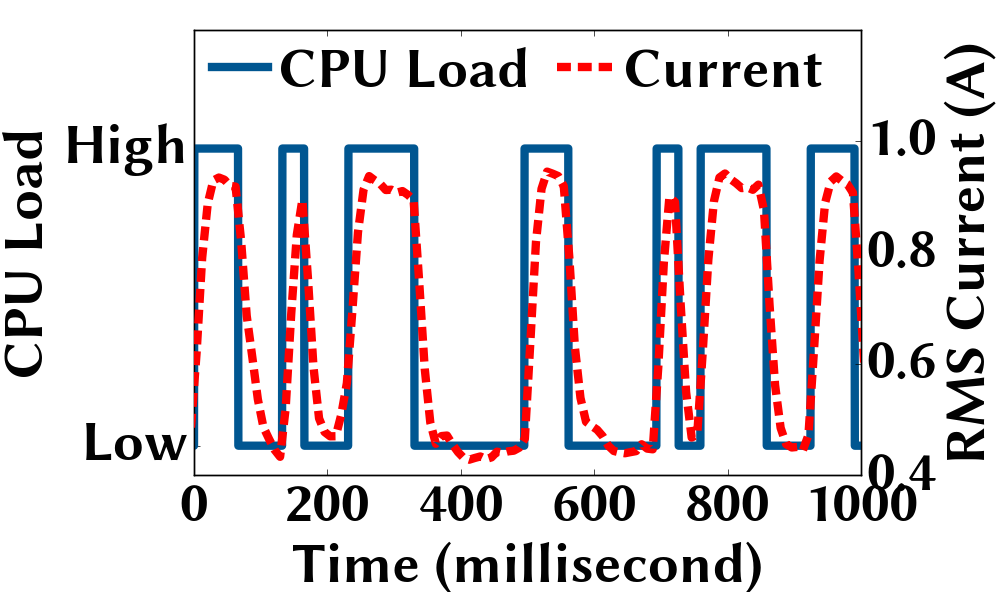}}\hspace{0.2cm}
	\subfigure[Frequency analysis of the received voltage signal]{\label{fig:powerVoltage_psd}\includegraphics[trim=0cm 0cm 0cm 0cm,clip,  width=0.23\textwidth,page=1]{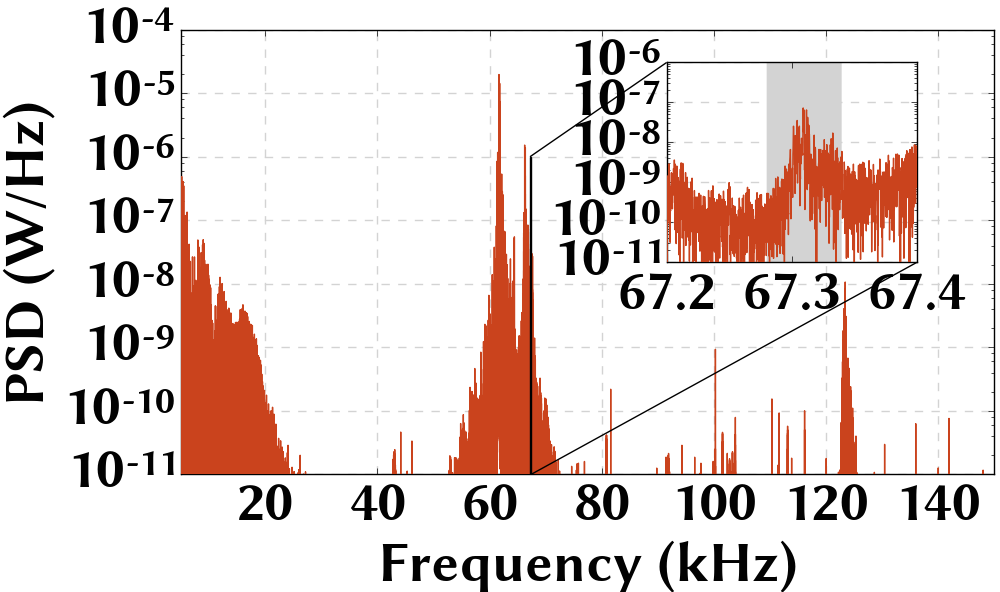}}
	\subfigure[Frequency spectrum of the received voltage signal]{\label{fig:powerVoltage_spectrum}\includegraphics[trim=0cm 0cm 0cm 0cm,clip,  width=0.23\textwidth,page=1]{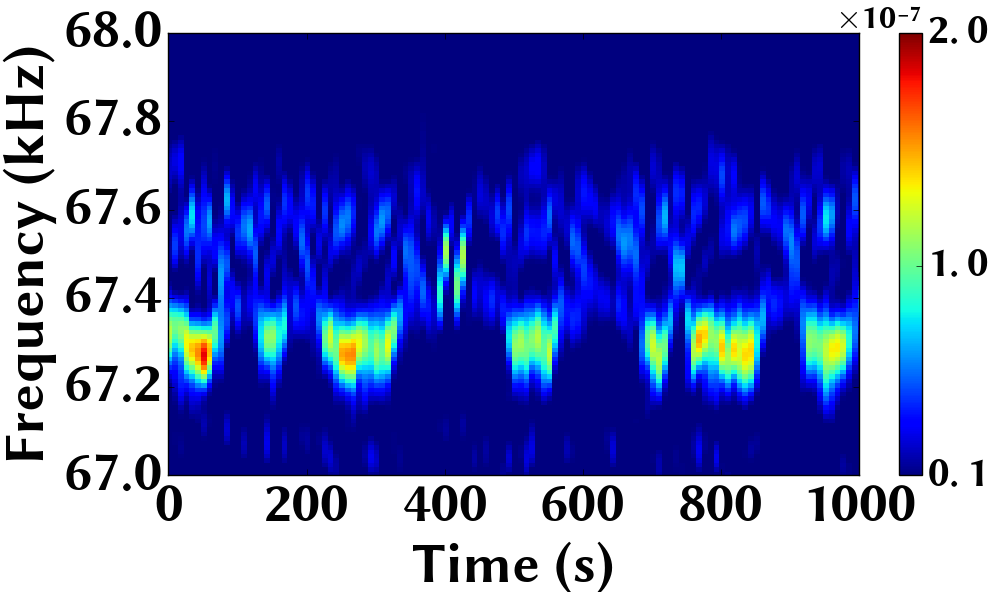}}\hspace{0.2cm}
	\subfigure[Filtered voltage signal and transmitter's current]{\label{fig:powerVoltage_voltage_current_amplitude}\includegraphics[trim=0cm 0cm 0cm 0cm,clip,  width=0.23\textwidth,page=1]{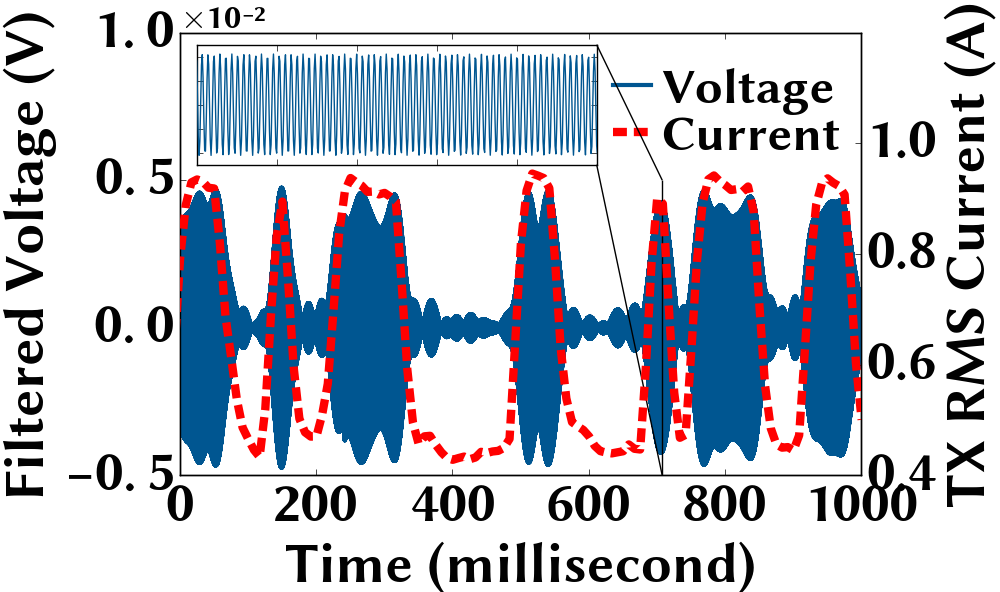}}
	\caption{Experiment in Lab \#1 with transmitter and receiver separated by 55ft.
		By applying a filter with passband of  $<$67.28kHz, 67.34kHz$>$, the amplitude of filtered voltage
		signals acquired by the receiver matches the transmitter's current amplitude.}
	\label{fig:data_TX_motivation}
\end{figure*}

\subsection{Extracting Transmitter's Current  based on PFC Switching Frequency}\label{sec:covert_channel_extracting_pfc}

The receiver's voltage signal in Eqn.~\eqref{eq:voltage_equation_overall} contains all the frequency components of the common source voltage  and currents
of all connected devices. That is, the high-frequency current ripples generated by PFC circuits
affect  voltage signals at any power outlet within the same power network,
which are referred to as switching noise \cite{pfcHandbook}.
Consequently, if we filter out all but the transmitter's high-frequency switching noises
from the receiver's voltage $V_r$, we are left with the transmitter's switching noise around
its PFC frequency and the switching noise amplitude is also highly correlated
with the transmitter's current $I_0$. This is achieved based on
the (approximate) orthogonality of PFC's switching noises.

Different computers typically have \emph{non}-overlapping PFC switching frequencies. In fact, our experiment
shows that even for computers with the same
configuration and manufactured by the same vendor, different computers
still have (slightly) different PFC switching frequencies due to manufacturing process variations and can simultaneously transmit data without much interference. Furthermore, the prominent PSD spikes
between 40kHz and 150kHz
do not interfere significantly
with harmonics (predominantly less than 20kHz) generated by other devices.
Therefore, in practice, the ripples in the receiver's voltage signal caused by the
target transmitter's switching noise do not suffer from significant interferences from other sources.

We empirically validate the feasibility of extracting the transmitter's current amplitude information. 
Our experiment is conducted in a lab with 30+ computers, where the transmitter
and receiver are plugged into two outlets located about 55 feet away from each other.
The details of the setup are presented in Section~\ref{sec:experiment_methodology}.
We vary the transmitting computer's current by varying
its CPU load  because compared to components such as hard disk and memory
chips, a computer's CPU has a high dynamic power that can be easily
adjusted by loading/unloading the CPU. Moreover, CPU is ubiquitously available in all computers
and needed by any running program.
As GPUs are power-consuming, a computer's current can also be significantly varied in a similar fashion by using increasing the utilization of a dedicated GPU. 
Note, however, that a GPU is less ubiquitous compared to a CPU, especially in ordinary office environments. Thus, throughout our study, we will only utilize CPU to change a computer's current for data exfiltration.


We show in Fig.~\ref{fig:powerVoltage_systemLoad} the transmitter's CPU load and current amplitude, which match with each other quite well. Then, we perform a frequency analysis
of the received voltage signal and show the result in Fig.~\ref{fig:powerVoltage_psd}.
We  see large frequency components between 40kHz and 80kHz (and sporadic higher-frequency spikes). These
are mainly due to different computers' PFC switching operations,
and the components around 67.3kHz are caused by our transmitter. 
The temporal variation of the PSD spikes created by the transmitter is shown in the frequency spectrum in Fig.~\ref{fig:powerVoltage_spectrum} where we can easily identify the transmitter's high current periods. 
Next, we filter the collected voltage signal
with a passband of $<$67.28kHz, 67.34kHz$>$ and show the filtered voltage
signal in Fig.~\ref{fig:powerVoltage_voltage_current_amplitude} where the filtered voltage signal resembles the current ripples (as shown in the zoom-in window).
The filtered voltage amplitude 
is close to zero during the low current periods because the PSD spikes shifts away from the 60Hz passband.

In summary, we have demonstrated that, with a proper band-pass filter,
\emph{the amplitude of the receiver's filtered voltage signal
	can 
	recover the transmitter's modulated current amplitude
	and hence be exploited for demodulation.}

\section{The Design of \ouralgSection}\label{sec:design}

In this section, we present
the design of \ouralg.
As shown in Fig.~\ref{fig:block_diagram_ouralg},
\ouralg  includes both a transmitter
(i.e., a desktop computer with implanted malware) and a receiver (i.e., any
voltage-collecting device plugged into a power outlet).

\subsection{Transmitter Design}\label{sec:design_transmitter}

Like in the prior literature on covert channels \cite{Security_AirGap_Bridgeware_ACM_Communications_2018_Guri:2018:BAM:3200906.3177230,Security_AirGap_Mobile_USB_Charging_ACNS_Mobile_CovertChannel_Spolaor2017NoFC,Security_AirGap_EMI_GSMem_Security_2015}, 
\ouralg focuses on the \emph{physical} process of data exfiltration ---
converting 1/0 bits into 
current amplitudes and
decoding it from a remote outlet.
That is, the implanted malware already collects
needed information (Section~\ref{sec:threat_model}) and encodes it
into 1/0 bit streams that are ready for {data framing} and {modulation}.
Next, we address the key design issues.

\begin{figure}[!t]
	\centering
	\subfigure[CPU load increase]{\label{fig:power_lag_rise}\includegraphics[trim=0cm 0cm 0cm 0cm,clip,  width=0.23\textwidth,page=1]{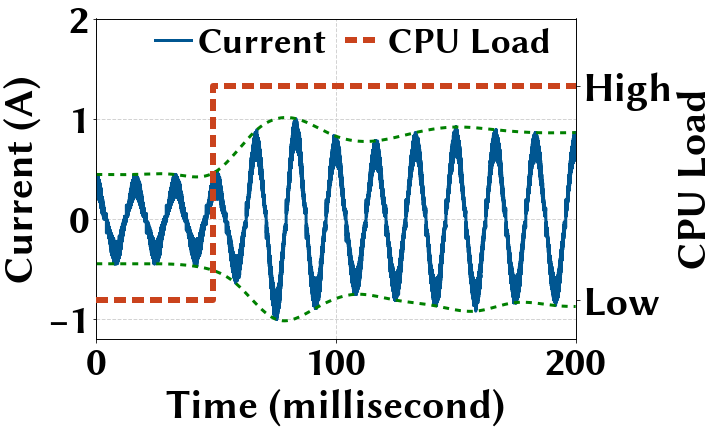}}\hspace{0.2cm}
	\subfigure[CPU load decrease]{\label{fig:power_lag_fall}\includegraphics[trim=0cm 0cm 0cm 0cm,clip,  width=0.23\textwidth,page=1]{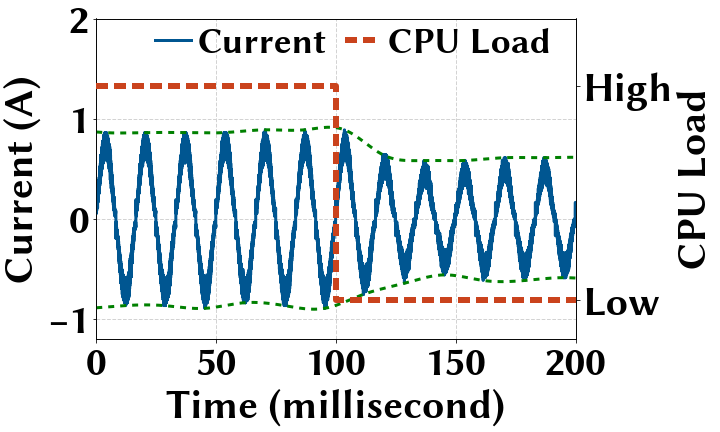}}
	\caption{Input current's response to CPU load changes. Due to power supply unit's internal control, the current does not change immediately after change in the CPU load.} 
	\label{fig:power_lag}
\end{figure}

\subsubsection{\textbf{Choosing symbol rate}}\label{sec:symbol_rate_inrush_current}

\ouralg  modulates the transmitter's current amplitude by varying the CPU load based on 1/0 bit values.
Thus, the achievable symbol rate
crucially depends on how fast
the current amplitude
changes in response to the CPU load.
While the CPU usage can be adjusted within a millisecond
or even faster \cite{UMich_Wenisch_PeakPowerModel_Server_SMPS_ISLPED_2010_Meisner:2010:PPM:1840845.1840911},
the current drawn by a computer may not instantly follow (i.e., lags) the CPU usage.
In our experiment with a Dell PowerEdge computer in Fig.~\ref{fig:power_lag_rise} shows that the current takes about 20$\sim$30 milliseconds to reach the peak after jump in the CPU load at around 50 milliseconds. In addition, because
of the feedback control adopted by a power supply unit, a sudden
CPU load change can create a current inrush followed by a current dip. As shown in Fig.~\ref{fig:power_lag_fall} , the current also lags by around 30 milliseconds when the CPU load sharply changes from high to low.


\begin{figure*}[!t]
	\centering
	\includegraphics[trim=0cm 14cm 1cm 0cm,clip,  width=1.0\textwidth,page=2]{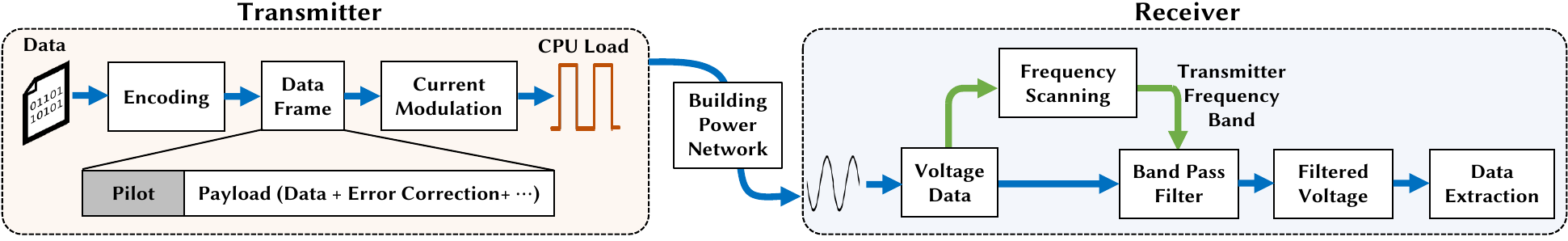}
	\caption{Block diagram of transmitter and receiver in \ouralg. The transmitter sends data to the receiver over the power network. }
	\label{fig:block_diagram_ouralg}
\end{figure*}

Based on our experiment results, we choose 33ms as the default symbol duration.
This means that, \emph{by
	modulating the transmitter's current amplitude,
	the achievable symbol rate in \ouralg is 30 symbols per second.}

\subsubsection{\textbf{Choosing  modulation mode}}\label{sec:design_transmitter_modulation}

For a fixed symbol rate, a higher bit rate can be achieved if each symbol carries more bits.
Here, we run experiments to see how many bits can be successfully mapped
into each symbol: $2^N$ discrete current/power levels are needed for $N$
bits per symbol. For $N=2$, we vary CPU loads at four levels
(0\%, 25\%, 75\% and 100\%, representing ``00'', ``01'', ``10'' and ``11'', respectively)
and see how input current
changes, given three different symbol lengths --- 33ms, 66ms, and 100ms.

\begin{figure}[!t]
	\centering
	\subfigure[Symbol length = 33ms]{\includegraphics[trim=0cm 0cm 0cm 0cm,clip,  width=0.15\textwidth,page=1]{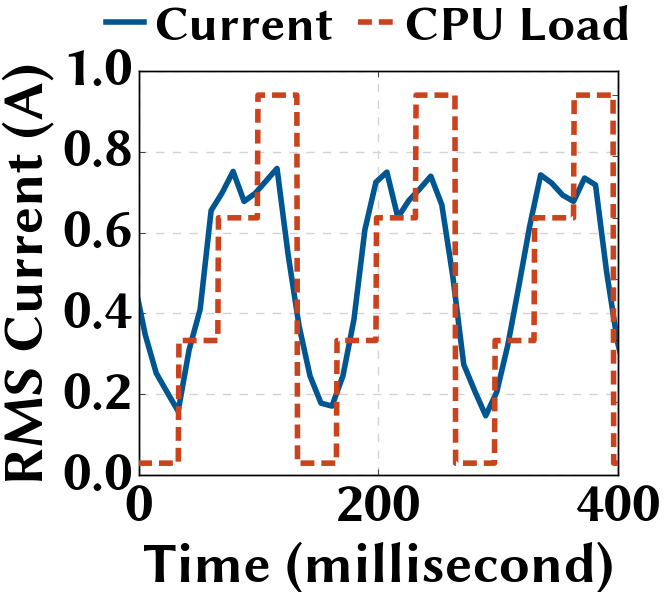}}\hspace{0.1cm}
	\subfigure[Symbol length = 66ms]{\includegraphics[trim=0cm 0cm 0cm 0cm,clip,  width=0.15\textwidth,page=1]{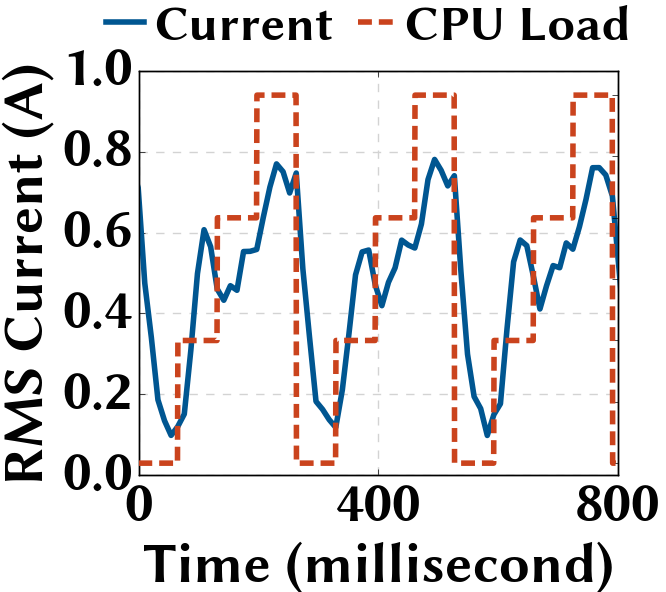}}\hspace{0.1cm}
	\subfigure[Symbol length = 100ms]{\includegraphics[trim=0cm 0cm 0cm 0cm,clip,  width=0.15\textwidth,page=1]{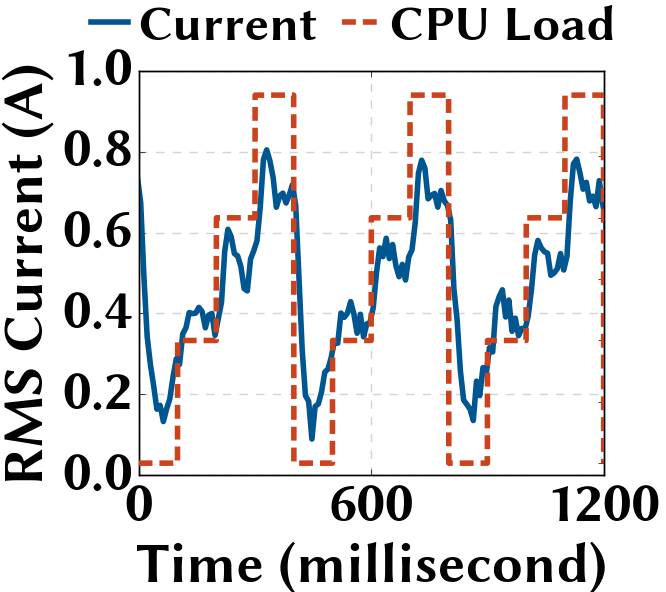}}
	\caption{Experiments for 2 bits per symbol on our Dell PowerEdge computer. With a symbol length
		of 100ms, there is sufficient time for the input current to steadily respond to CPU load changes.}\label{fig:multi_level}
\end{figure}


We show the results in Fig.~\ref{fig:multi_level} and see that when
each symbol lasts 33ms, there is a mismatch between the CPU
load (representing 2 digital bits) and the resulting current amplitude,
thus clearly leading to a very high symbol error rate. As explained in Section~\ref{sec:symbol_rate_inrush_current},
this mismatch is mainly due to the complex feedback control mechanisms and large
capacitors inside a computer's power supply unit.
If the symbol length
increases to 66ms, there is still insufficient time for the current amplitude
to yield a steady response. 
The current amplitude correlates well with the CPU load  when the symbol length
increases to 100ms, but this means that the effective bit rate is only 20 bits/second,
which is even lower than 30 bits/second achieved by a binary modulation with a symbol length of 33ms. With
$N>2$ bits per symbol, there is an even poorer correlation between the CPU load
and input current, unless the symbol length is sufficiently large.

As a result, we choose \emph{binary} modulation 
and use high and low currents to represent ``1'' and ``0'', respectively.
Concretely, we build upon the existing
literature 
\cite{Security_AirGap_Fansmitter_Acoustic_arXiv_DBLP:journals/corr/GuriSDE16,Security_AirGap_Mobile_USB_Charging_ACNS_Mobile_CovertChannel_Spolaor2017NoFC}
and design a simple current modulator as described in Algorithm~\ref{Alg:power_modulator}
in Appendix~\ref{sec:appendix_current_modulation}.
The current modulator program
takes 1/0 bit information as the input and
runs some dummy calculations (e.g., generating random numbers) to load the CPU and change the computer's input current.

\subsubsection{\textbf{Choosing the frame length}}

Like in many communications systems \cite{Goldsmith_WirelessCommunications_2005},
\ouralg groups 1/0 bit sequences into frames, each beginning with a
pilot sequence (Section~\ref{sec:design_pilot_symbol}).
As shown in Fig.~\ref{fig:block_diagram_ouralg},
following the pilot symbols is the actual payload that
contains uncoded bits or coded bits using error correction techniques.

In our context, the transmitter's PFC switching frequency is unknown to the receiver
and may vary over time, albeit slowly. Throughout each frame, however,
the transmitter's PFC switching frequency should remain relatively constant.
%
As shown in the cumulative density function (CDF) and the frequency spectrum in Fig.~\ref{fig:PFC_frequency_variation}, the PFC switching frequency does not vary by more than 50Hz within 5 seconds.
On the other hand,
for a good bit detection, 
the receiver's frequency band for filtering voltage signals can be only 60Hz.
Thus, we conservatively choose a frame length of 100 symbols, resulting in frame duration of 3.3 seconds.


\subsubsection{\textbf{Choosing pilot symbols}}\label{sec:design_pilot_symbol}

In wireless communications, pilot symbols are 
symbols mutually known to both the transmitter and receiver and
inserted at the beginning of each frame, allowing the receiver to estimate
the channel state and synchronize data reception \cite{Goldsmith_WirelessCommunications_2005}.
In our context,
pilot symbols are needed by the receiver to identify
the transmitter's signals 
and  the start of a data frame  (details
in Section~\ref{sec:design_receiver}).
Nonetheless, the length of a pilot sequence needs to be properly chosen.
A too short sequence  may not be enough for the attacker
to accurately acquire the transmitter's PFC switching frequency,
whereas a too long sequence takes up an unnecessarily high overhead.
In our experiments in Section~\ref{sec:experiment_result_different_settings}, we find that a
6-bit pilot sequence (``110010'' in our study) strikes a good balance, while we also examine
4-bit and 8-bit pilot sequences (Table~\ref{table:diff_cases}).
Note that, for some scenarios (e.g., when
there are multiple transmitting computers), a longer pilot sequence
may be needed to uniquely identify each transmitting computer
and better locate the corresponding frequency band.
However, in general, there is no need to change the pilot sequence for different buildings or power suppliers.

To summarize, considering the hardware constraints and behavior of the PFC switching frequency, we choose in our default design a symbol period of 33 milliseconds, 1 bit per symbol, 100 bits in each data frame with a frame duration of 3.3 seconds, and a 6-bit pilot sequence ``110010''.

\subsection{Receiver Design}\label{sec:design_receiver}

On the receiver side,
\ouralg calculates the average amplitude of filtered voltage signals for every bit length and compares it against a detection threshold for deciding a 1/0 bit value.
There are three major steps: (1) finding a passband for the filter;
(2) identifying the start
of a data frame; and (3) demodulating the extracted signals
into 1/0 data bits.

%
%

\subsubsection{\textbf{Finding the filter's passband}}

\begin{algorithm}[!t]
	\caption{Finding a Passband for Filtering Voltage Signals}
	\begin{algorithmic}[1]
		\State Input: Voltage signal $V(t)$ of a pilot length, symbol length $T_b$, pilot sequence $\mathbf{B}_p$, max window  $F_{\max}$ kHz, increment $F_{inc}$ kHz
		\For {$lb=$20 $\sim$ (150-$F_{\max}$) kHz with increment $F_{inc}$}
		\For {$ub=lb+F_{inc}$ $\sim$ $lb+F_{\max}$ with increment $F_{inc}$}
		\State $\tilde{V}(t)_{lb,ub} \leftarrow$ $V(t)_{<lb,ub>}$ \;\;\;//filtering
		\State $E(t)_{{lb,ub}} \leftarrow$ Envelop of $\tilde{V}(t)_{lb,ub}$
		\State Extract bits $\mathbf{B}^*$ from $E(t)_{lb,ub}$ based on $T_b$
		\State Error vector $\mathbf{err}(t)_{lb,ub}$ $\leftarrow$  $\mathbf{B}^*$ XOR $\mathbf{B}_p$ 
		\EndFor
		\EndFor
		\If {$\mathbf{err}(t)_{lb,ub} == \mathbf{0}$}  \;\;//pilot sequence found
		\State Return $<lb,ub>$ 
		\EndIf
	\end{algorithmic}
	\label{Alg:pilot_scan}
\end{algorithm}

In practice, the transmitter's PFC switching frequency
is unknown to the receiver.
Based on a predetermined pilot sequence, we propose a scanning process to find a frequency
passband for filtering received voltage signals and retaining the prominent PSD
spikes generated by the transmitter. The scanning process
is described in Algorithm~\ref{Alg:pilot_scan}, where two nested loops scan through the frequency bands from 20 kHz to 150 kHz with a moving window of variable size.
The notion of $<lb,ub>$ means a band-pass filter
with $lb$ and $ub$ being the lower and upper cutoff frequencies, respectively.
When $<lb,ub>$ appears as a subscript,
it means that a signal passes through a band-pass filter $<lb,ub>$.

In each inner loop, the amplitude/envelop of filtered voltage signal is extracted
and then evenly sliced into bit-length pieces. 
Then, the average amplitude for each piece is used for deciding
binary bit values. Specifically,
we set the
the mean of bit-wise average amplitudes of the filtered signal
as the binary bit decision threshold: a bit-wise average amplitude
higher than the threshold is decoded as ``1'', and ``0'' otherwise. Then,
by comparing the extracted bits with the 
pilot sequence,
we return the passband that yields no errors for pilot detection.

As the PFC switching frequency does not vary significantly over time,
we only need to scan through a wide frequency range (e.g., from 20kHz to 150kHz)
once. \emph{After the initial scan, the receiver
	only needs to quickly fine tune its filter's passband over a much narrower range} (e.g., 500Hz around the previously-found passband, instead of 20kHz to 150kHz) to compensate for runtime switching frequency
offsets. Moreover, provided that voltage signals are acquired and stored,
bit extraction can be done offline,
and hence the scanning complexity is not an issue.
In fact,
with $F_{inc}=0.01$kHz, Algorithm~\ref{Alg:pilot_scan}
only takes less than 5 minutes in Matlab 
on a laptop.

\begin{algorithm}[!t]
	\caption{Identifying Frame Start and Bit Threshold}
	\begin{algorithmic}[1]
		\State Input: Pilot length $T_p$, bit length $T_b$, pilot sequence $\mathbf{B}_p$
		\Loop { at current time $t$}
		\State Obtain filtered voltage $\tilde{V}(t,t-T_p)$
		\State $E(t,t-T_p) \leftarrow$ Signal envelop of $\tilde{V}(t,t-T_p)$
		\State Extract bits $\mathbf{B}^*$ from $E(t,t-T_p)$ based on $T_b$
		\State Error vector $\mathbf{err}(t)$ $\leftarrow$  $\mathbf{B}^*$ XOR $\mathbf{B}_p$ \;\;//bit-wise		
		\If {$\mathbf{err}(t) == \mathbf{0}$}  \;\;//pilot sequence found
		\State Return payload starting time $t$, and bit detection threshold (average of $E(t,t-T_p)$)
		\EndIf
		\EndLoop
	\end{algorithmic}
	\label{Alg:frame_scan}
\end{algorithm}

\subsubsection{\textbf{Identifying the start of a frame}}

While the filter's passband does not vary significantly
due to the slow variation of PFC switching frequencies,
the bit detection threshold depends on the non-controllable environment
(e.g., other computers with similar
PFC switching frequencies can cause interferences)
and hence can change quickly at runtime.
We propose a time-domain scanning process to identify the bit detection threshold
and start of each data frame based on the predetermined pilot sequence.
The scanning process is presented in Algorithm~\ref{Alg:frame_scan}.

\subsubsection{\textbf{Extracting data bits}}
After identifying a pilot sequence, 
the receiver can extract actual payload bits using the bit detection
threshold returned by Algorithm~\ref{Alg:frame_scan}:
if a bit-wise average amplitude of the filtered signal is
higher than the threshold, then the corresponding payload bit is decided as ``1'', and ``0'' otherwise.

\section{Evaluation}

This section presents  experiment results
to validate the practical feasibility of \ouralg,
highlighting 
that \ouralg achieves an effective rate of up to 28.48 bits/second to
a receiver located in another room about 90 feet ($\approx27.4$ meters) away.

\subsection{Methodology}\label{sec:experiment_methodology}



\begin{figure}[!t]
	\centering
	\includegraphics[trim=1cm 4.5cm 11.5cm 3cm,clip,  width=0.45\textwidth,page=1]{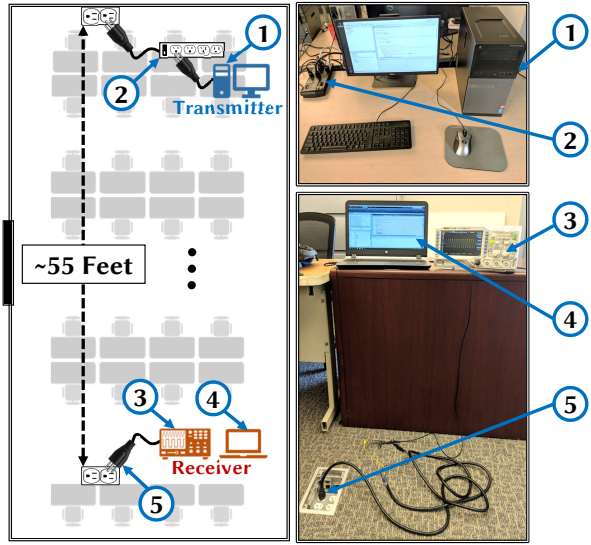}
	\caption{Experiment setup at Lab \#1 in Building~A. \raisebox{.5pt}{\textcircled{\raisebox{-.9pt} {1}}} Transmitter. \raisebox{.5pt}{\textcircled{\raisebox{-.9pt} {2}}} Power strip. \raisebox{.5pt}{\textcircled{\raisebox{-.9pt} {3}}} ADC
		for voltage signal acquisition (oscilloscope in our experiments). \raisebox{.5pt}{\textcircled{\raisebox{-.9pt} {4}}} Laptop for voltage
		filtering and bit detection.  \raisebox{.5pt}{\textcircled{\raisebox{-.9pt} {5}}} Receiver's power outlet.}\label{fig:layout_lab}
\end{figure}

\textbf{Experiment setup.} As listed in Table~\ref{table:diff_configurations}, we test seven computers with different operating systems/configurations, including four recently purchased from two top vendors (Dell and Acer) which
collectively account for more than 25\% of the global PC market
share \cite{Market_PC_IDC_Q2_2018}, an iMac (27 inch), and two custom-built computers with top-brand power supply units (Corsair and EVGA).
They all have CCM type PFC in their power supply units. Our collection of desktops is representative of commonly used models in our target case of modern enterprise office setup.

These computers are located
in four different labs/offices in two different buildings (referred to as A and B, respectively).
As  our default location, Lab \#1 in Building~A is a large shared lab space housing
about 30 students with 30+ \emph{active} desktop computers,
where the transmitter's and receiver's
power outlets are approximately 55 feet away from each other.
Thus, Lab \#1 represents an environment with high
power line noises.
The layout of Lab \#1 is illustrated in Fig.~\ref{fig:layout_lab}.
In addition, we also run experiments in another two labs (\#2 with about 10 students
and \#3 with about 15 students) in Building~A.
Our experiment setup in Building~B is illustrated in Fig.~\ref{fig:layout_office},
where the transmitter and receiver are located
in two different rooms 
about 90 feet away from each other.

\textbf{Transmitted data.} Like other studies on covert channels \cite{Security_AirGap_Bridgeware_ACM_Communications_2018_Guri:2018:BAM:3200906.3177230,Security_AirGap_Mobile_USB_Charging_ACNS_Mobile_CovertChannel_Spolaor2017NoFC,Security_AirGap_EMI_GSMem_Security_2015},
\ouralg focuses on stealthy exfiltration of already collected information. Thus,
for illustration, we use 8-bit ASCII values of the string ``password123'' followed by six random bits as the payload data and ``110010'' as the 6-bit pilot sequence. Each frame has 100 bits.
We also randomly generate 50 data frames for each experiment.
Each symbol carries one bit and lasts 33ms.
Like the existing covert channel literature,
error correction coding 
is orthogonal and omitted in our experiments.


\textbf{Current amplitude modulation.} To modulate the current amplitude (Section~\ref{sec:design_transmitter_modulation}),
we implement a current modulator following Algorithm~\ref{Alg:power_modulator}
which can be incorporated into existing malware for data exfiltration.
The program is written in Java
and uses multiple nested  $\sin()$ and $\cos()$ computations on a random floating number
to vary the CPU load. We use multi-threading to increase the impact on the CPU. However, as the default case, we do not use any CPU pinning and allow regular OS scheduling. 

%
%
%

\textbf{Voltage measurement.} To collect voltage signals, 
we use
a Rigol 1074z oscilloscope as a proxy ADC circuit.
To remove the voltage's dominant
60Hz component and improve the signal acquisition
precision, we follow \cite{Enev:2011:TVP:2046707.2046770}
and insert a RC high-pass filter with $\sim$10kHz cutoff frequency between the power outlet and the oscilloscope.
We use 500 kilo-samples per seconds (kSa/s) as the default sampling rate.
While we use a bulky oscilloscope for convenience, the receiver only
needs to hide a small ADC circuit (e.g., using ATmega Microcontroller) to acquire and digitize
the voltage signal (Section~\ref{sec:threat_model}).

%






\textbf{Frequency analysis and filters.} 
We
use Matlab to perform Fourier analysis and signal filtering. In real-world implementation,
the receiver may also use a digital signal processing chip
to filter voltage signals and demodulate bit values in real time.


\textbf{Metrics.} We calculate the bit error rate (i.e.,
percentage of bit errors) and effective transmission rate
measured in bits/second (i.e., actual payload bit rate, excluding
pilot symbols and erroneous bits). We also list in Table~\ref{table:diff_configurations} the
distance between the transmitter's and receivers' power outlets.




\subsection{Evaluation Results}\label{sec:experiment_results}

We now present our evaluation results,
highlighting that \ouralg achieves \emph{stealthy} data exfiltration
from desktop computers. We first present a snapshot of data exfiltration using \OurAlg followed by the case where four transmitters are simultaneously sending data to a single receiver. We then evaluate \OurAlg under different background applications, different numbers of CPU cores used by \OurAlg, and different pilot lengths. We also test how the CPU states affect \OurAlg, and the effectiveness of \OurAlg in data exfiltration without a line of sight. Finally, we present results with computers from different manufacturers.

\subsubsection{\textbf{A snapshot of data exfiltration}}

\begin{figure*}[!t]
	\centering
	\includegraphics[trim=0cm 0cm 0cm 0cm,clip,  width=0.96\textwidth,page=1]{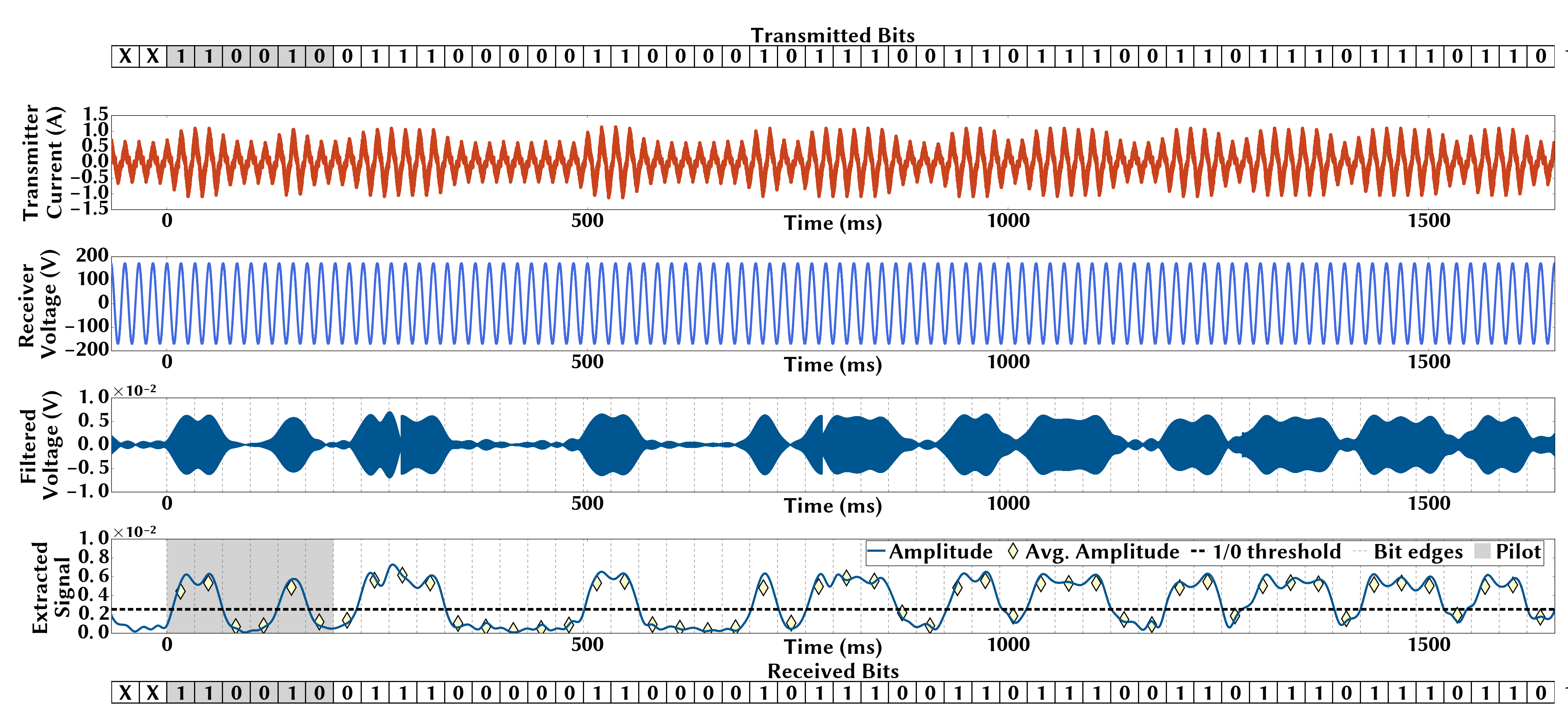}
	\caption{Dell Optiplex computer: A snapshot of data exfiltration. The receiver's voltage signal is filtered with a passband of $<$67.28kHz, 67.34kHz$>$ and the bit-wise average amplitudes of filtered voltage signals reveal the transmitted bits.}\label{fig:detection_demo}
\end{figure*}


We show in Fig.~\ref{fig:detection_demo} 
the different stages of \ouralg for
data exfiltration from our Dell Optiplex computer with 4 CPU cores. For clarity, we only show 
the first 50 bits in a data frame. At the top, we show the frame bits with the pilot sequence highlighted in a gray shade. The ``X''s prior to the pilot sequence indicate ``no data''. Also, the time index ``0'' indicates the start of the frame. We then show the transmitter's current, which is modulated by varying the CPU
load. We can see that the current amplitude changes with the transmitted bits.



Then,  we show the receiver's unfiltered voltage signal, which is affected
by grid voltage variations as well as all other loads sharing the same power
network and hence  barely reveals any useful information. Next, we show the filtered voltage signal with a passband of $<$67.28 kHz, 67.34 kHz$>$ identified by Algorithm~\ref{Alg:pilot_scan}. The envelop/amplitude of the filtered voltage signal is extracted, and then the average amplitude is used for bit detection. From the pilot sequence, we identify the frame's starting point and set the bit detection threshold accordingly (Algorithm~\ref{Alg:frame_scan}).
The bit-wise average amplitudes are then demodulated into the received bits --- an average amplitude above the detection threshold is considered as ``1'', and
``0'' otherwise. 
In this experiment, the received bits perfectly match the transmitted bits without any errors,
resulting an effective payload bit rate of 28.48 bits/second.

\begin{figure}[!t]
	\centering	
	\includegraphics[trim=0cm 6.5cm 0cm 6.5cm,clip, width=0.48\textwidth]{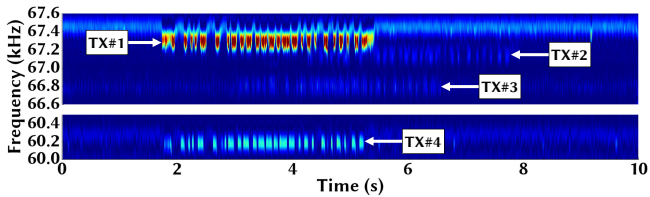}
	\caption{Four transmitters (TX\#1$\sim$4) simultaneously sending data to a single receiver. Bit error rates of TX\#1, TX\#2, TX\#3, and TX\#4 are 0.0\%, 6.8\%, 1.1\%, and 0.0\%, respectively. The higher bit error rate for TX\#2 is due to the partial overlap with TX\#1. }\label{fig:spectrum_voltage_multi}
\end{figure}

\subsubsection{\textbf{Simultaneous transmissions
}}\label{sec:multi_TX}

We simultaneously use multiple computers as transmitters to demonstrate that the PSD spikes from different computers do not interfere with each other.
We conduct this experiment in Lab\#1 using three Dell OptiPlex desktops with identical power supply units and one Dell XPS desktop with a different supply unit. Note that, while we use four computers as transmitters, there are other computers (30+) in the building that also generate switching noise spikes due to their PFC circuits.

Fig.~\ref{fig:spectrum_voltage_multi} shows the frequency spectrum of the voltage signal from the receiver while the corresponding data frame extractions are shown in Appendix~\ref{sec:apendix_multi_TX}. In the top figure, we show the frequency spectrum of the three transmitters with identical power supply units whose PSD spikes are close to each other (within the frequency band $<$66.6kHz, 67.6kHz$>$). The bottom figure shows the frequency spectrum of the other transmitter which generates PSD spikes around 60.1kHz. Each transmitter sends a single data frame at different times. We can clearly identify the data frames of the four transmitters in the frequency spectra. We use four 60Hz band-pass filters with center frequencies 60.13kHz, 67.09kHz, 67.3kHz, and 66.72kHz to extract the transmitted data frames. As shown in Appendix~\ref{sec:apendix_multi_TX}, we achieve data exfiltration with a maximum bit error rate less than 7\% for any transmitter.

\begin{table}[]
	\centering
	\small
	\caption{Summary of Data Exfiltration from Dell Optiplex}
	\label{table:diff_cases}
	\begin{tabular}{|l|c|c|c|}
		\hline
		\multicolumn{1}{|c|}{\textbf{Scenario}}  & \textbf{\begin{tabular}[c]{@{}c@{}}Bit Error\\ Rate\end{tabular}} & \textbf{\begin{tabular}[c]{@{}c@{}}Bits Per\\ Second\end{tabular}} & \textbf{\begin{tabular}[c]{@{}c@{}}Detection\\ Results\end{tabular}} \\ \hline
		Default (4 cores) & 0.0\% & 28.48 & Fig.~\ref{fig:detection_demo}\\ \hline
		With YouTube streaming & 2.3\% & 27.82 & Fig.~\ref{fig:detection_youtube} \\ \hline
		With MS Word running & 0\% & 28.48 & Fig.~\ref{fig:revision_detection_background_word_res} \\ \hline
		With web browsing & 0\% & 28.48 & Fig.~\ref{fig:revision_detection_background_wEB_res} \\ \hline
		With HDD file transfer & 3.5\% & 27.48 & Fig.~\ref{fig:revision_detection_background_io_res}\\ \hline
		With ML training & 1.67\% & 28.00& Fig.~\ref{fig:revision_detection_background_learning_res} \\ \hline
		Loading 1 CPU core & 8.9\% & 25.94 & Fig.~\ref{fig:detection_1core} \\ \hline
		Loading 2 CPU cores & 2.5\% & 27.77 & Fig.~\ref{fig:detection_2core}\\ \hline
		Loading 3 CPU cores & 0.0\% & 28.48 & Fig.~\ref{fig:detection_3core} \\ \hline
		Using 4-bit pilot sequence& 3.3\% & 28.13 & Fig.~\ref{fig:detection_4bit_pilot} \\ \hline
		Using 8-bit pilot sequence& 0.0\% & 27.88 & Fig.~\ref{fig:detection_8bit_pilot} \\ \hline
	\end{tabular}
\end{table}

\subsubsection{\textbf{Data exfiltration under different settings}}\label{sec:experiment_result_different_settings}

Now, we vary the default settings listed in Section~\ref{sec:experiment_methodology}.
The results are summarized in Table~\ref{table:diff_cases}, while the figures
are shown in Appendix~\ref{sec:appendix_experiment_dell_optiplex}.

First,  we run a concurrent program by playing ``See You Again'' on YouTube 
on a Google Chrome browser, which
is one of the most viewed videos \cite{YouTube_SeeYouAgain_Video}.
As video streaming can be fairly CPU intensive\footnote{Our test desktop does not have a dedicated GPU to offload video processing.}, running YouTube adds random variations to the transmitter's power. It also reduces the difference between transmitted "0"s and "1"s, resulting in a 2.3\% bit error rate.
We also run experiments using MS Word, web browsing, file transfer, and machine learning training as background applications resulting in 0\%, 0\%, 3.5\%, and 1.67\% bit error rates, respectively. In the MS word experiment, we mimic user behavior by repeatedly opening a new file, typing a few lines of texts and then saving the file.  For the web browsing experiment, we open new popular websites (e.g., GMail and Facebook), scroll through the page content, and follow links to other pages. For the file transfer experiment, we transfer a 5GB file from one HDD drive to another in our desktop computer running Windows 10. For the machine learning experiment, we repeat training Tensorflow in Python with 6000 samples from  the MNIST data set taking around fives minutes to finish \cite{lecun1998gradient}.
From the results from Table~\ref{table:diff_cases} we see that under a diverse set of background applications running simultaneously, \OurAlg still maintains a low bit error rate.


Second, we use CPU pinning to restrict the
number of cores that are assigned to the modulation program in \ouralg. Reducing number of cores increases the bit error rate because it limits how much the modulation program can vary the transmitter's current. Nonetheless, we find that even by loading only one CPU core (the weakest transmission), \ouralg achieves an effective rate of 25.94 bits/second. In Table~\ref{table:diff_cases} we omit the four core case which is our default case with 0\% bit error rate.

Finally, we consider 4-bit (``1101'') and 8-bit (``11001010'') pilot sequences. We see that
when using a 4-bit pilot sequence ``1101'',
the bit detection threshold may not be properly set due to lack of enough
pilot symbols, but it has lower overhead than the 8-bit pilot and \ouralg
still achieves 28.13 bits/second.

\begin{table*}[]
	\centering
	\small
	\caption{Summary of Experiments on Seven Different Computers.}\label{table:diff_configurations}
	\resizebox{\textwidth}{!}{%
		\begin{tabular}{|c|c|c|c|c|c|c|c|c|c|c|}
			\hline
			\textbf{\begin{tabular}[c]{@{}c@{}}Transmitting\\Computer\end{tabular}}& \textbf{Configuration} & \textbf{\begin{tabular}[c]{@{}c@{}}Operating\\ System\end{tabular}} & \textbf{\begin{tabular}[c]{@{}c@{}}Power Supply\\ Unit\end{tabular}} & \textbf{Year} & \textbf{\begin{tabular}[c]{@{}c@{}}PFC\\ Switching\\ Frequency\end{tabular}} & \textbf{Location} & \textbf{\begin{tabular}[c]{@{}c@{}}TX-RX\\ Distance\end{tabular}} & \textbf{\begin{tabular}[c]{@{}c@{}}Bit\\ Error\\ Rate\end{tabular}} & \textbf{\begin{tabular}[c]{@{}c@{}}Bits\\ Per\\ Second\end{tabular}} & \textbf{\begin{tabular}[c]{@{}c@{}}Detection\\ Results\end{tabular}} \\ \hline
			
			\textbf{\begin{tabular}[c]{@{}c@{}}Dell Optiplex\\ 9020\end{tabular}} & \begin{tabular}[c]{@{}c@{}}Core i7-4790, \\ 16 GB\end{tabular} & Windows 10 & \begin{tabular}[c]{@{}c@{}}Dell-L290EM-01 300W\\ by Lite-on Tech. Co.\end{tabular} & 2015 & $\sim$67.3 kHz & \begin{tabular}[c]{@{}c@{}}Lab \#1\\ (Building A)\end{tabular} & $\sim$55 feet & 0.0\% & 28.48 & Fig.~\ref{fig:detection_demo} \\ \hline
			
			\textbf{\begin{tabular}[c]{@{}c@{}}Dell PowerEdge\\ R630\end{tabular}} & \begin{tabular}[c]{@{}c@{}}Dual Xeon\\ E52640, 32GB\end{tabular} & \begin{tabular}[c]{@{}c@{}}Ubuntu\\ Server 14.04\end{tabular} & \begin{tabular}[c]{@{}c@{}}Dell-E495E-S1 495W\\ by Astek Intl.\end{tabular} & 2016 & $\sim$65.8 kHz & \begin{tabular}[c]{@{}c@{}}Office\\ (Building B)\end{tabular} & $\sim$90 feet & 0.0\% & 28.48 & Fig.~\ref{fig:detection_office_other_room} \\ \hline
			
			\textbf{\begin{tabular}[c]{@{}c@{}}Dell XPS\\ 8920\end{tabular}} & \begin{tabular}[c]{@{}c@{}}Core i7-7700, \\ 16 GB\end{tabular} & Windows 10 & \begin{tabular}[c]{@{}c@{}}Dell-460AM-03 385W\\ by Delta Electronics Inc.\end{tabular} & 2017 & $\sim$60.1 kHz & \begin{tabular}[c]{@{}c@{}}Lab \#1\\ (Building A)\end{tabular} & $\sim$55 feet & 0.0\% & 28.48 & Fig.~\ref{fig:detection_multi_TX_05} \\ \hline
			
			\textbf{Acer G3-710} & \begin{tabular}[c]{@{}c@{}}Core i7-7700, \\ 16 GB\end{tabular} & \begin{tabular}[c]{@{}c@{}}Ubuntu \\ 16.04\end{tabular} & ACER 750W & 2016 & $\sim$63.5 kHz & \begin{tabular}[c]{@{}c@{}}Lab \#2\\ (Building A)\end{tabular} & $\sim$20 feet & 10.1\% & 25.60 & Fig.~\ref{fig:detection_acer} \\ \hline
			
			\textbf{\begin{tabular}[c]{@{}c@{}}Custom \\ Built \#1\end{tabular}} & \begin{tabular}[c]{@{}c@{}}Core i7-7700,\\ 16GB\end{tabular} & Windows 10 & \begin{tabular}[c]{@{}c@{}}Corsair 850W\\ RM850x-RPS0110\end{tabular} & 2018 & $\sim$91.2 kHz & \begin{tabular}[c]{@{}c@{}}Lab \#1\\ (Building A)\end{tabular} & $\sim$55 feet & 8.1\% & 26.17 & Fig.~\ref{fig:detection_corsair}\\ \hline
			
			\textbf{\begin{tabular}[c]{@{}c@{}}Custom\\ Built \#2\end{tabular}} & \begin{tabular}[c]{@{}c@{}}Core i7-7700K,\\ 16 GB\end{tabular} & \begin{tabular}[c]{@{}c@{}}Ubuntu\\ 16.04\end{tabular} & \begin{tabular}[c]{@{}c@{}}EVGA 850W\\ Supernova 850G2\end{tabular} & 2016 & $\sim$67.7 kHz & \begin{tabular}[c]{@{}c@{}}Lab \#3\\ (Building A)\end{tabular} & $\sim$15 feet & 9.2\% & 25.85 & Fig.~\ref{fig:detection_kim} \\ \hline
			
			\multirow{2}{*}{\textbf{\begin{tabular}[c]{@{}c@{}}Apple iMac\\ Model A1419\\(27-inch)\end{tabular}}} & \multirow{2}{*}{\begin{tabular}[c]{@{}c@{}}Core i5-3470S, \\ 8 GB\end{tabular}} & \multirow{2}{*}{\begin{tabular}[c]{@{}c@{}}macOS\\ 10.13.3\end{tabular}} & \multirow{2}{*}{\begin{tabular}[c]{@{}c@{}}Apple 300W\\PA13112A1\\(for 2012-2017 models)\end{tabular}} & \multirow{2}{*}{2015} & \multirow{2}{*}{$\sim$101 kHz} & \multirow{2}{*}{\begin{tabular}[c]{@{}c@{}}Lab \#1\\ (Building A)\end{tabular}} & \multirow{2}{*}{$\sim$55 Feet} & \begin{tabular}[c]{@{}c@{}}16\%\\ (50ms/sym)\end{tabular} & 15.79 & \multirow{2}{*}{Fig.~\ref{fig:detection_iMac} } \\ \cline{9-10}
			&  &  &  &  &  &  &  & \begin{tabular}[c]{@{}c@{}}2\%\\ (100ms/sym)\end{tabular} & 9.21 &  \\ \hline
			
		\end{tabular}
	}
\end{table*}

\subsubsection{\textbf{Impact of CPU scaling on the transmitter}}
\begin{figure}[!t]
	\centering
	\subfigure[Transmitter power consumption]{\label{fig:power_different_CPU_state_main}\includegraphics[trim=0cm 0cm 0cm 0cm,clip,  width=0.23\textwidth,page=1]{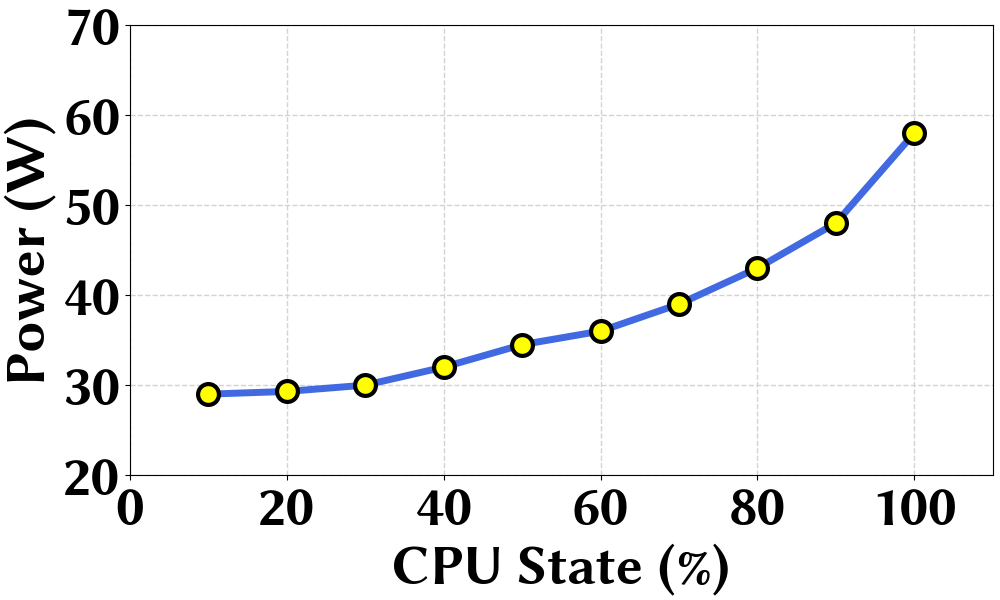}} \hspace{0.1cm}
	\subfigure[\OurAlg's data exfiltration accuracy]{\label{fig:performance_different_CPU_state_main}\includegraphics[trim=0cm 0cm 0cm 0cm,clip,  width=0.23\textwidth,page=1]{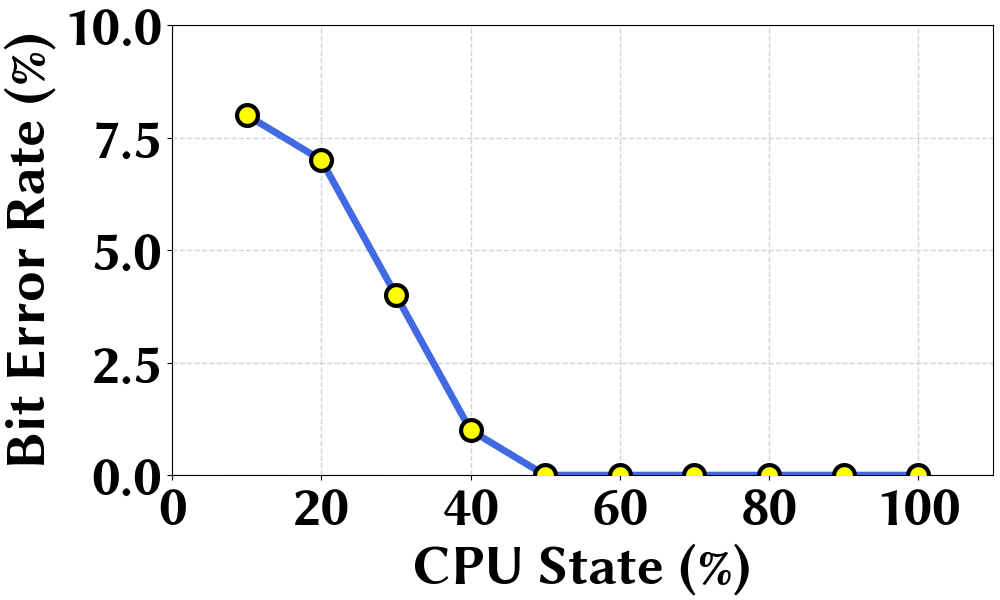}}
	\caption{The impact of CPU power state on the bit error rate of \OurAlg and the transmitter's (Dell Optiplex computer) power consumption. In our Dell Optiplex, 10\% and 100\% CPU states correspond to CPU frequencies of 0.79GHz and 3.68GHz, respectively.}\label{fig:impact_CPU_states_main}
	
\end{figure}

We tune the CPU scaling of the transmitter by changing the ``maximum processor state'' in the Windows power management system. We vary the maximum processor state from 10\%  to 100\% which, in our test computer, corresponds to CPU frequency from 0.79 GHz to 3.68 GHz. We use the default settings and transmit 100 frames under each different CPU scaling and record the average power consumption of the transmitter using a WattsUp power meter. Figs.~\ref{fig:power_different_CPU_state_main} and \ref{fig:performance_different_CPU_state_main} show the bit error rates and power consumption for the different CPU scaling. Similar to our experiment with the number of cores for \OurAlg, we see a higher bit error rate when the transmitter consumes less power due to reduced CPU speed.
Note that, While
dynamic CPU scaling at runtime is supported by
modern CPUs, it is more commonly applied in data centers with sophisticated power-performance control, where
energy saving is a crucial concern. For typical enterprise
environments, DVFS is not applied and instead, desktop computers
commonly run at the high-performance mode (i.e., 100\% maximum processor state, which is also the default setting in Windows 10).

\subsubsection{\textbf{Data exfiltration without line of sight}}

\begin{figure}[!t]
	\centering
	\includegraphics[trim=0.5cm 13.5cm 12cm 3cm,clip,  width=0.48\textwidth,page=1]{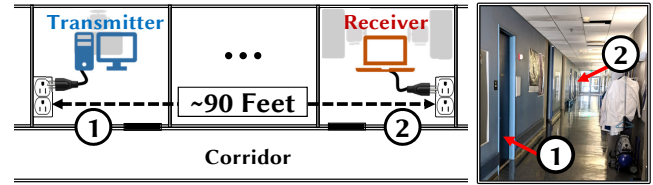}
	\caption{Experiment setup in  Building~B. \raisebox{.5pt}{\textcircled{\raisebox{-.9pt} {1}}} Transmitter's room. \raisebox{.5pt}{\textcircled{\raisebox{-.9pt} {2}}} Receiver's room. No line of sight between the transmitter and the receiver.}\label{fig:layout_office}
\end{figure}

As shown in Fig.~\ref{fig:layout_office}, we run an experiment in Building~B where the transmitter and receiver are plugged into two outlets located
in two different rooms which are approximately 90 feet away from each other.
We use a Dell PowerEdge computer as
the transmitter which, albeit using a different
switching frequency, adopts same PFC design
as our Dell Optiplex computer.
We show a snapshot of extracted signals at the receiver in Fig.~\ref{fig:detection_office_other_room} in Appendix~\ref{sec:appendix_no_LOS}.
We see that, compared to 
Fig.~\ref{fig:detection_demo}, the
distinction between bit ``1'' and bit ``0'' in terms
of the average amplitude of filtered voltage is less significant.
Nonetheless, even with a transmitter-receiver distance of 90 feet
and across different rooms, \ouralg still achieves error-free stealthy data exfiltration
with a rate of 28.48 bits/second.

\subsubsection{\textbf{Data exfiltration from other computers}}\label{sec:result_other_computer}


To further evaluate \ouralg, we run more experiments on four additional computers:
an Apple iMac, 
an
Acer, and two custom-built systems with top-brand power supply units (Corsair and EVGA).
These computers are located in different labs  in Building~A.
For the iMac, we could achieve a maximum of 15.79 bits/second due the iMac's slow response to power change and wider frequency signature (discussed further in Appendix~\ref{sec:appendix_imac}).
We summarize our results in Table~\ref{table:diff_configurations}, while the details
are presented in Appendices~\ref{sec:appendix_other_computers} and \ref{sec:appendix_imac}.

In summary,  while certain conditions may degrade the effectiveness of \OurAlg, our experiments on different computers and under different settings
confirm that \ouralg can exploit a building's power network as a covert channel
for stealthy data exfiltration from 
desktop computers without using a conventional communication network.

\subsection{Discussions}\label{sec:result_discussion}

We now discuss \ouralg from the following important aspects.

\textbf{Comparison with PLC adapters.}
Without dedicated physical powerline communications (PLC) adapters,
\ouralg still achieves information transfer over power networks.
This is due to the PFC's capability of generating prominent high-frequency
quasi-orthogonal switching noises.
During
our experiments, we have found that the amplitudes of PFC-induced
switching noise spikes and  high-frequency
voltage signals (in MHz range) modulated by a PLC adapter
(NETGEAR PowerLINE 1000) are in the same order of magnitude.
Thus, in practice, the achievable transmission range of
\ouralg  is expected to be similar to
that of a PLC adapter
(typically up to a few tens of meters) \cite{PowerLineCommunications_Survey_JSAC_2016_Cano:2016:SAP:2963140.2963846}.

\textbf{Missed frames and bits.}
In any covert channels  \cite{Security_AirGap_EMI_GSMem_Security_2015,Security_AirGap_Mobile_USB_Charging_ACNS_Mobile_CovertChannel_Spolaor2017NoFC,Security_AirGap_Electric_PowerHammer}, some frames may not be successfully received due to erroneous pilot
and/or payload symbols, and the receiver is not able to notify the transmitter
due to the unidirectional covert channel.
To trade efficiency for reliability,
the transmitter may send each data frame multiple times
and/or apply error correction coding \cite{Goldsmith_WirelessCommunications_2005}.

\textbf{Higher bit rate and limit.}
Like in the existing power-based data exfiltration literature \cite{Security_AirGap_Mobile_USB_Charging_ACNS_Mobile_CovertChannel_Spolaor2017NoFC,Security_AirGap_Electric_PowerHammer}, we empirically demonstrate the achievable bit rate of \ouralg. Nonetheless, we provide conjectures on two possible approaches to further improving the achievable bit rate for data exfiltration. First, we may possibly improve the achievable bit rate if privileged access to the target computer's PFC is granted. Specifically, the PFC's feedback gain may be altered to increase the responsiveness of the power supply unit, i.e., the power supply unit follows changes in the CPU utilization/power consumption more closely. This can be achievable since most modern computer power supplies utilize digital control to adjust the feedback gain \cite{pfcHandbook,pfcST}. Second, given privileged access to the computer's power supply unit, another complementary approach is to modulate the PFC's switching frequency for data exfiltration. 
The digitally-controlled PFC circuit allows dynamically setting the switching frequency through its control program \cite{zhang2004digital}.
However, this approach may still be restricted by how often we can change the switching frequency. Moreover, frequency modulation requires wider bandwidth and hence can be susceptible to greater interferences from other devices/computers.

\textbf{Sources of bit errors.} Various factors can introduce bit
errors during data exfiltration, including the source signal
strength (affected by the amount of transmitter' power
consumption that can be modulated by the malware and
the PFC design), signal propagation path and fading (affected
by the relative location/distance of transmitter and receiver,
line impedance, building's power network topology),
interferences from other devices with similar PFC switching frequencies,
among others. 
While it is challenging, if not impossible,
to theoretically quantify the impact of different factors
on the resulting bit errors for a given transmitter-receiver pair, one can \emph{qualitatively} conclude
based on standard  bit error analysis
for additive white-Gaussian noise channels \cite{Goldsmith_WirelessCommunications_2005}
that a lower bit error rate can be achieved by increasing the source
signal strength, reducing the signal propagation fading, and/or mitigating interferences.
These are also reflected by our above empirical results.

\textbf{Scalability of NoDE.}
The scalability of the simultaneous exfiltration depends largely on whether or not switching noise spikes
generated by different computers overlap with each other. Next, we discuss the following three different cases ---
orthogonal switching noises spikes, overlapping switching noise spikes, and practical
scenarios.

$\bullet$ \emph{When switching noise spikes
	of different computers are perfectly frequency-orthogonal.}
In this case, data exfiltrations from different transmitting computers can be viewed
as independent, without much inference from each other.
If we conservatively assume that the switching noise spike
of a computer occupies a frequency band of 500Hz (400Hz sidelobes around
the most prominent spike plus 100Hz guard band) and
that the noise spikes of all transmitting computers are perfectly
frequency-orthogonal, then \ouralg
can achieve simultaneous data exfiltrations from \emph{up to}
200 computers over the frequency range of 50--150kHz, which
is the range for typical switching frequencies of PFC circuits \cite{pfcHandbook}.

$\bullet$ \emph{When switching noise spikes
	of different computers are overlapping in frequencies.}
In this case, simultaneous data exfiltrations become challenging, as in the case
of any communications systems \cite{Goldsmith_WirelessCommunications_2005}.
Thus,
different transmitting computers need to access the covert channel
at different times.
This is not restrictive, since a target computer may not be always
sending data over our covert channel.

$\bullet$ \emph{Practical scenario.}
In practice, the likelihood of overlapping PFC-induced noise spikes is not very high because the switching frequency is not tightly regulated in the power supplies.
Thus, under a scale of up to a few tens of computers, we expect that the switching
noise spikes of some computers may partially overlap, while most
noise spikes do not overlap (Fig.~\ref{fig:spectrum_voltage_multi}).
If other background computers have overlapping
switching noise spikes with a target computer, their switching noises can
be viewed as quasi-static background noises and do not significantly
affect data exfiltrations from the target computer.
For example, even on a single computer with simultaneous background  applications,
we show in Table~\ref{table:diff_cases} that \ouralg still can successfully
exfiltrate data.

\emph{To summarize}, 
the likelihood of having overlapping PFC-induced noise spikes among two computers is not very high in practice, thus allowing simultaneous data
exfiltrations.
Nonetheless, when two or more target computers have the overlapping
switching noise spikes, they need to transmit information
at different times using \ouralg.

\section{Defense Mechanism}\label{sec:defense}

Three major  approaches exist to defend against \ouralg --- eliminating PFC-induced
switching noises, preventing the switching noises entering the power network, and suppressing malware activities. The first two approaches involve hardware implementation and/or modification, while the last approach can be implemented primarily in software.


\subsection{Eliminating PFC-induced Switching Noises}
Completely eliminating PFC noise would require re-designing of computers'
power supply units with fundamentally different PFC strategy. However, it is non-trivial to find alternative solutions to replace the existing mature designs of power supply units without compromising energy efficiency. Moreover, such a change will require an industry-wide upgrade which is not likely to occur anytime soon.
Alternatively, a stricter EMI regulation can be imposed to include components less than 150kHz.

\begin{figure}[!t]
	\centering
	\subfigure[UPS powered computer]{\label{fig:defense_ups}\includegraphics[width=0.23\textwidth]{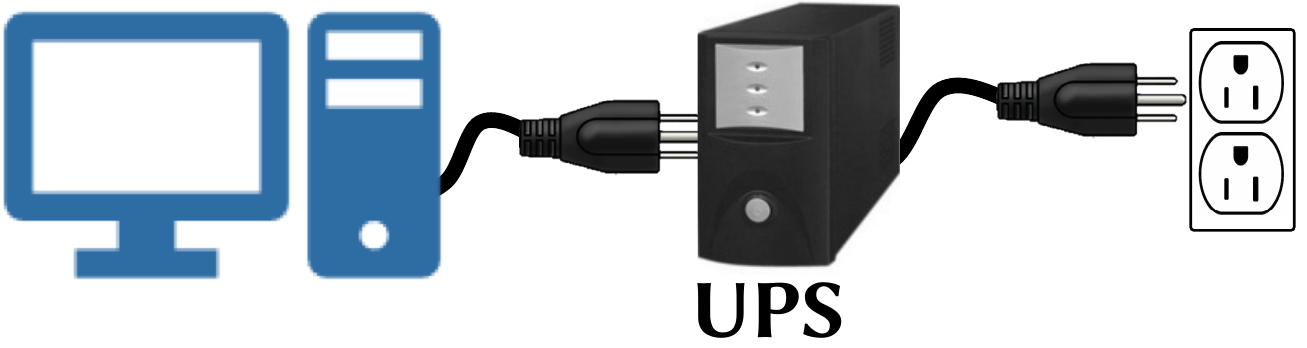}}\hspace{0.1cm}
	\subfigure[Power line noise filter]{\label{fig:defense_filter}\includegraphics[width=0.23\textwidth]{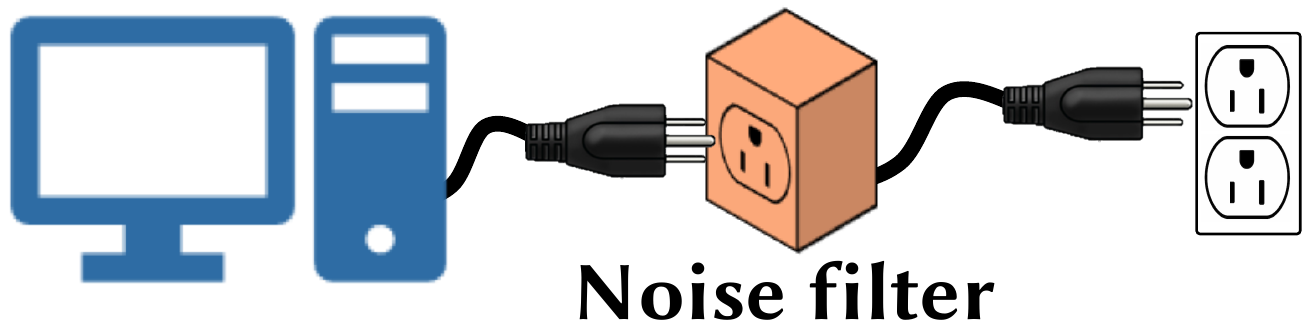}}
	\caption{Possible defense mechanisms to prevent PFC switching noise from entering the power network. (a) Powering a computer through a UPS. (b) Adding power line noise filters.}
\end{figure}
\subsection{Preventing Switching Noise From Entering Power Networks}
An intuitive defense against \OurAlg is to power a computer through a UPS instead of directly connecting it to a power outlet, and thereby restricting the PFC noise from entering the power network. In such a case, as shown in Fig.\ref{fig:defense_ups}, the UPS sits between the desktop power supply and the power outlet (i.e., power network). However, an UPS does not necessarily provide electrical isolation from the power network. Instead, it ``bypasses'' the utility power to its connected devices during normal operation. The UPS acts as an alternate power source when the power supplied through the utility is interrupted in the event of voltage drop or complete power losses. To illustrate this, we connect our Dell PowerEdge computer to a 600VA CyberPower UPS  and conduct our data exfiltration experiment in Building~B. As shown in Fig.~\ref{fig:detection_result_ups_all}, we have ``zero'' bit error, which matches our previous experiments without the UPS. Hence, an UPS-powered computer does not necessarily mean it is immune to the threat of \ouralg, let alone the added UPS cost.

Another defense  is to insert a power line noise filters between computers and power outlets, as shown in Fig.~\ref{fig:defense_filter}. The filters are commercially available to use together with household/office appliances for reducing interference and better facilitating power line communications. However, they mainly reduce the amplitude  of appliance-generated noise entering
the power network, without complete elimination. To demonstrate this, we run an experiment on our Dell PowerEdge computer plugged into a power line noise filter (X10 XPPF \cite{PFC_power_line_filter}) in Building~B. The resulting received signal is shown in Fig.~\ref{fig:detection_server_plc_filter},
from which we see that the signal amplitude of the PFC-induced switching noise spike is degraded by more than a factor of 10 compared to the case without any filters (Fig.~\ref{fig:detection_office_other_room}).
Thus, while not entirely prohibiting the transmitter's
switching noise spikes from entering the power network, the power line noise filter can significantly attenuate the amplitude of spikes, reducing the effective transmission distance.

\subsection{Suppressing Malware Activities}
As data exfiltration is done by varying the CPU load to modulate the transmitter's overall power consumption, a possible defense is to randomly vary the CPU load to de-correlate the overall power consumption with information bits. In the prior literature, hardware-based techniques have been developed to randomize the power consumption and obfuscate the power signature of instructions executed in devices \cite{moradi2015assessment,benini2003energy,guneysu2011generic,das2017high}. While they are efficient in terms of the power overhead for randomization, such hardware-based techniques are typically tailored for devices with specific functions such as cryptography
and not suitable for commodity processors. On the other hand, power randomization can also
be achieved by software-based approaches.

\begin{figure*}[!t]
	\centering
	\includegraphics[width=0.7\textwidth]{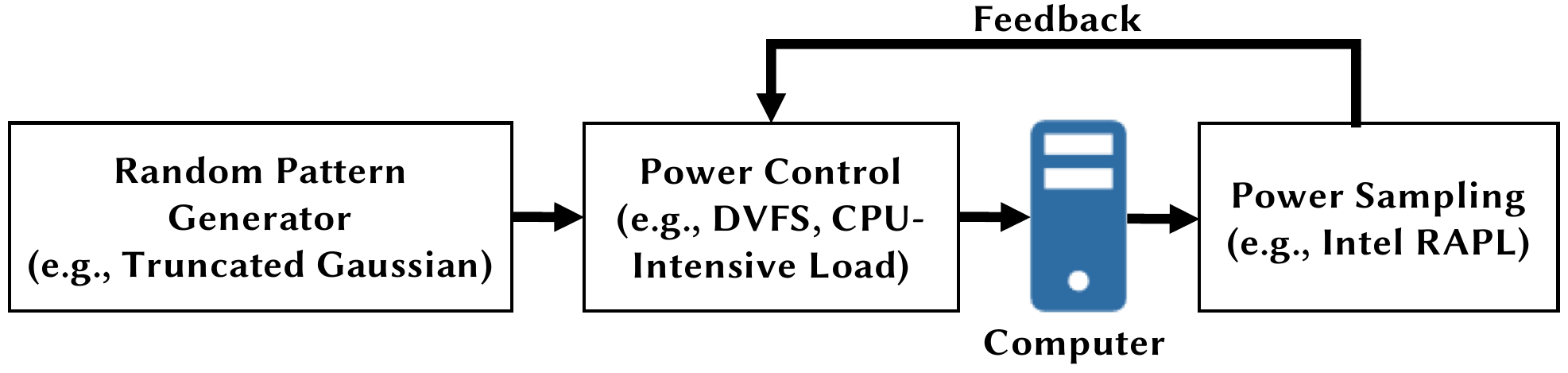}
	\caption{The building blocks of \rPower  which defends against \OurAlg by randomizing the computer's power consumption.} \label{fig:defense_randomizer_gaussian}
\end{figure*}

Concretely, we evaluate software-based approaches for power randomization with two different implementations --- \rNoise and \rPower. In \rNoise, we design a program that launches CPU-intensive computations at random times to add random power consumption. In \rPower, we follow the state-of-the-art technique to randomize the overall power consumption using a feedback loop \cite{pothukuchi2019maya}. As illustrated in Fig.~\ref{fig:defense_randomizer_gaussian}, \rPower samples the CPU power and uses combinations of CPU speed scaling (DVFS) and CPU-intensive computation to find-tune the power to follow the random pattern generator. The key difference between these two approaches is that \rNoise mainly adds random power noise to the existing power consumption patterns, whereas \rPower randomizes the overall power by more proactively controlling the CPU. 

\begin{figure}[!t]
	\centering
	\subfigure[Different CPU loading intervals]{\label{fig:performance_power_noise}\includegraphics[width=0.23\textwidth]{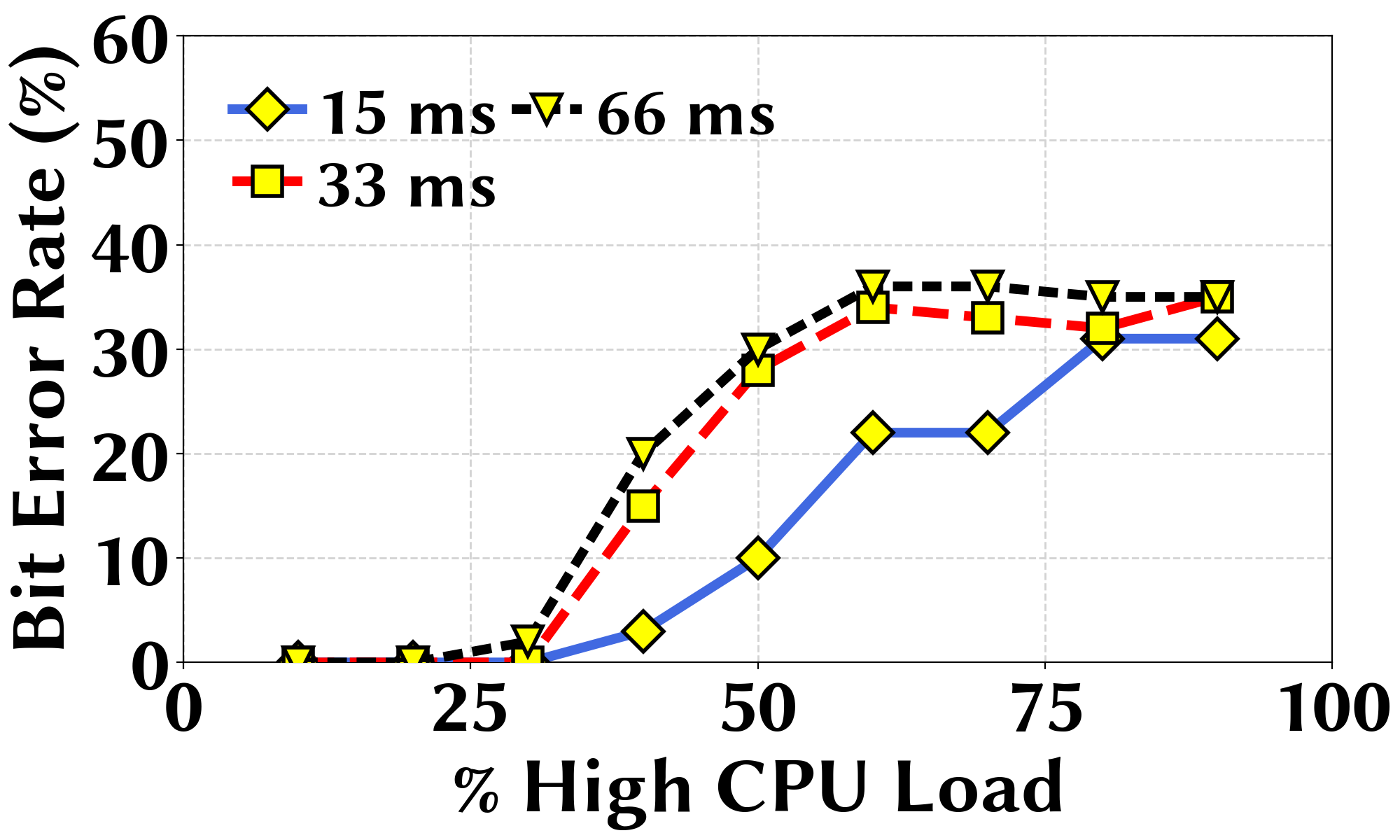}}\hspace{0.1cm}
	\subfigure[Different numbers of CPU cores]{\label{fig:performance_power_different_core}\includegraphics[width=0.23\textwidth]{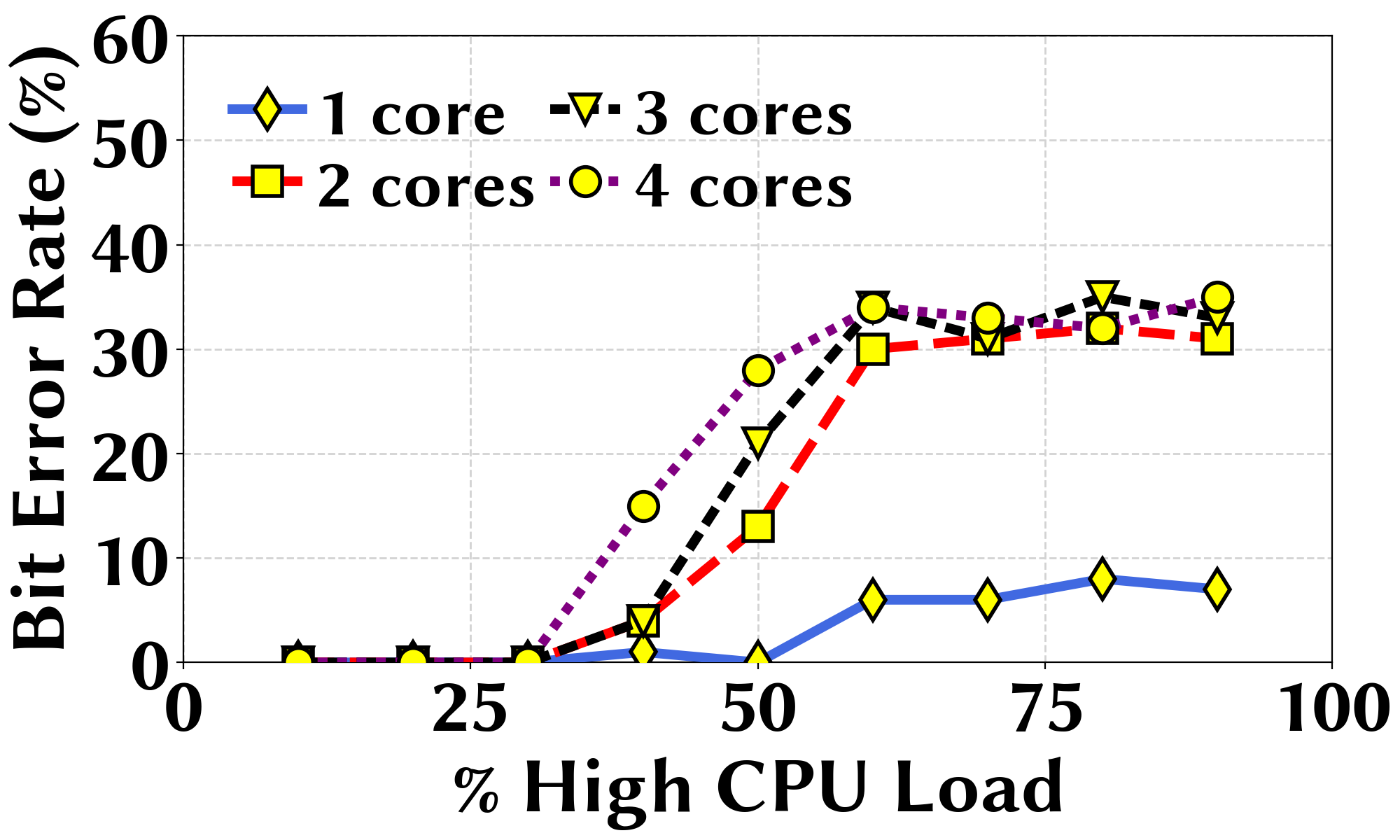}}\hspace{0.1cm}
	\caption{(a) and (b) Impact of \rNoise on \OurAlg's bit error rate for different settings.}
\end{figure}

\rNoiseBf. We test \OurAlg's performance under varying settings of  \rNoise's time interval of  added CPU loads, percent time of high CPU load, and the number of CPU cores used. Fig.~\ref{fig:performance_power_noise} shows \OurAlg's bit error rates as we increase the percent time of high CPU load from 10\% to 90\% for three different loading intervals (15/33/66 milliseconds). We see a general trend that, regardless of the interval length, an increasing percentage of CPU high load affects \OurAlg more. We also see that 15 milliseconds loading interval is worse than both 33 and 66 milliseconds intervals because it does not create a sustained high CPU load.
On the other hand, we see from Fig.~\ref{fig:performance_power_different_core} that loading more CPU cores by \rNoise also increases \ouralg's bit error rate.

\begin{figure}[!t]
	\centering
	\includegraphics[width=0.48\textwidth]{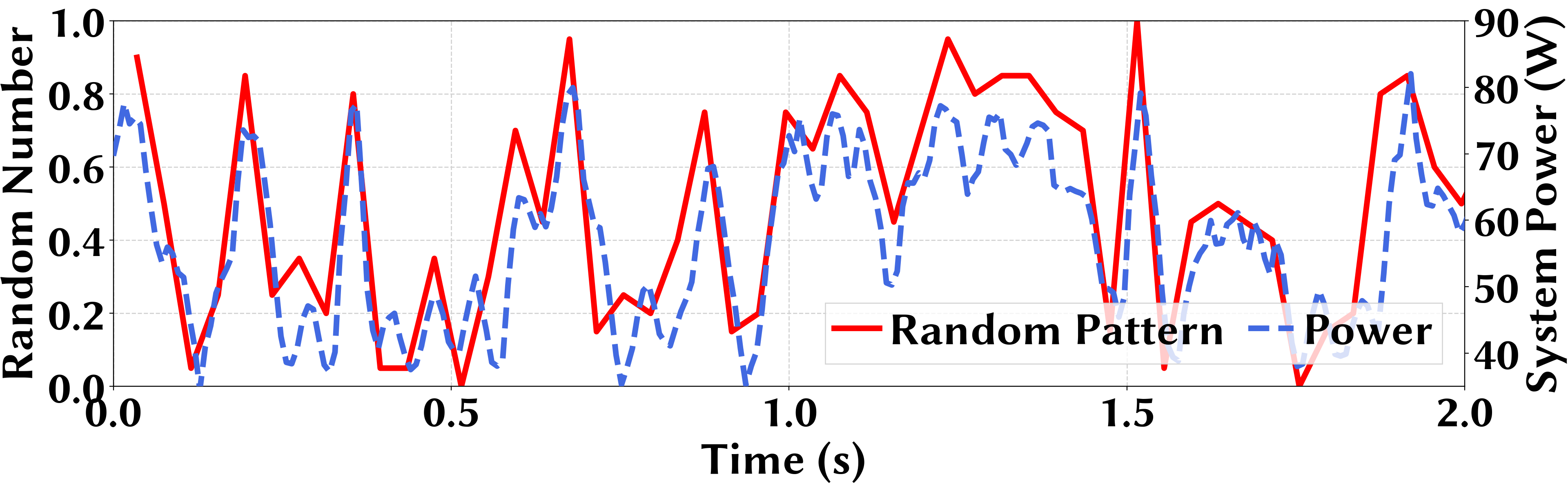}
	\caption{Illustration of the power variation following the target random number with a 0.5 mean and 0.5 standard deviation.}	\label{fig:gaussianRandom_RandomVsPower}
\end{figure}

\rPowerBf. We implement \rPower in Ubuntu and use Intel RAPL interface for sampling the power \cite{intel_RAPL,intel_power_gadget}, \texttt{cpufreq} for DVFS \cite{cpufreq}, and repeated floating-point operations as CPU-intensive computation. While we implement \rPower on Linux, software-based power monitoring and DVFS in other systems (e.g., Windows) are also available \cite{intel_power_gadget,amd_uProf,powercfg,cpufreq}.
In our experiment, the target power is determined by random numbers generated following a Gaussian distribution where we discard values smaller than 0 and greater than 1, resulting in a truncated Gaussian distribution.
The values of 0 and 1 corresponding to 35W and 85W in our experiment, respectively.
The mean and variance are user-set inputs to \rPower,
and the resulting probability density function can be expressed as
\begin{eqnarray}
f(x;\mu,\sigma) = \begin{cases}
\frac{1}{\Phi(\mu,\sigma) \sigma\sqrt{2\pi}}e^{-\left(\frac{x-\mu}{\sigma}\right)^2}  , &\text{if } 0\leq x \leq 1,\\
0, &\text{otherwise},
\end{cases}
\end{eqnarray}
where $\Phi(\mu,\sigma) = \frac{1}{\sigma\sqrt{2\pi}}\int_{0}^{1}e^{-\left(\frac{x-\mu}{\sigma}\right)^2} dx$, $\mu$ is the mean, and $\sigma$ is the standard deviation of the untruncated Gaussian distribution.
Based on the feedback, the CPU control block sets the appropriate CPU speed using DVFS and the amount of CPU-intensive workload to follow the randomly set power consumption target. In Fig.~\ref{fig:gaussianRandom_RandomVsPower}, we show a snapshot of the random number sequence and corresponding power consumption of \rPower running with a mean of 0.5 and a standard deviation of 0.5 for a truncated Gaussian distribution. We update the randomly set target power every 40 milliseconds. We see that the power consumption closely follows the supplied random number with minor deviations at times mainly due to the computer power supply's internal controls. While we use a truncated Gaussian distribution, this approach can be adopted with other probability distributions. In our evaluation, \rPower works perfectly against \OurAlg, and we cannot even identify the pilot sequence to extract the transmitted bits.

\begin{figure*}[!t]
	\centering
	\subfigure[\rNoise: power overhead]{\label{fig:performance_power_all}\includegraphics[width=0.3\textwidth]{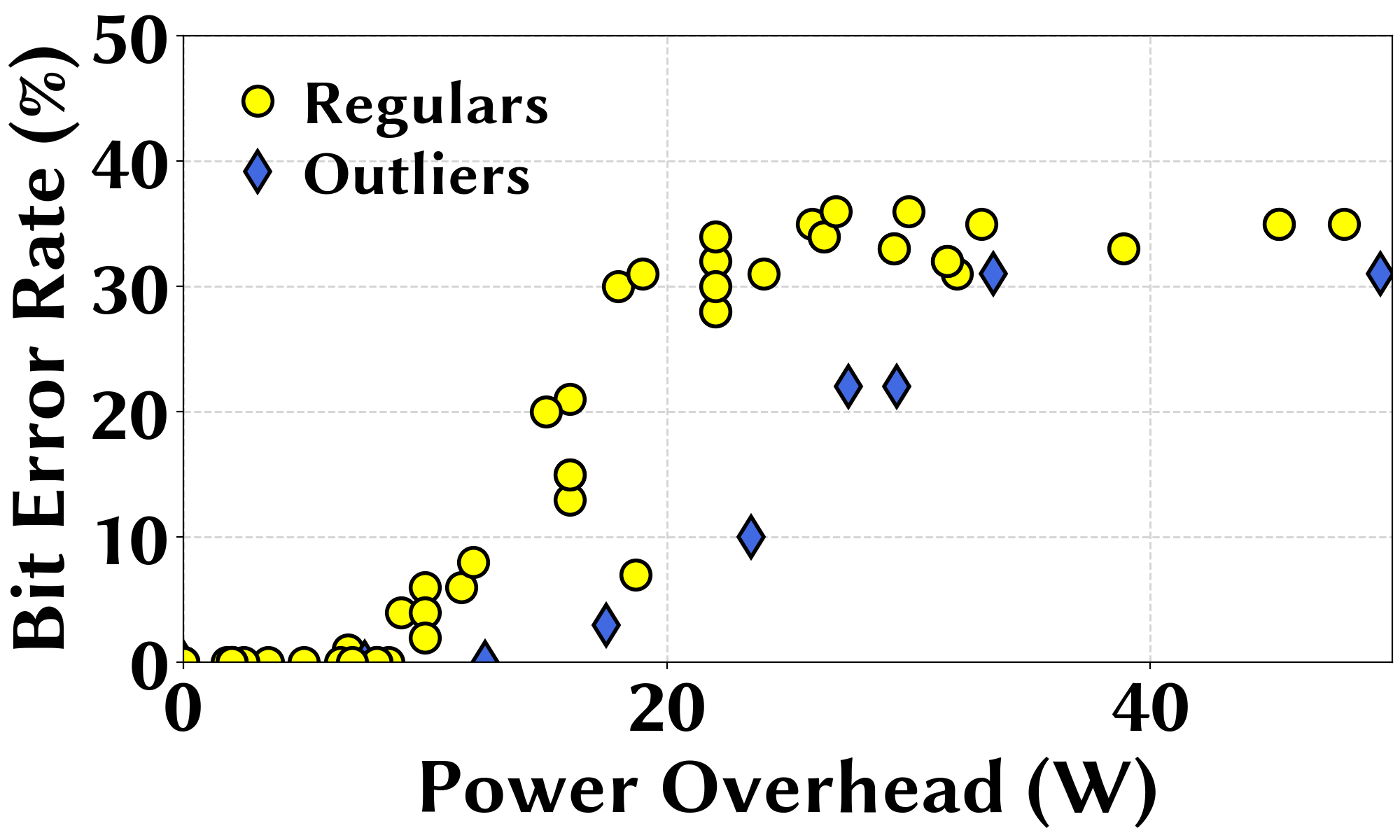}}\hspace{0.1cm}
	\subfigure[\rPower: power overhead]{\label{fig:gaussianRandom_powerOverhead_std05}\includegraphics[width=0.3\textwidth]{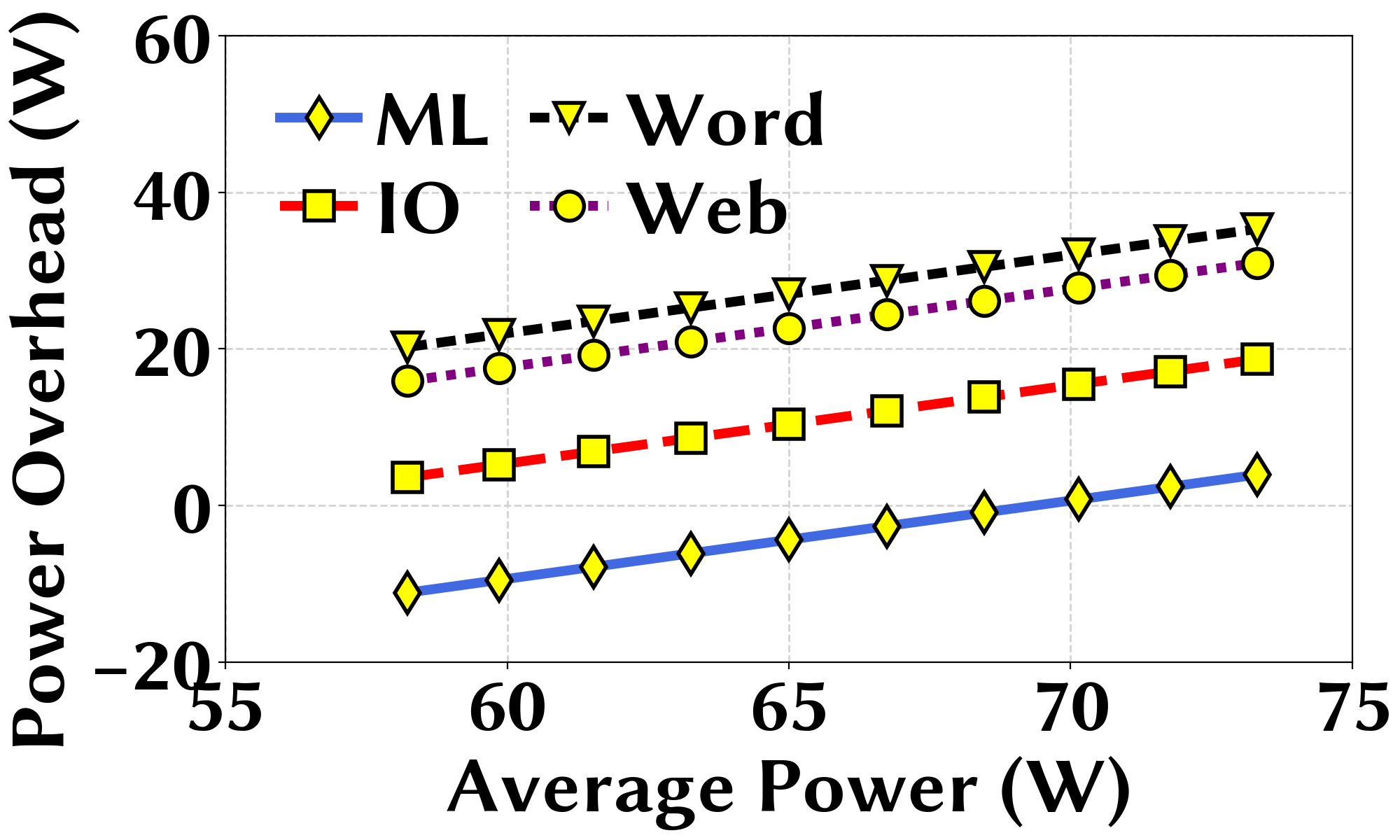}}\hspace{0.1cm}
	\subfigure[\rPower: perf. overhead]{\label{fig:gaussianRandom_perfOverhead_std05}\includegraphics[width=0.3\textwidth]{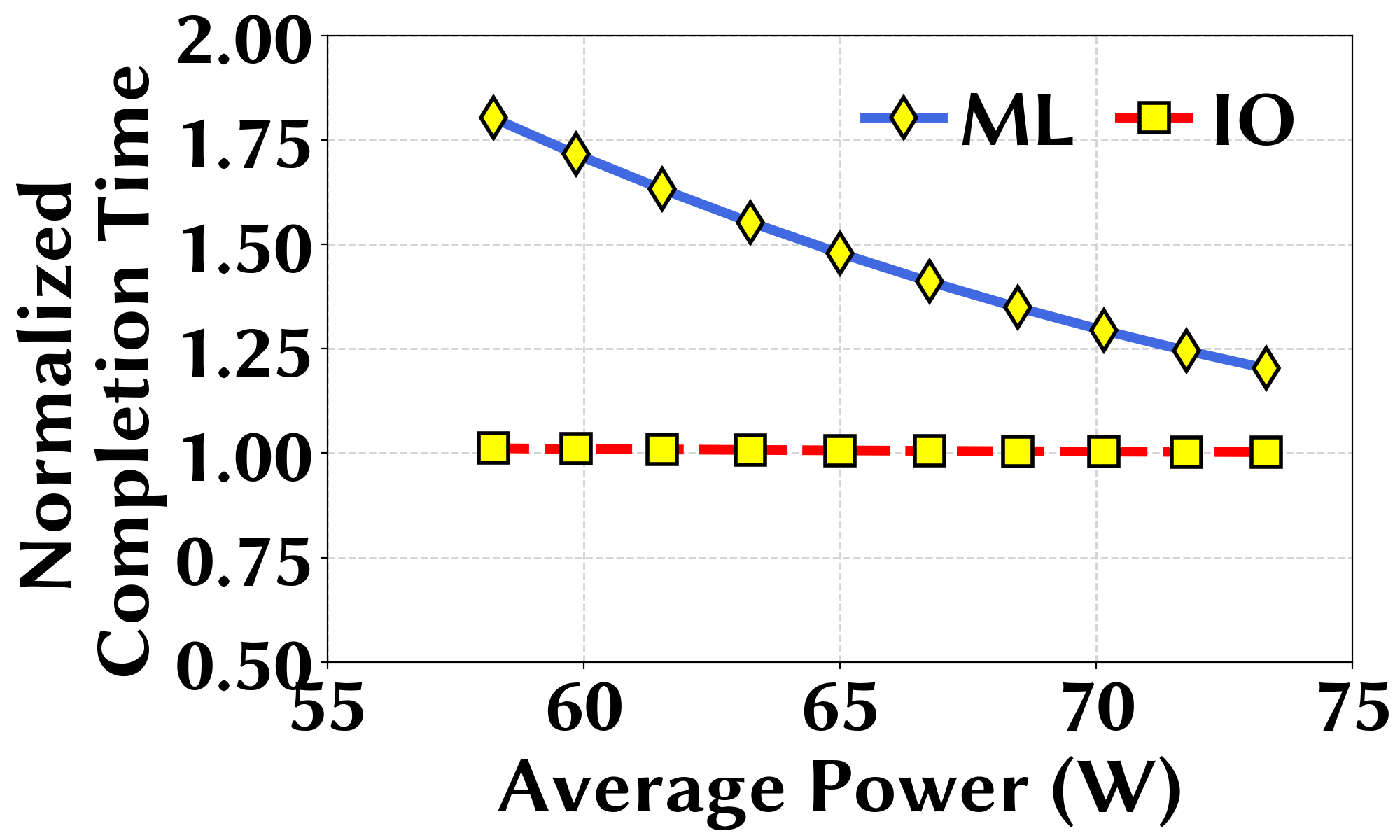}}
	\caption{(a) The power overhead of \rNoise at different levels of effectiveness (i.e., \OurAlg's bit error rates). The 15 ms CPU load interval setting is marked as outliers. (b) Power overhead of \rPower for different target average power. (c) Performance overhead of \rPower. Completion time is normalized to the case without \rPower.}
\end{figure*}

\textbf{Overhead.} Both \rNoise and \rPower add overhead to the system to defend. Since \rNoise injects random power noise, it results in additional power consumption by the computer. To have a detailed view of \rNoise's overhead, we show the power overhead under different cases in Fig.~\ref{fig:performance_power_all} with the 15-millisecond results as outliers. We calculate the power overhead by running \rNoise without any transmission and subtracting the idle power ($\sim$28W) from the average power consumption. The key message from Fig.~\ref{fig:performance_power_all} is that \rNoise can significantly affect \OurAlg's performance (> 30\% bit error rate) when it injects more than 20W of random power consumption. Nonetheless, in relation to the 28W idle power of our test computer, this amounts to a 70\% overhead.

For \rPower, we have both power overhead when CPU-intensive workloads are launched to increase the overall power, and performance overhead when CPU speed is throttled to reduce the overall power. To evaluate the power and performance overheads, we run \rPower with four different applications --- Machine learning training (ML), large file transfer in hard disk (IO), word processing (Word), and web browsing (Web). Figs.~\ref{fig:gaussianRandom_powerOverhead_std05} and \ref{fig:gaussianRandom_perfOverhead_std05} show the change in power and performance overheads as we change the average power consumption target by changing the mean of the truncated Gaussian random distribution. We calculate the power overhead by subtracting the average power consumption without \rPower from the average power consumption with \rPower. For the performance overhead, we take the ML training and file transfer completion times with \rPower and normalize the values with respect to those without \rPower. We exclude Word and Web from the performance overhead evaluation since they do not have completion times like ML or IO applications. Note that during our experiments, we do not experience any significant perceivable impact of \rPower on the performance of word processing and web browsing.

We see that the power overheads increase with increasing average power consumption. Also, \rPower's power overhead with perfect defense is lower than \rNoise's power overhead at higher bit error rates. More importantly, however, we see that different applications have different power overheads for the same average power, with ML having a negative power overhead when the average power is less than 70W. This is due to each application's power requirement. \rPower running along an application with a low power requirement mostly adds power to follow its power target, whereas \rPower needs to apply CPU throttling more frequently to reduce power when the underlying application requires more power. This is also why ML has a negative power overhead, indicating that due to frequent throttling \rPower has reduced the average power lower than the otherwise average power requirement of ML without \rPower. From Fig.~\ref{fig:gaussianRandom_perfOverhead_std05}, it is also evident that  frequent throttling at lower average power targets causes a higher performance overhead for the ML application. For the IO application, we do not see a significant performance variation since CPU throttling does not severely affect the file transfer speed. The key take-away from the results is that the overhead for \rPower depends on the underlying application and, for power-hungry applications, reducing the power overhead comes at the expense of performance degradation. A favorable balance between power and performance overheads can be attained through careful choice of the random number distribution parameters (e.g., mean).

\textbf{\rNoiseBf vs \rPowerBf}. \rPower is more effective against \OurAlg than \rNoise, since \rPower offers a perfect defense against \OurAlg and incurs a lower power overhead as well, especially for power-hungry background applications. However, \rPower incurs possible performance overheads due to the CPU throttling. It may also require additional OS privileges and/or accesses for CPU throttling and instrumentation for power sampling.

In addition to the aforementioned pro-active countermeasures, a reactive defense approach would be to identify and remove data exfiltration malware by monitoring application behavior or computer CPU utilization. However,  the constant emergence of new malware remains as, and will continue being, one of the greatest threats faced by computers
\cite{Security_Malware_ThreatReport_McAfee_March_2018,Security_Threat_Survey_SANS_2017,Security_Malware_Forecast_SophosLabs_2018,Security_MalwareTrends_GDataSecurityBlog_2017}.
On the other hand, utilizing the knowledge of \ouralg for data exfiltration, one can design a power network voltage monitoring system that continuously scans through the voltage
signals for suspicious switching noise spike patterns in the high frequency.
A potential drawback of this approach is the  computational burden
to continuously monitor a large frequency spectrum
since PFC-induced switching noise is generated by every computer power supply unit.

{In summary,} we see that different countermeasures against \OurAlg have their own merits and hurdles. Based on our study, as a hardware-based defense, we recommend the installation
of power noise filters because of its attenuation on  PFC-induced
switching noises. As for the software-based technique, we recommend power randomization due to its effectiveness.

\begin{table*}[!t]
	\caption{Summary of Data Exfiltration Attacks with Different Media.}
	\vspace{-0.3cm}
	\label{table:exfiltration_summary}
	\resizebox{0.8\textwidth}{!}{%
		\begin{tabular}{|l|l|l|l|}
			\hline
			\textbf{Medium}                                                     & \textbf{Proposed Design}                                                                                                                                                                                                                                                                                                                                                                                                                                                                                                                                                                            & \textbf{Bit Rate}         & \textbf{Effective Distance}                                              \\ \hline
			Acoustic                                                            & \begin{tabular}[c]{@{}l@{}}HDD noise \cite{Security_AirGap_Acoustic_DiskFiltration_DiskNoise_ESORICS_2017}, Fan noise \cite{Security_AirGap_Fansmitter_Acoustic_arXiv_DBLP:journals/corr/GuriSDE16}, Mesh network \cite{Security_AirGap_Acoustic_Ultrasound_MeshNetworks_NotIsraeli_Journal_2013},\\ RSA key extraction \cite{genkin2014rsa}, Computer speakers \cite{Security_AirGap_Acoustic_Ultrasound_FPS_Canada_2014}\\
				Gyroscope modulation using ultrasound \cite{Security_CovertInterference_PhoneGryoscope_WOOT_2016_Farshteindiker:2016:PHS:3027019.3027025}\end{tabular} & 0.25$\sim$140 bits/s             & 0$\sim$11 meters                                                              \\ \hline
			Thermal                                                             & Computer generated heat \cite{Security_AirGap_Thermal_BitWhisper_CSF_2015_Guri:2015:BCS:2859845.2859982}                                                                                                                                                                                                                                                                                                                                                                                                                                                                                             & 0.0022 bits/s                    & 0.4 meters                                                               \\ \hline
			\begin{tabular}[c]{@{}l@{}}Electromagnetic\\ emanation\end{tabular} & \begin{tabular}[c]{@{}l@{}}CRT monitor EMI radiation\cite{kuhn1998soft}, Memory bus \cite{Security_AirGap_EMI_GSMem_Security_2015},\\ Extracting cryptographic key \cite{genkin2015stealing,Camurati:2018:SCE:3243734.3243802,Genkin:2016:EKE:2976749.2978353}\end{tabular}                                                                                                                                                                                                                                                                                           & 2$\sim$512 bits/s                & 0$\sim$20 meters                                                              \\ \hline
			Magnetic                                                            & \begin{tabular}[c]{@{}l@{}} Hard drive head \cite{Security_AirGap_Magnetic_Mobile_ASP_DAC_2016_Matyunin2016CovertCU}, Escaping Faraday cage \cite{Security_AirGap_Magnetic_ODINI_FaradyCage_Bypass}\end{tabular}                                                                                                                                                                                                                                                                                                                                                                  & 4$\sim$40 bits/s                 & 0.15$\sim$1.5 meters                                                              \\ \hline
			Optical                                                             & \begin{tabular}[c]{@{}l@{}}Equipment status LED \cite{Security_AirGap_Optimal_Keyboard_LED_Loughry:2002:ILO:545186.545189}, HDD LED \cite{Security_AirGap_Optical_HardDrive_LED_Guri2017LEDitGOL}\end{tabular}                                                                                                                                                                                                                                                                                                                                                      & 4$\sim$56 kbits/s                & Line of sight                                                            \\ \hline
			Power                                                               & \begin{tabular}[c]{@{}l@{}} Power consumption \cite{Security_AirGap_Electric_PowerHammer}, \\ Mobile's charging power \cite{Security_AirGap_Mobile_USB_Charging_ACNS_Mobile_CovertChannel_Spolaor2017NoFC},\\ Key extraction from mobile's power analysis \cite{Genkin:2016:EKE:2976749.2978353}\end{tabular}                                                                                                                                                                                                                                                                           & \begin{tabular}[c]{@{}l@{}}2$\sim$1000 bits/s\\ (projected)\end{tabular} & \begin{tabular}[c]{@{}l@{}}Length of\\ power/charging\\cable\end{tabular} \\ \hline
		\end{tabular}
	}
\end{table*}

\section{Related Work}

There have been a plethora of studies on data exfiltration under a threat model where an adversary tries to extract information from a tightly secured computer system without using traditional data transfer protocols (e.g., network).
The
key idea is to encode information in 
\emph{physical} attributes (e.g., the heat generated
by
a computer \cite{Security_AirGap_Thermal_BitWhisper_CSF_2015_Guri:2015:BCS:2859845.2859982}) to carry it to an external receiver (e.g., temperature sensor). Meanwhile,
decoding changes of these physical attributes does
not require any cyber access to the target system, thus bypasses the system's defense and forms covert channels for stealing information.
Alternatively, a secure system may spill its secrets by inadvertently influencing an externally visible physical property (i.e., a side channel) \cite{genkin2014rsa,genkin2015stealing,Genkin:2016:EKE:2976749.2978353}.

Table~\ref{table:exfiltration_summary} summarizes the physical medium, key design attributes, transmission rates, and effective distances of the recently proposed data exfiltration attacks. Compared to the existing research, \ouralg achieves a reasonably high bit rate of 28.48 bits/s. Most acoustic covert/side channels cannot achieve a transfer rate higher than \OurAlg, except for \cite{Security_AirGap_Acoustic_Ultrasound_FPS_Canada_2014} (140 bits/s) which requires the target to be equipped with a speaker and the receiver be in the same room as the target computer. \cite{Security_AirGap_Acoustic_Ultrasound_FPS_Canada_2014} can reach up to 67,000 bits/s only 
in the audible range (20Hz$\sim$20kHz) at the expense of high detection possibility. On the other hand, both electromagnetic emanation and magnetic covert/side channel can achieve similar transfer rates as \ouralg but have much shorter effective distances. \cite{Security_AirGap_EMI_GSMem_Security_2015} can reach a commendable speed of 1000 bits/s but requires professional-grade receiver hardware with a high sophistication.

Among all, optical covert channels attain the best transmission rates because of the extremely fast response time of LEDs. But, they require the receiver to be in the line of sight of the transmitter (e.g., be in the same room). Also, because of the high bit rates, photodiodes 
need to be used as the receiver, further restricting the effective distance even within the line of sight. In contrast, \ouralg can have the transmitter and receiver in two different rooms that are 27.4 meters away from each other without line of sight.

Another important aspect of \OurAlg is that unlike other studies which cannot achieve both their highest bit rate and longest distance for the same settings (i.e., increasing distance decreases bit rate), \ouralg works at the 27.4 meters distance without compromising its rate of 28.48 bits/s at 0\% bit error.
While Table~\ref{table:exfiltration_summary} is not an exhaustive list, it provides important insights into the potential and limitation of various physical covert/side channels-based data exfiltration attacks.

\section{Concluding Remarks}

In this paper, we studied data exfiltration from a desktop
computer in an enterprise environment,
and proposed \ouralg to achieve stealthy information transfer over a building's power network without
using any PLC adapters. \ouralg exploits
high-frequency switching noises caused by the
PFC circuits built into all of today's computers and achieves an effective rate of 28.48 bits/second
with a distance of 90 feet (27.4 meters) without line
of sight.
We validated \OurAlg's data exfiltration capability under different settings and hardware configurations. We also showed that
certain configurations such as CPU speed scaling may reduce \OurAlg's data exfiltration rates. In addition, we  offered some insights into the limitations and open issues of our proposed system.
Finally, we outlined a few possible defenses and suggested both hardware-based and software-based techniques.

\bibliographystyle{plain}

\newpage

\appendix

\section*{\Large Appendix}

\section{Related Notions}\label{sec:appendix_related_notions}
\textbf{Power spectrum/ frequency spectrum:} Power spectrum disintegrates a signal into its frequency components and show the power of each frequency component. To illustrate the changes in happening in different frequency components of a signal, the power spectrum or frequency spectrum is typically shown over time with heat maps.  

\textbf{Passband:} In signal processing, filters are applied on a signal to attenuate undesired frequency components while let \emph{pass} the useful frequency components. The passband refers to the frequency range that a filter allows to pass through. The passband is identified using the lower and upper cutoff frequencies. When a filter allows a specific frequency band to pass, it is called a band-pass filter. When the passband starts at zero (i.e., lower cutoff = 0 Hz), it is called a low-pass filter. When the passband ends at infinity (i.e., upper cutoff - infinity Hz) it is called a high-pass filter.

\textbf{Harmonics:} Harmonics are frequency components at multiples of the fundamental frequency. In power system, harmonics are produced when the 50Hz/60Hz sinusoidal voltage or current gets distorted by non-linear loads (such as SMPS). Harmonics in the power system create unwanted losses in power transmission. 

\section{Voltage from a Power Outlet}\label{sec:appendix_voltage_variation}

\begin{figure}[!h]
	\centering
	\subfigure[A snapshot of voltage]{\label{fig:voltage_variation_trace}\includegraphics[trim=0cm 0cm 0cm 0cm,clip,  width=0.23\textwidth,page=1]{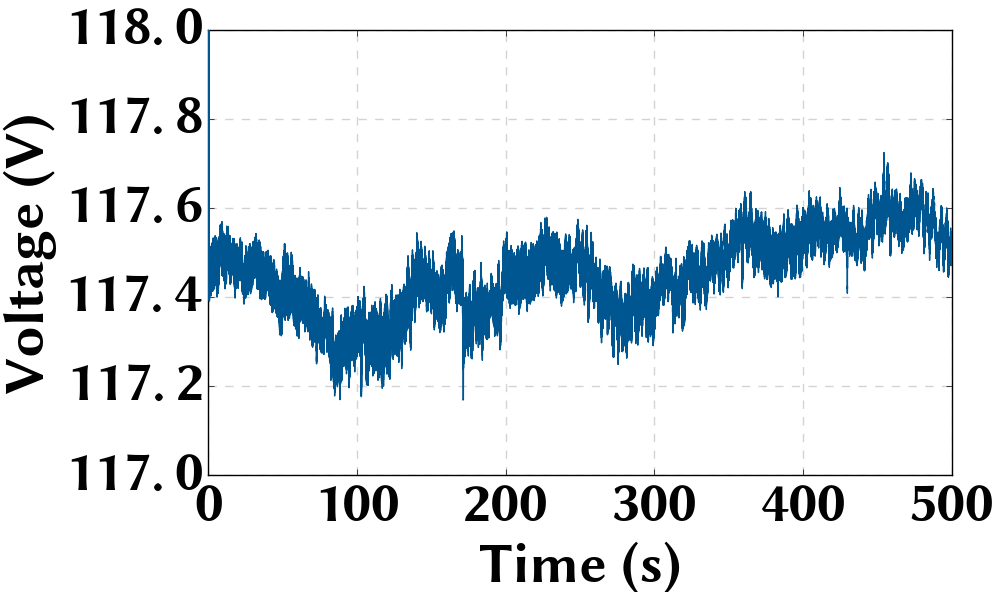}}\hspace{0.2cm}
	\subfigure[PMF of voltage]{\label{fig:voltage_variation_pmf_daily}\includegraphics[trim=0cm 0cm 0cm 0cm,clip,  width=0.23\textwidth,page=1]{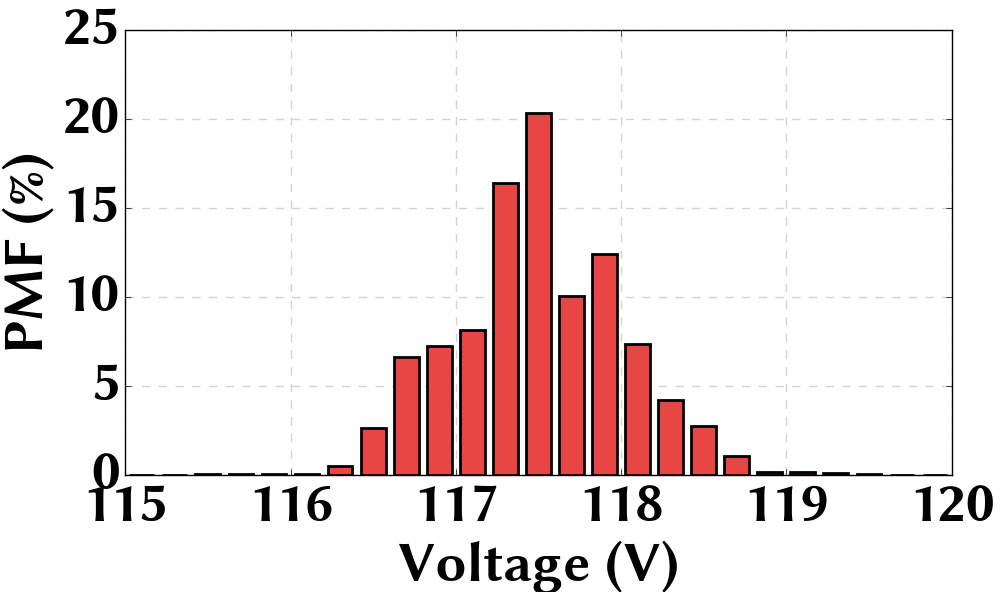}}
	\caption{The voltage of a power outlet varies over time.}\label{fig:utility_voltage}
\end{figure}

We show in Fig.~\ref{fig:voltage_variation_trace} a snapshot of the voltage trace collected from a power outlet in our lab. It can be seen
that the supplied voltage varies
by more than 500mV within just a few minutes. We further
show the probability mass function (PMF) of a 24-hour voltage distribution in
Fig.~\ref{fig:voltage_variation_pmf_daily}, demonstrating a nearly 5V variation
in the actual supplied voltage.

\section{Computer's Power Supply Unit}

%
%

\subsection{Conduction Modes for PFC}\label{sec:appendix_conduction_mode}

In Table~\ref{table:pfc_designs}, we show a summary of
major conduction modes for PFC circuits used in today's computers.

\begin{table}[!h]
	\caption{Summary of Major Conduction Modes for PFC \cite{pfcHandbook}}
	\label{table:pfc_designs}
	\footnotesize
	\resizebox{0.5\textwidth}{!}{%
		\vspace{-0.2cm}
		\begin{tabular}{|c|c|c|c|c|c|}
			\hline
			\textbf{\begin{tabular}[c]{@{}c@{}}Conduction\\Mode\end{tabular}} & \textbf{\begin{tabular}[c]{@{}c@{}}Power\\ Rating\end{tabular}}   &\textbf{\begin{tabular}[c]{@{}c@{}}Current\\Waveform\end{tabular}} & \textbf{\begin{tabular}[c]{@{}c@{}}Frequency\\Analysis\end{tabular}} & \textbf{Property}\\ \hline
			\begin{tabular}[c]{@{}c@{}}Continuous\\ Conduction\\ Mode \textbf{(CCM)}\end{tabular} & \textgreater{}300W & \begin{minipage}{.06\textwidth}
				\centering
				\includegraphics[width=0.8\textwidth]{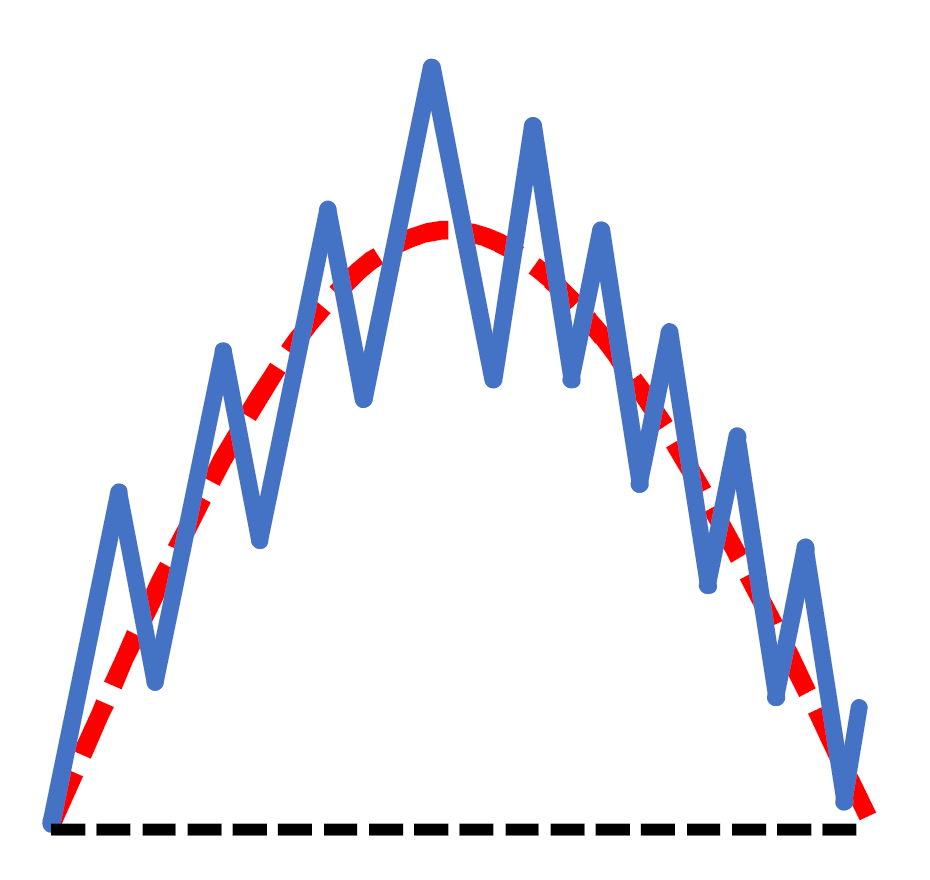}
			\end{minipage} &  \begin{minipage}{.06\textwidth}
				\centering
				\includegraphics[width=1\textwidth]{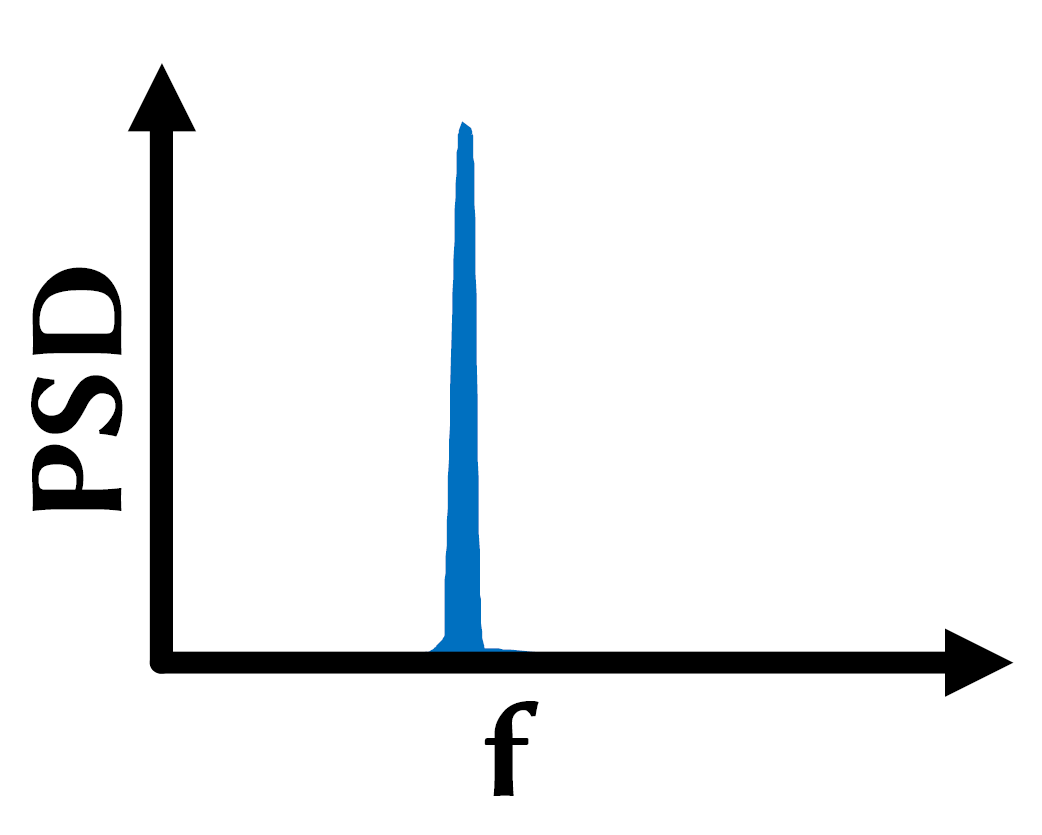}
			\end{minipage}
			& \begin{tabular}[l]{@{}l@{}}- Fixed frequency\\- Large inductor\\ - Lowest peak current\end{tabular}
			\\  [1ex]\hline
			\begin{tabular}[c]{@{}c@{}}Discontinuous\\ Conduction\\ Mode \textbf{(DCM)}\end{tabular} & \textless{}300W &    \begin{minipage}{.06\textwidth}
				\centering
				\includegraphics[width=0.82\textwidth]{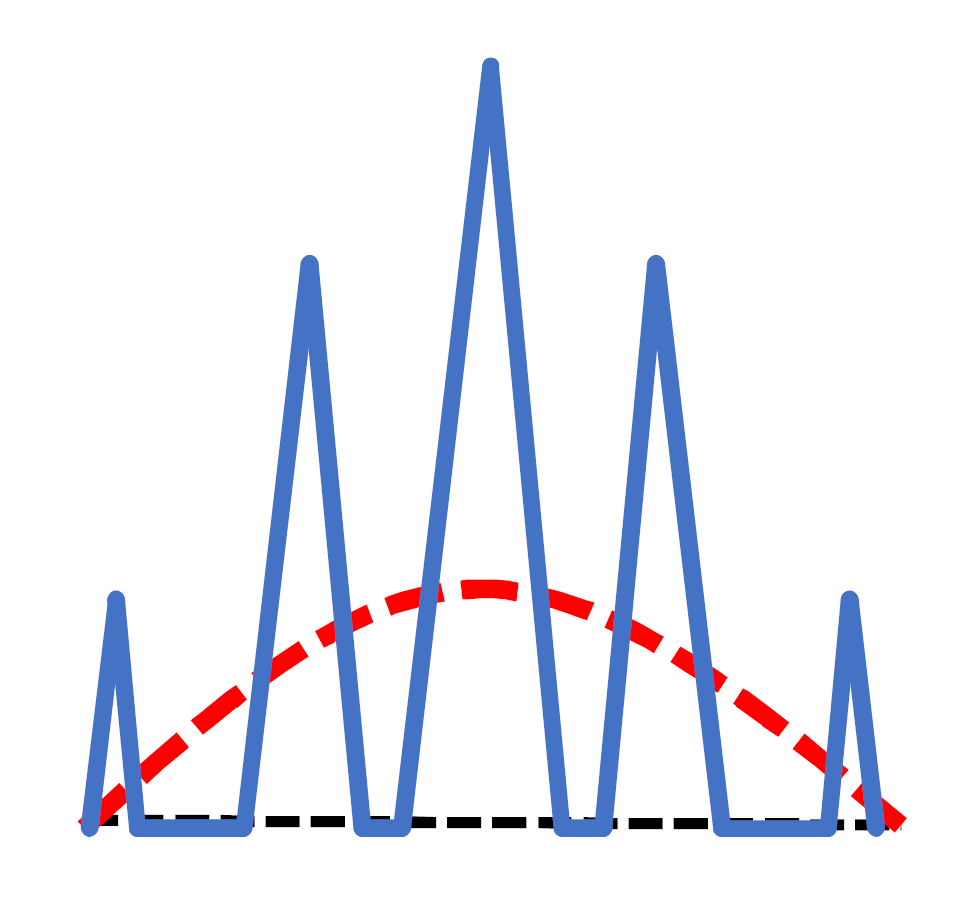}
			\end{minipage} &  \begin{minipage}{.06\textwidth}
				\centering
				\includegraphics[width=1\textwidth]{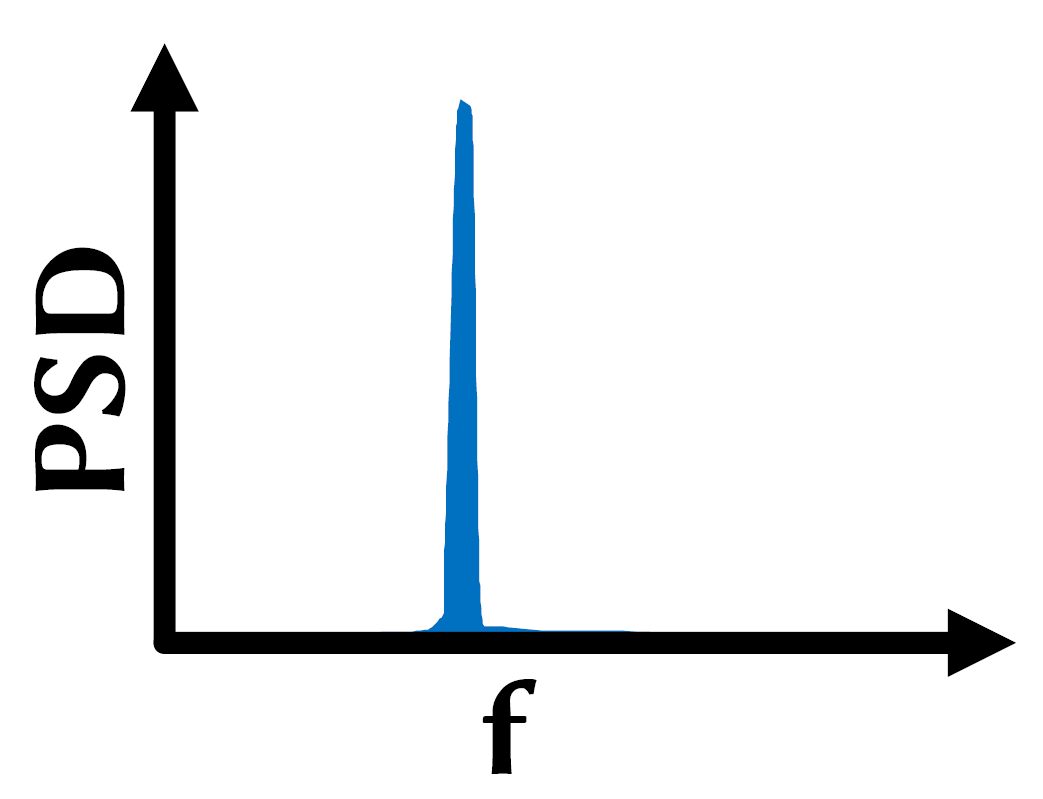}
			\end{minipage}
			& \begin{tabular}[l]{@{}l@{}}- High peak current\\- Reduced  inductance\\ - Good stability\end{tabular}
			\\ [1ex]\hline
			\begin{tabular}[c]{@{}c@{}}Critical\\ Conduction\\ Mode \textbf{(CrCM)}\end{tabular} & \textless{}300W &  \begin{minipage}{.06\textwidth}
				\centering
				\includegraphics[width=0.85\textwidth]{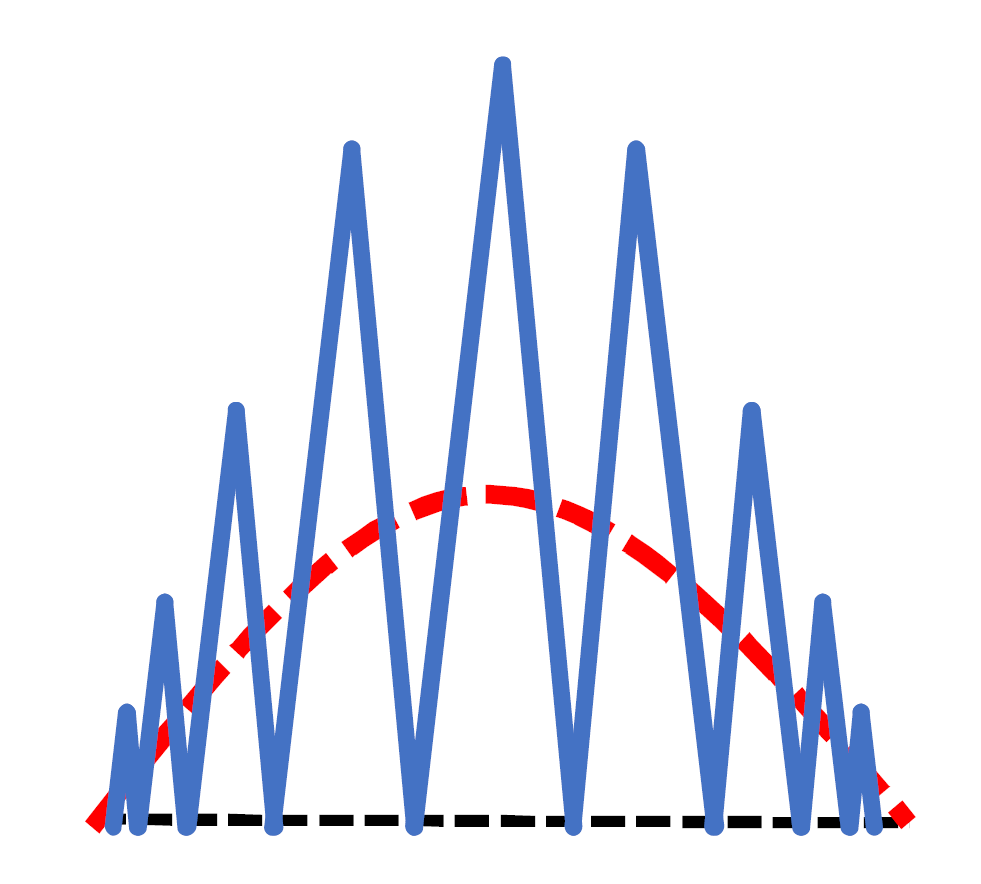}
			\end{minipage} &  \begin{minipage}{.06\textwidth}
				\centering
				\includegraphics[width=1\textwidth]{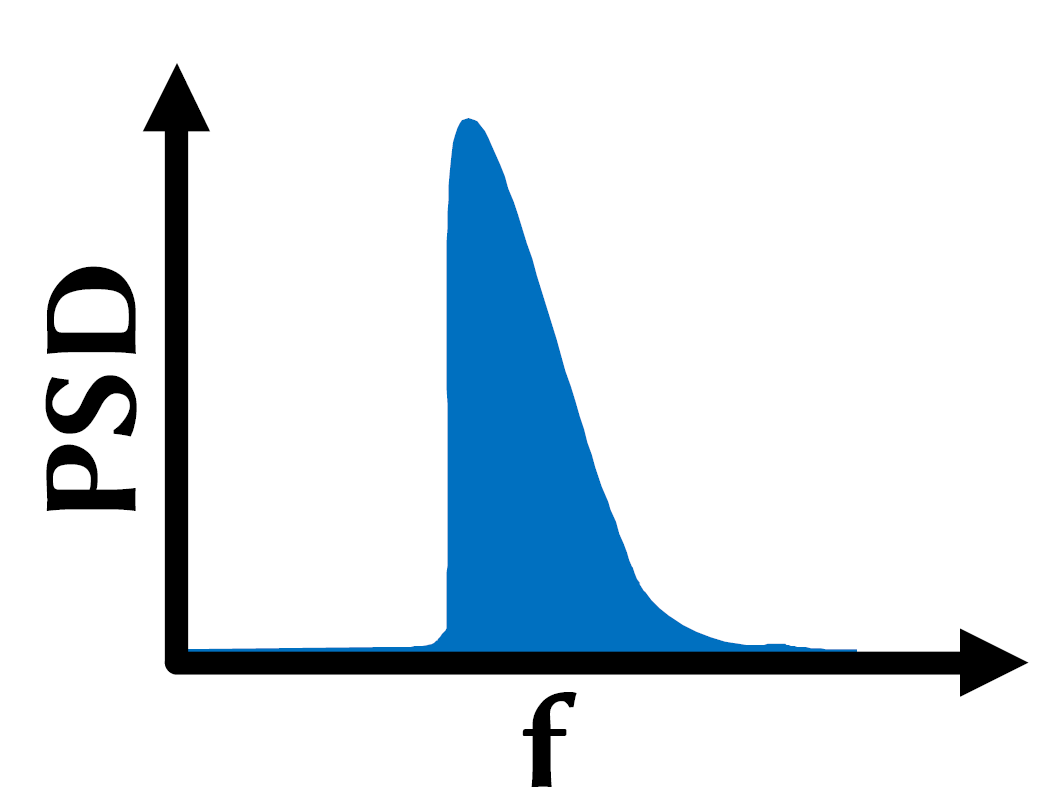}
			\end{minipage}
			& \begin{tabular}[l]{@{}l@{}}- Varying frequencies\\- High peak current\end{tabular}
			\\ [1ex]\hline
		\end{tabular}
	}
\end{table}

\subsection{PFC Switching Frequency Variation}\label{sec:appendix_pfc_variation}

\begin{figure}[!h]
	\centering
	\subfigure[CDF]{\label{fig:cdf_of_frequency_variation}\includegraphics[trim=0cm 0cm 0cm 0cm,clip,  width=0.23\textwidth,page=1]{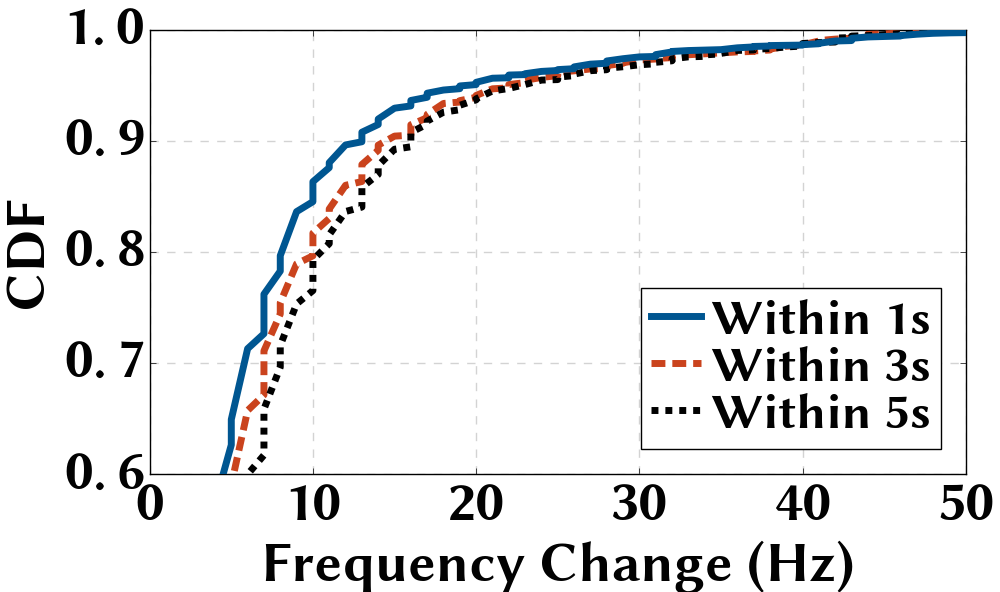}}
	\subfigure[Frequency spectrum]{\label{fig:PFC_frequency_psd_spectrum}\includegraphics[trim=0cm 0cm 0cm 0cm,clip,  width=0.23\textwidth,page=1]{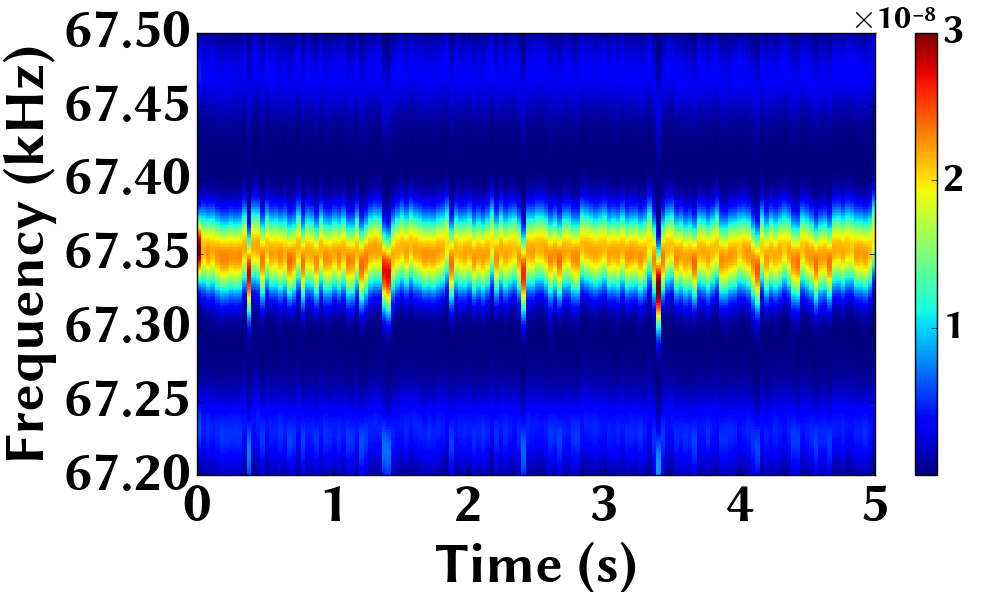}}
	\caption{PFC switching frequency variation over time.}
	\label{fig:PFC_frequency_variation}
\end{figure}

We show the cumulative density function (CDF) of our Dell Optiplex's PFC switching frequency variation in Fig.~\ref{fig:cdf_of_frequency_variation}. We see that the PFC
switching frequency varies no more than 50Hz within 5 seconds. In Fig.~\ref{fig:PFC_frequency_psd_spectrum}, we also show the frequency spectrum over a 5 second window and confirm that there is only a small variation in the PFC switching frequency.

\subsection{Current Drawn by Microsoft Surface Book}\label{sec:appendix_current_microsoft}
\begin{figure}[!h]
	\centering
	\subfigure[Current and voltage waveforms]{\label{fig:pfc_without}\includegraphics[trim=0cm 0cm 0cm 0cm,clip,  width=0.23\textwidth,page=1]{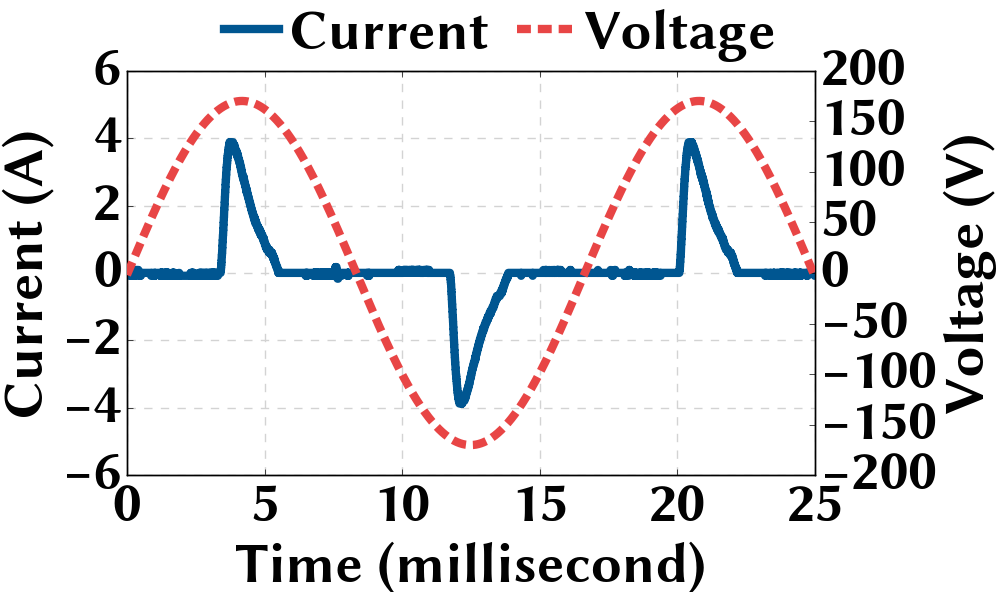}}\hspace{0.2cm}
	\subfigure[Frequent analysis of current]{\label{fig:pfc_without_psd_log}\includegraphics[trim=0cm 0cm 0cm 0cm,clip,  width=0.23\textwidth,page=1]{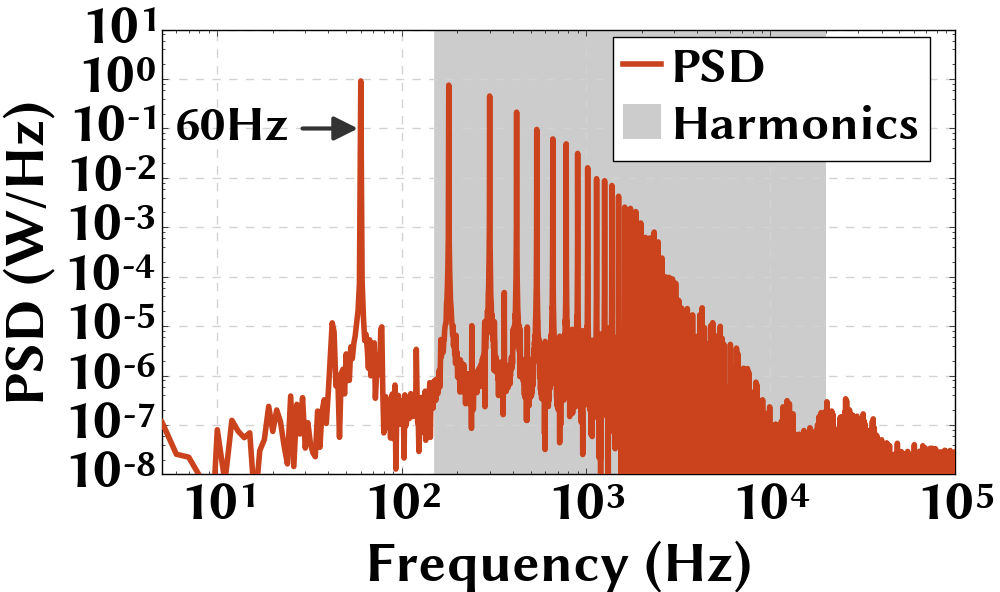}}
	\caption{The current drawn by a Microsoft Surface Book contains a significant amount of harmonics.}\label{fig:pfc_impact_harmonics}
\end{figure}

We show
the measured current drawn by a Microsoft Surface Book with a 65W power rating in Fig.~\ref{fig:pfc_without}
and the frequency analysis of the current in Fig.~\ref{fig:pfc_without_psd_log}.
We see that the current waveform looks by no means like
the sine voltage waveform and the total harmonic components are even stronger
than the 60Hz component in terms of power spectral density (PSD). Beyond 20kHz, the intensity of
harmonics decreases to an extremely low level, leaving only background noise.

\section{Algorithm for Current Modulator}\label{sec:appendix_current_modulation}

We design a simple current modulator as described in Algorithm~\ref{Alg:power_modulator}.
The current modulator program
takes 1/0 bit streams as the input and
runs some dummy calculations (e.g., generating random numbers) to load the CPU and change the computer's input current.
\begin{algorithm}[!h]
	\caption{Current Modulator}
	\begin{algorithmic}[1]
		\State Input: Bit stream $\mathbf{B}$ and symbol duration $T$
		\For {every bit $B_i$}
		\If {$B_i == 1$}
		\State Run dummy calculations for $T$ milliseconds
		\Else
		\State Idle for $T$ milliseconds
		\EndIf
		\EndFor
	\end{algorithmic}
	\label{Alg:power_modulator}
\end{algorithm}

\section{Data Exfiltration Results}

\subsection{Experiment on Dell Optiplex Computer}\label{sec:appendix_experiment_dell_optiplex}

\begin{figure*}[!h]
	\centering
	\subfigure[Youtube, bit error rate = 2.3\%, bit rate = 27.82 bits/s]{\label{fig:detection_youtube}\includegraphics[width=0.46\textwidth]{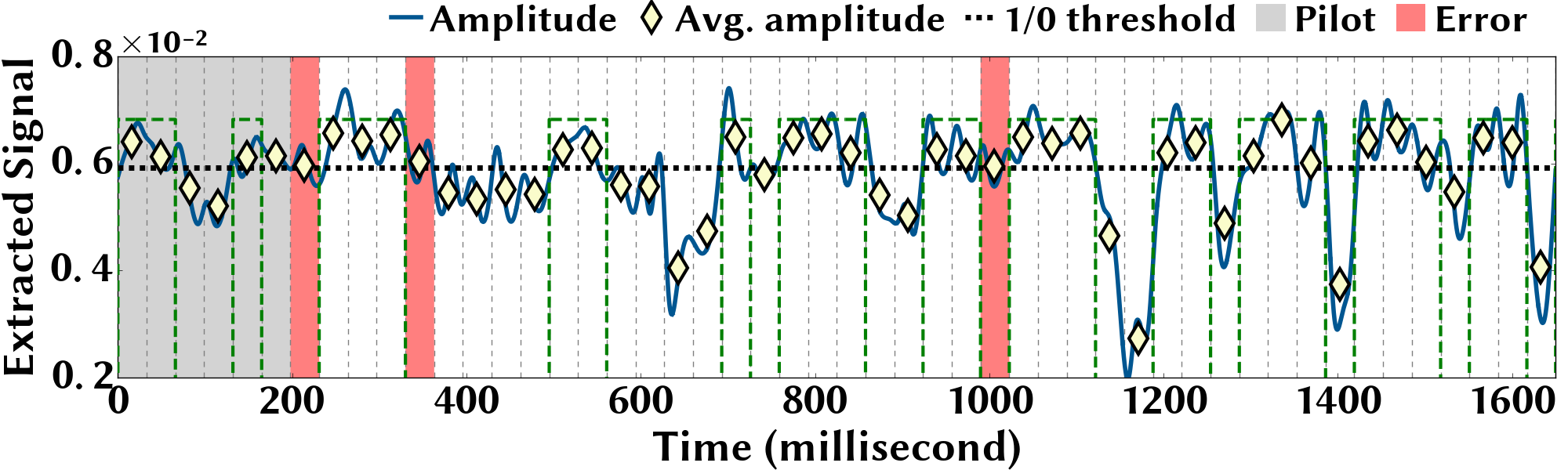}}\hspace{0.1cm}
	\subfigure[MS Word, bit error rate = 0\%, bit rate = 28.48 bits/s]{\label{fig:revision_detection_background_word_res}\includegraphics[width=0.46\textwidth]{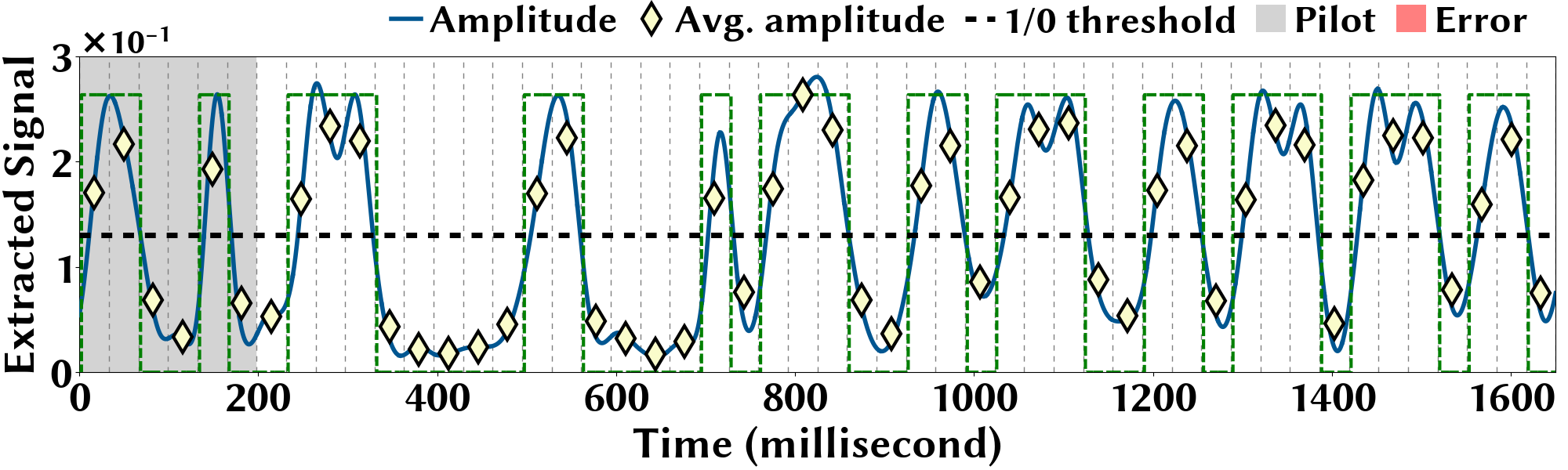}}
	\subfigure[Web browser, bit error rate = 0\%, bit rate = 28.48 bits/s]{\label{fig:revision_detection_background_wEB_res}\includegraphics[width=0.46\textwidth]{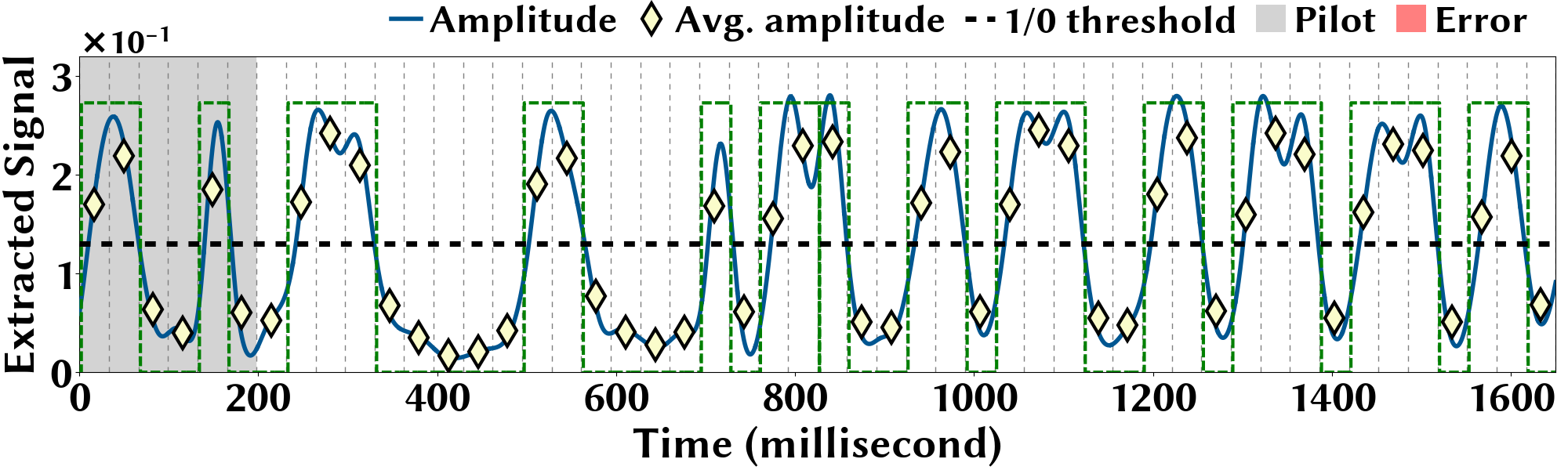}}\hspace{0.1cm}
	\subfigure[File transfer, bit error rate = 3.5\%, bit rate = 27.48 bits/s]{\label{fig:revision_detection_background_io_res}\includegraphics[width=0.46\textwidth]{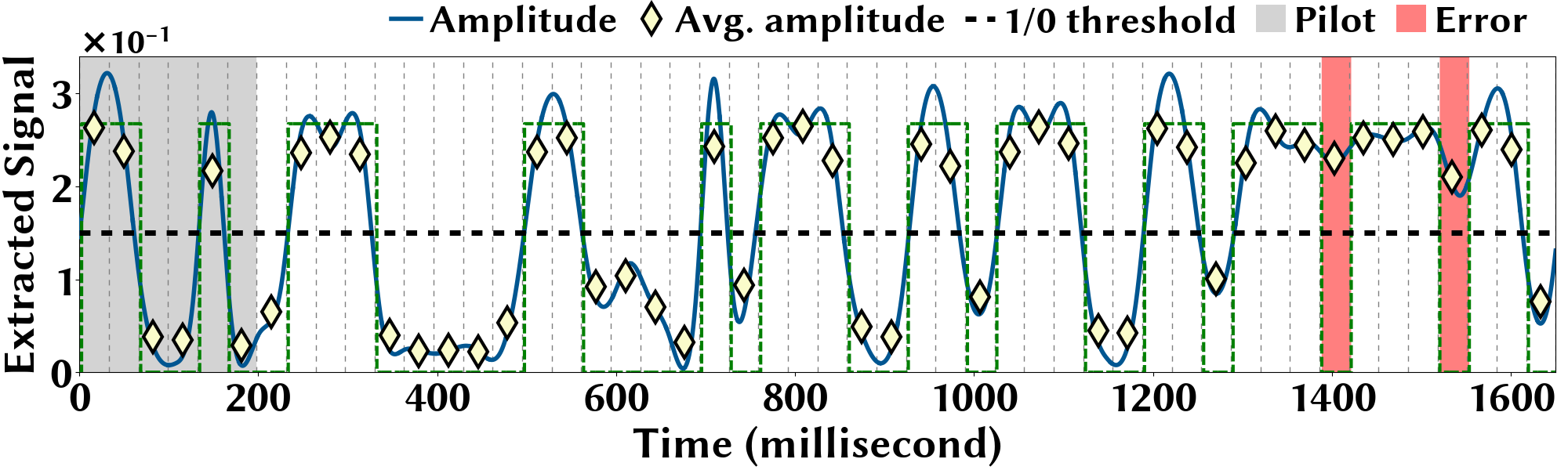}}
	\subfigure[Machine learning, bit error rate = 1.67\%, bit rate = 28 bits/s]{\label{fig:revision_detection_background_learning_res}\includegraphics[width=0.46\textwidth]{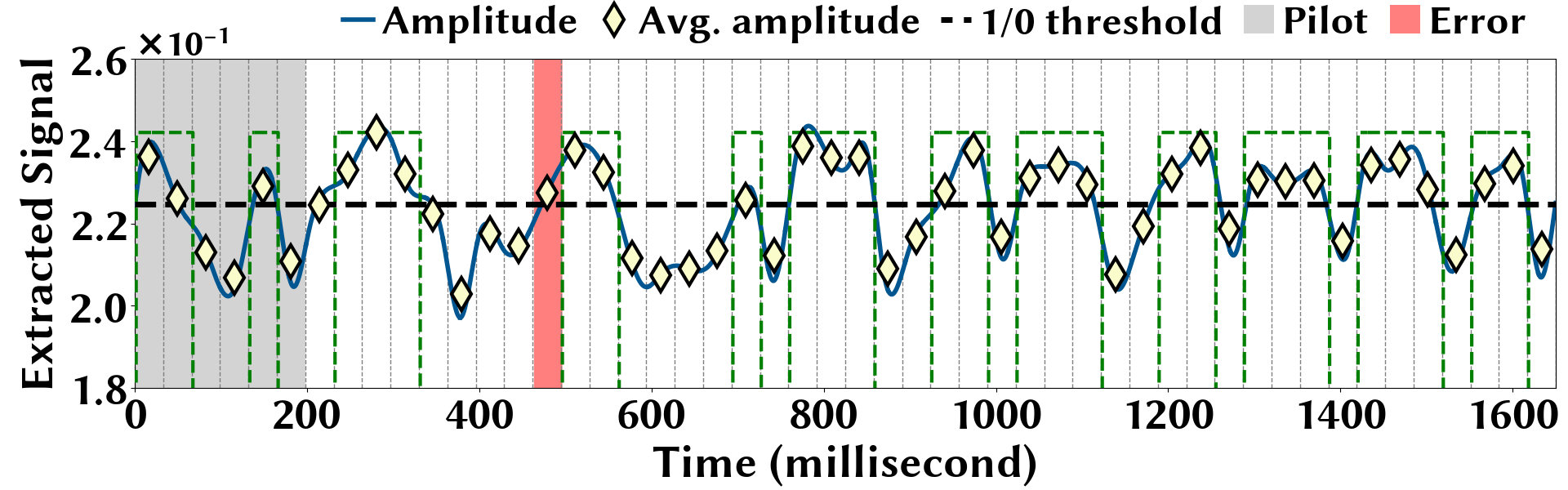}}
	\vspace{-0.2cm}	
	\caption{Different background applications.}\label{fig:revision_detection_background}
	\vspace{-0.0cm}	
\end{figure*}

\textbf{\OurAlg with different background application.}
Now, we run experiments on our Dell Optiplex computer
under settings different from the default one.
First, to run a concurrent program to mimic user's normal activity,
we play ``See You Again'' on YouTube on
a Google Chrome browser, which
is one of the most viewed videos \cite{YouTube_SeeYouAgain_Video}.
We also run experiments using MS Word, web browsing, file transfer, and machine learning training as background applications resulting in 0\%, 0\%, 3.5\%, and 1.67\% bit error rates, respectively. In the MS word experiment, we mimic user behavior by repeatedly opening a new file, typing a few lines of texts and then saving the file.  For the web browsing experiment, we open new popular websites (e.g., GMail and Facebook), scroll through the page content, and follow links to other pages. For the file transfer experiment, we transfer a 5GB file from one HDD drive to another in our desktop computer running Windows 10. For the machine learning experiment, we repeat training Tensorflow in Python with 6000 samples from  the MNIST data set taking around fives minutes to finish \cite{lecun1998gradient}. We show snapshots of the detections in Fig.~\ref{fig:revision_detection_background}.

\begin{figure*}[!h]
	\centering
	\subfigure[1 core. 8.9\% bit error rate,
	and 25.94 bits/second.]{\label{fig:detection_1core}\includegraphics[width=0.46\textwidth]{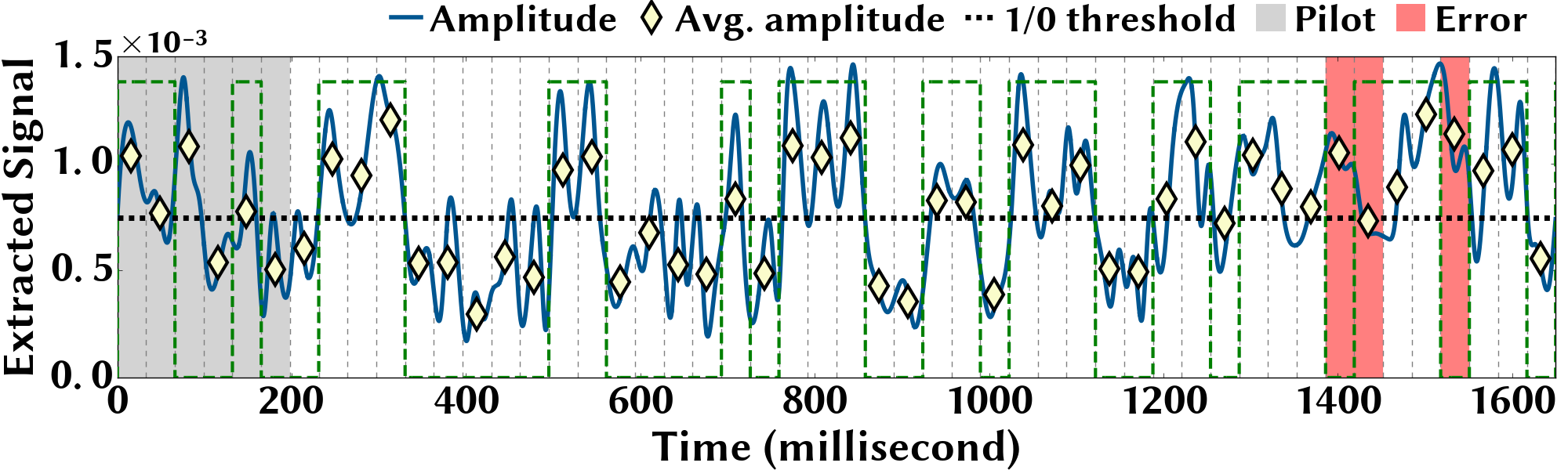}}\hspace{0.1cm}
	\subfigure[2 cores. 2.5\% bit error rate, 27.77 bits/second.]{\label{fig:detection_2core}\includegraphics[width=0.46\textwidth]{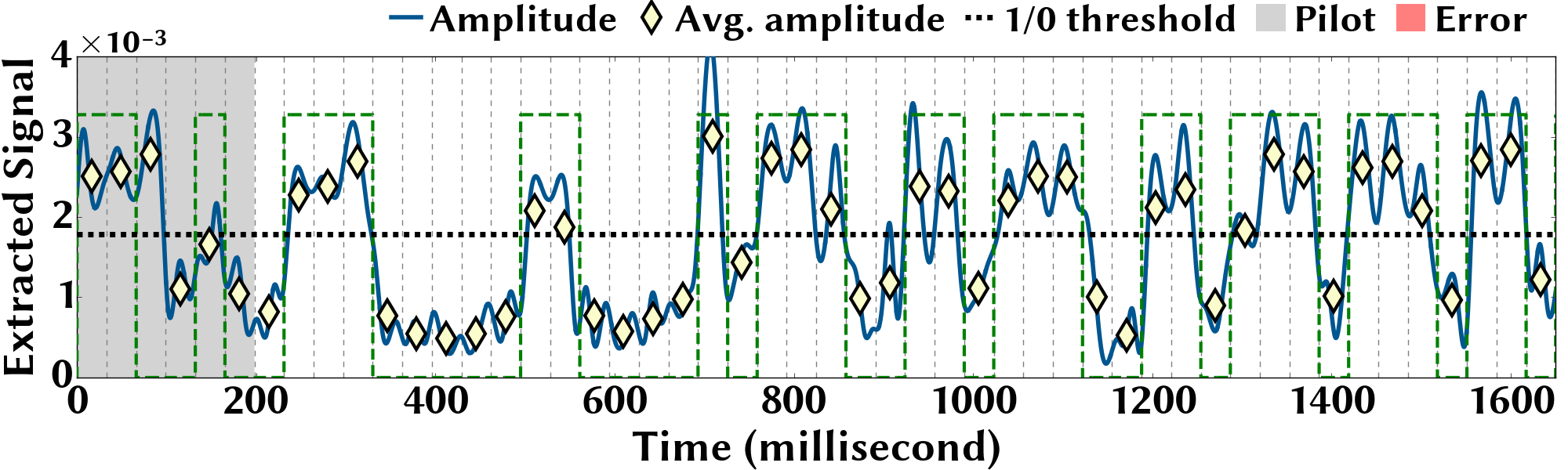}}
	\subfigure[3 cores. 0.0\%, 28.48 bits/second.]{\label{fig:detection_3core}\includegraphics[width=0.46\textwidth]{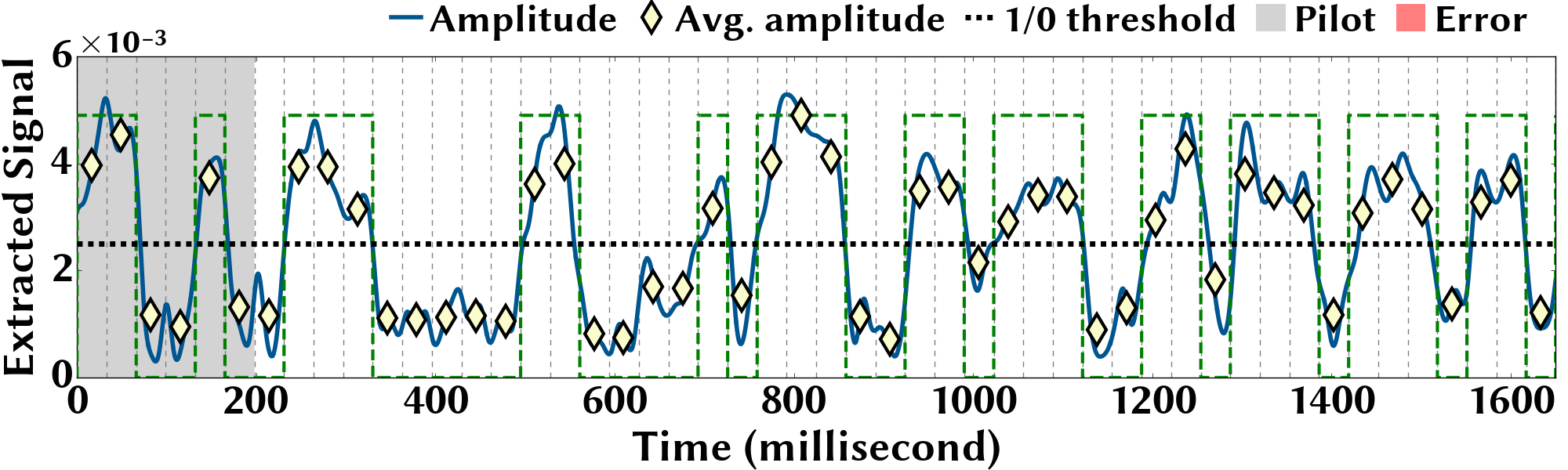}}
	\vspace{-0.2cm}	
	\caption{Dell Optiplex computer with different numbers
		of CPU cores assigned to the modulation program in \ouralg. The receiver filters its received voltage signals with passband of $<$67.28kHz, 67.34kHz$>$.}
	\vspace{-0.0cm}	
	\label{fig:appendix_dell_optiplex_cores_all}
\end{figure*}

\textbf{\OurAlg with different number of CPU cores.}
Next, we restrict the number of cores that are assigned to the modulation program in \ouralg, and show the experimental results in Fig.~\ref{fig:appendix_dell_optiplex_cores_all}.

\begin{figure*}[!h]
	\centering
	\subfigure[4-bit pilot. A snapshot of data exfiltration.]{\label{fig:detection_4bit_pilot}\includegraphics[width=0.44\textwidth]{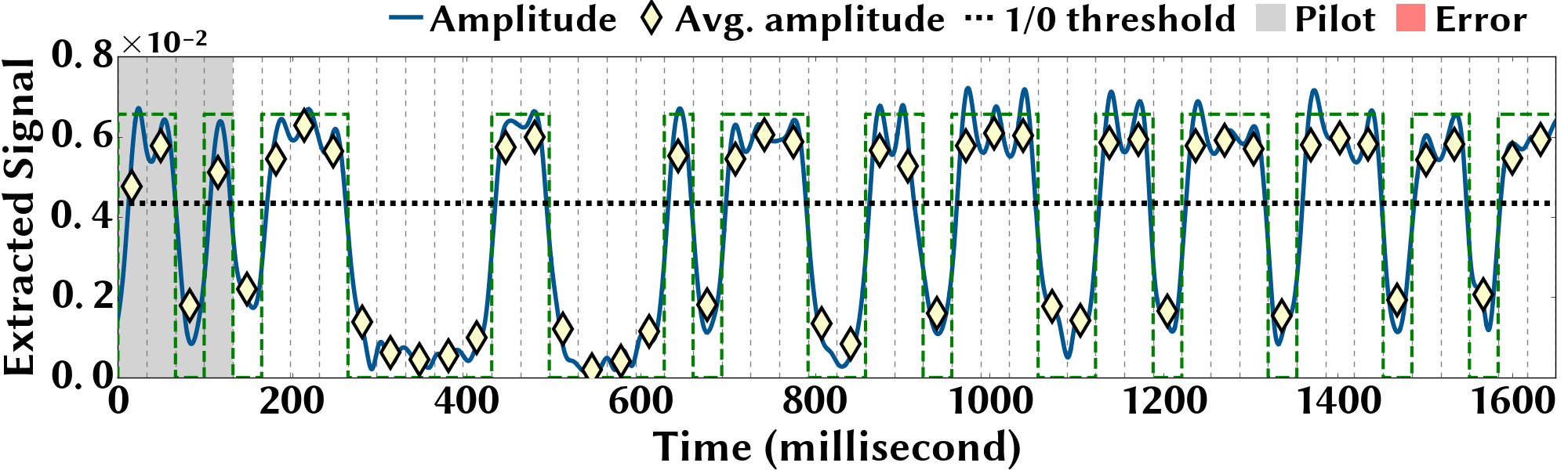}}\hspace{0.1cm}
	\subfigure[8-bit pilot. A snapshot of data exfiltration.]{\label{fig:detection_8bit_pilot}\includegraphics[width=0.44\textwidth]{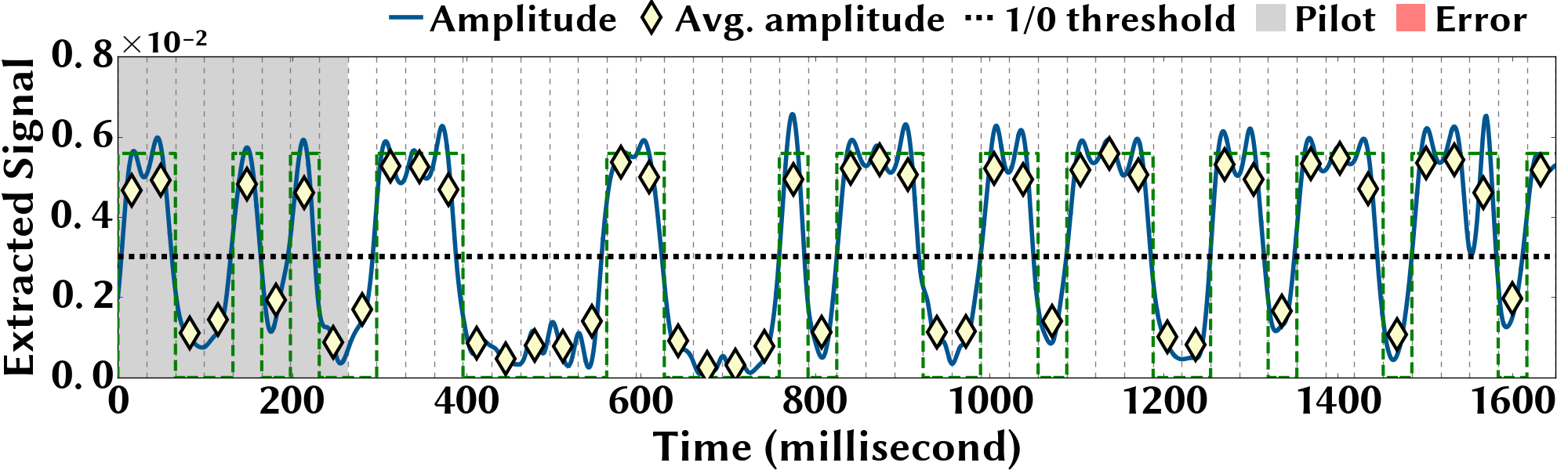}}
	\vspace{-0.2cm}	
	\caption{Dell Optiplex computer with different pilot sequences. The receiver filters its received voltage signals with passband of $<$67.28kHz, 67.34kHz$>$.}
	\vspace{-0.0cm}
	\label{fig:appendix_dell_optiplex_pilot_sequence_all}
\end{figure*}

\textbf{\OurAlg with different pilot lengths.} Finally, we consider 4-bit (``1101'') and 8-bit (``11001010'') pilot sequences, and show the experiment results in Fig.~\ref{fig:appendix_dell_optiplex_pilot_sequence_all}.

\subsection{Experiment Using Multiple Transmitters}\label{sec:apendix_multi_TX}

\begin{figure*}[!h]
	\centering
	\subfigure[TX\#1. 0.0\% bit error rate.]{\includegraphics[width=0.46\textwidth]{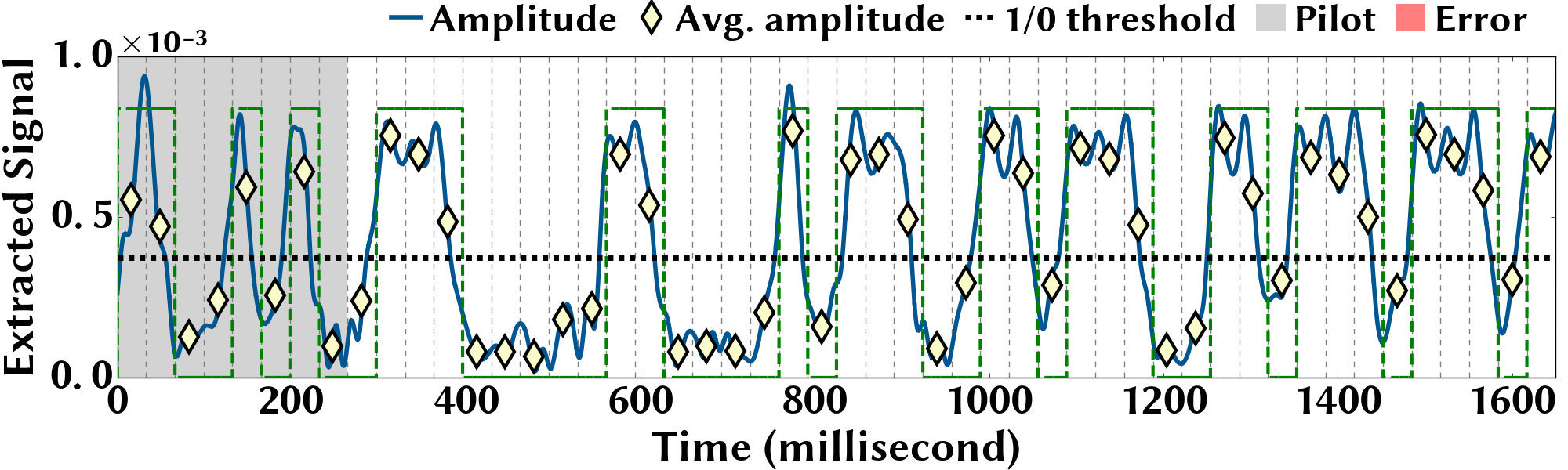}}\hspace{0.1cm}
	\subfigure[TX\#2. 6.8\% bit error rate.]{\includegraphics[width=0.46\textwidth]{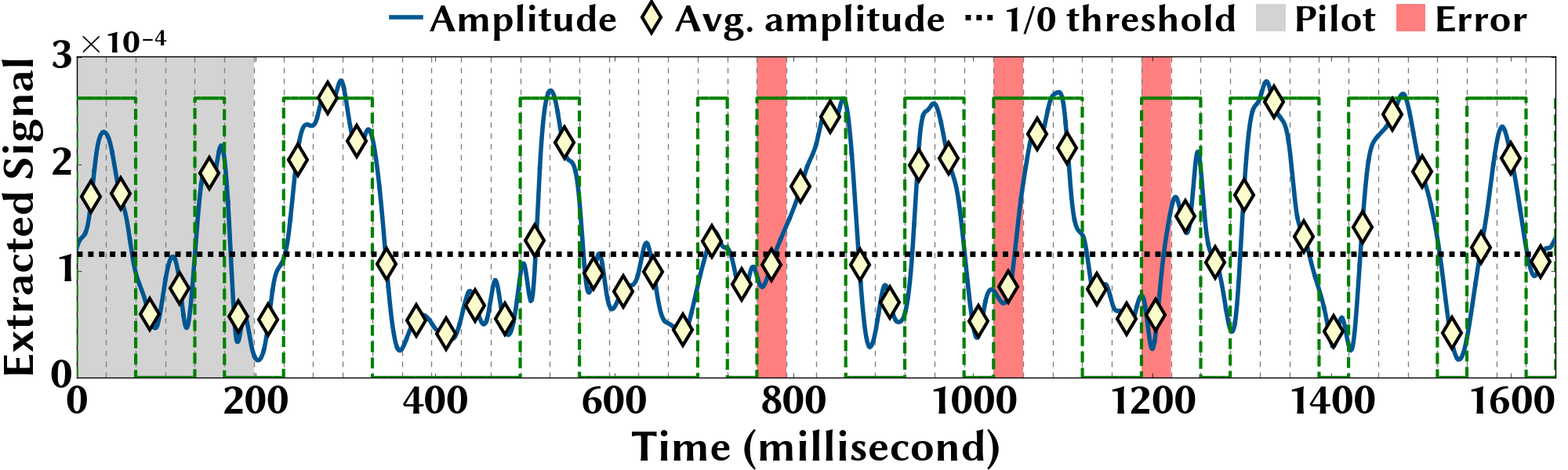}}
	\subfigure[TX\#3. 1.1\% bit error rate.]{\includegraphics[width=0.46\textwidth]{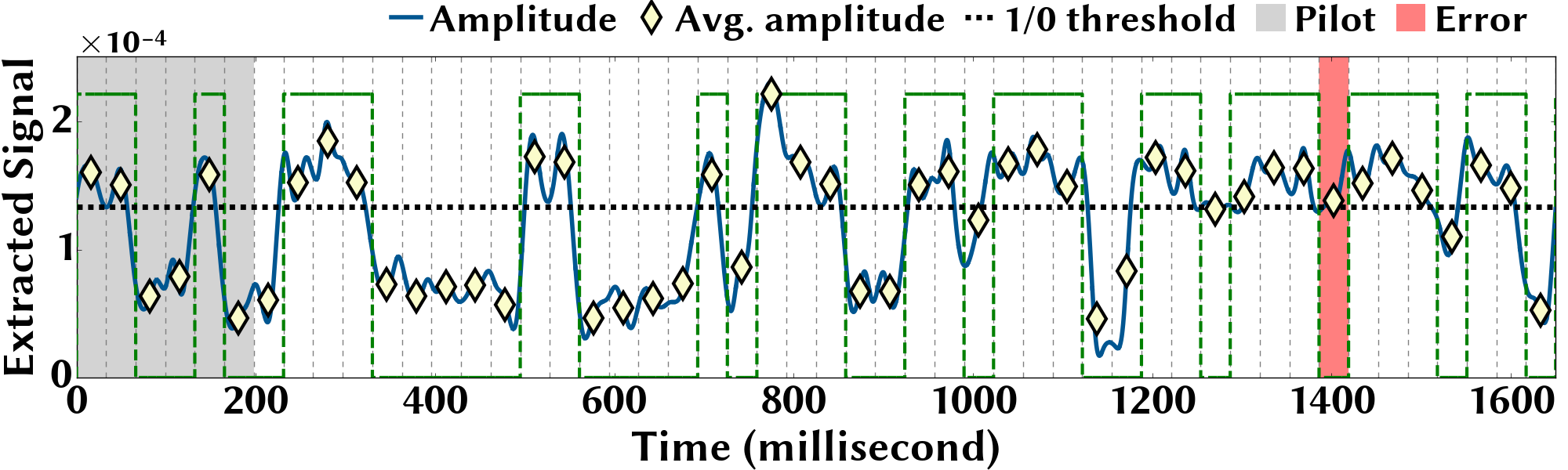}}
	\subfigure[TX\#4.  0.0\% bit error rate.]{\label{fig:detection_multi_TX_05}\includegraphics[width=0.46\textwidth]{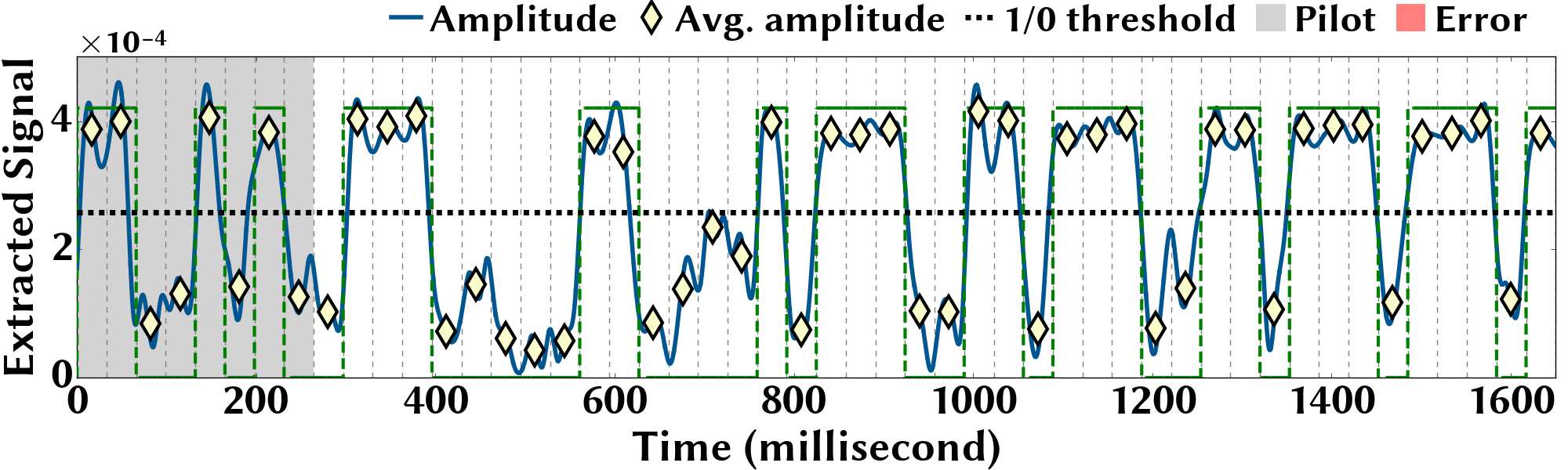}}
	\vspace{-0.2cm}	
	\caption{Snapshots of data exfiltration with multiple transmitters.}\label{fig:detection_multi_TX}
	\vspace{-0.0cm}	
\end{figure*}

Fig.~\ref{fig:detection_multi_TX} shows the snapshot of data exfiltration for different transmitters from our multi-transmitter experiment. Our results show 0\% error for TX\#1 and TX\#4 while 6.8\% and 1.1\% error for TX\#2 and TX\#3, respectively.

\subsection{Impact of CPU scaling on the transmitter}

\begin{figure*}[!h]
	\centering
	\subfigure[10\% maximum processor state, 8\% bit error rate, and 26.2 bits/second]{\includegraphics[width=0.46\textwidth]{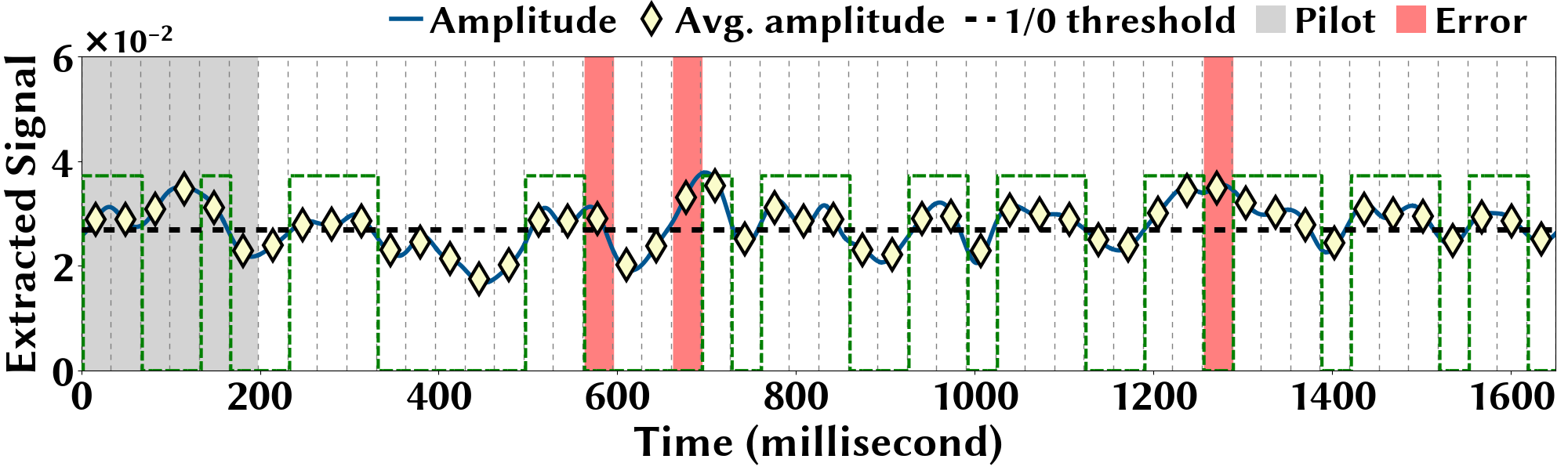}}\hspace{0.1cm}
	\subfigure[50\% maximum processor state, 0\% bit error rate, and 28.48 bits/second]{\includegraphics[width=0.46\textwidth]{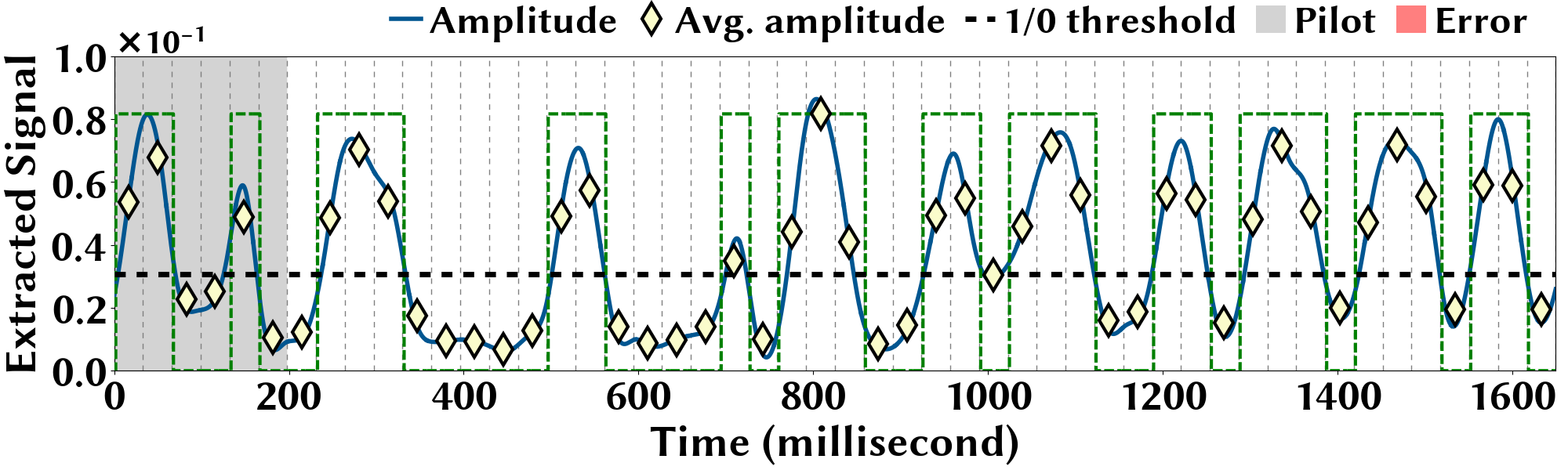}}
	\subfigure[100\% maximum processor state,0\% bit error rate, and 27.48 bits/second]{\includegraphics[width=0.46\textwidth]{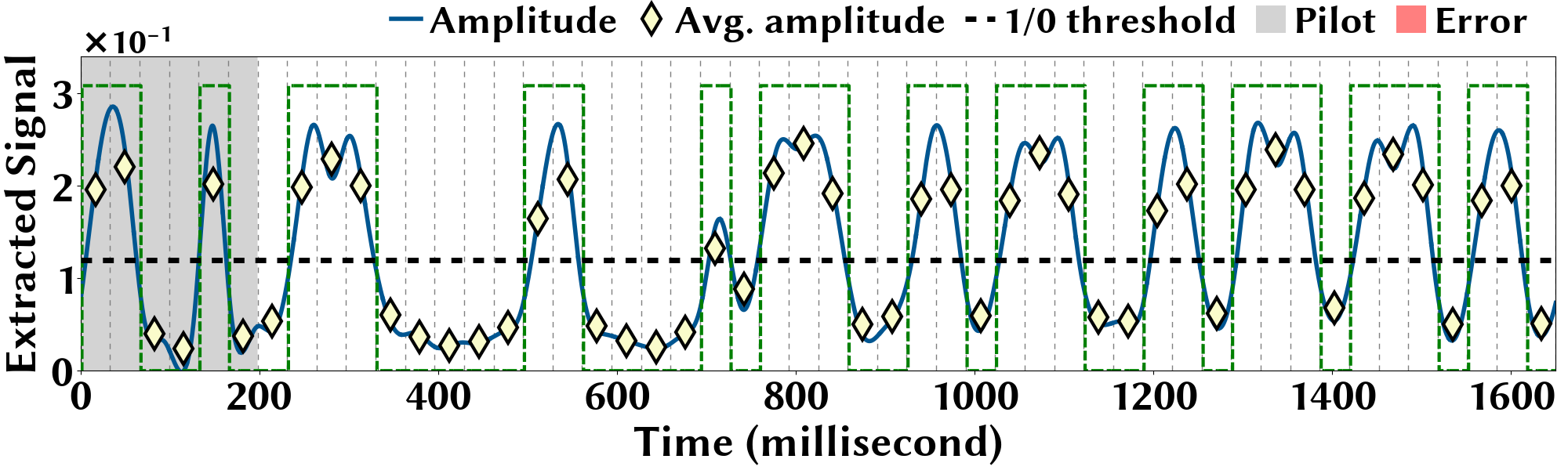}}
	\vspace{-0.2cm}	
	\caption{Dell Optiplex computer with different CPU's maximum
		power states.}\label{fig:detection_cpu_state}
	\vspace{-0.0cm}	
\end{figure*}

Snapshots of detection results for 10\%, 50\%, and 100\% CPU states are shown in Fig.~\ref{fig:detection_cpu_state}.

\subsection{Experiment Without Line of Sight}\label{sec:appendix_no_LOS}

\begin{figure*}[!h]
	\centering
	\includegraphics[trim=0cm 0cm 0cm 0cm,clip,  width=0.46\textwidth]{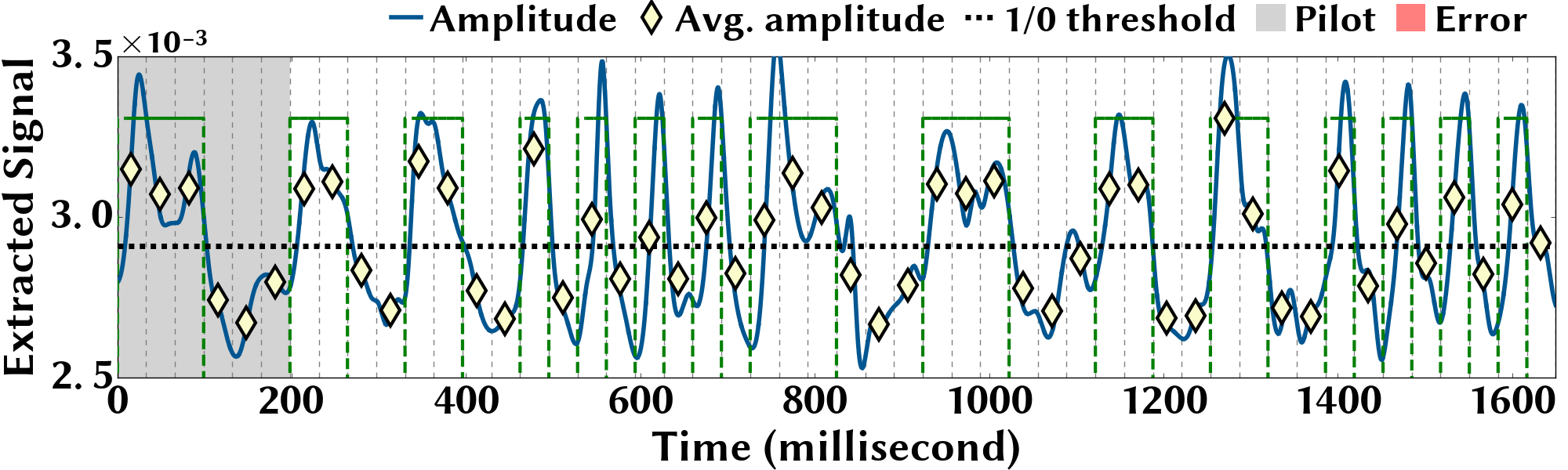}
	\vspace{-0.3cm}
	\caption{Dell PowerEdge computer in Building~B with no line of sight between the receiver and transmitter. The voltage signal is filtered
		with a passband of $<$65.77kHz, 65.83kHz$>$.}\label{fig:detection_office_other_room}
	\vspace{-0.0cm}
\end{figure*}

Fig.~\ref{fig:detection_office_other_room} shows the snapshot of data exfiltration when the receiver and transmitters are placed in two separate rooms 90 feet away from each other in Building B.

\subsection{Experiment on Other Computers}\label{sec:appendix_other_computers}

\begin{figure*}[!h]
	\centering
	\subfigure[Acer]{\label{fig:detection_acer} \includegraphics[width=0.44\textwidth]{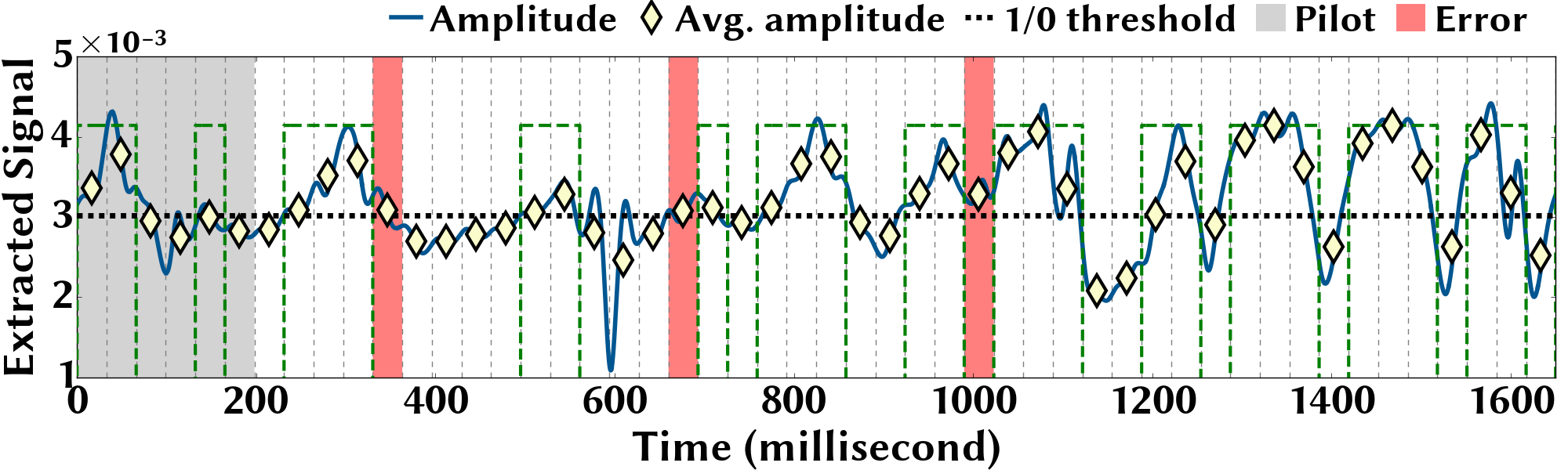}}\hspace{0.1cm}
	\subfigure[Custom built \#1]{\label{fig:detection_corsair} \includegraphics[width=0.44\textwidth]{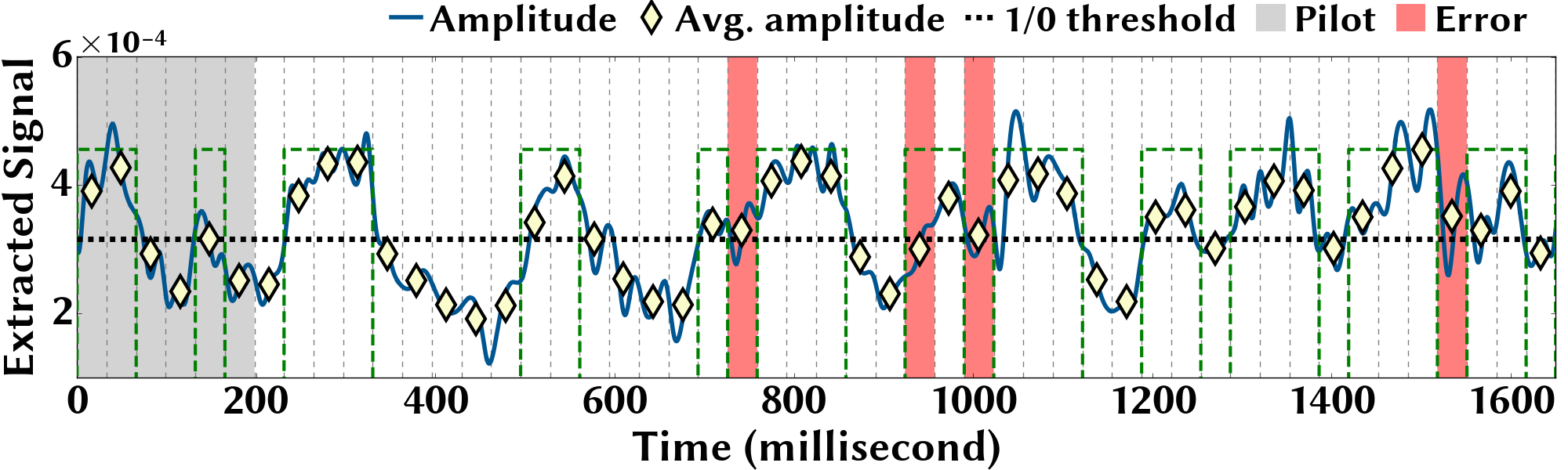}}
	\subfigure[Custom built \#2]{\label{fig:detection_kim} \includegraphics[width=0.44\textwidth]{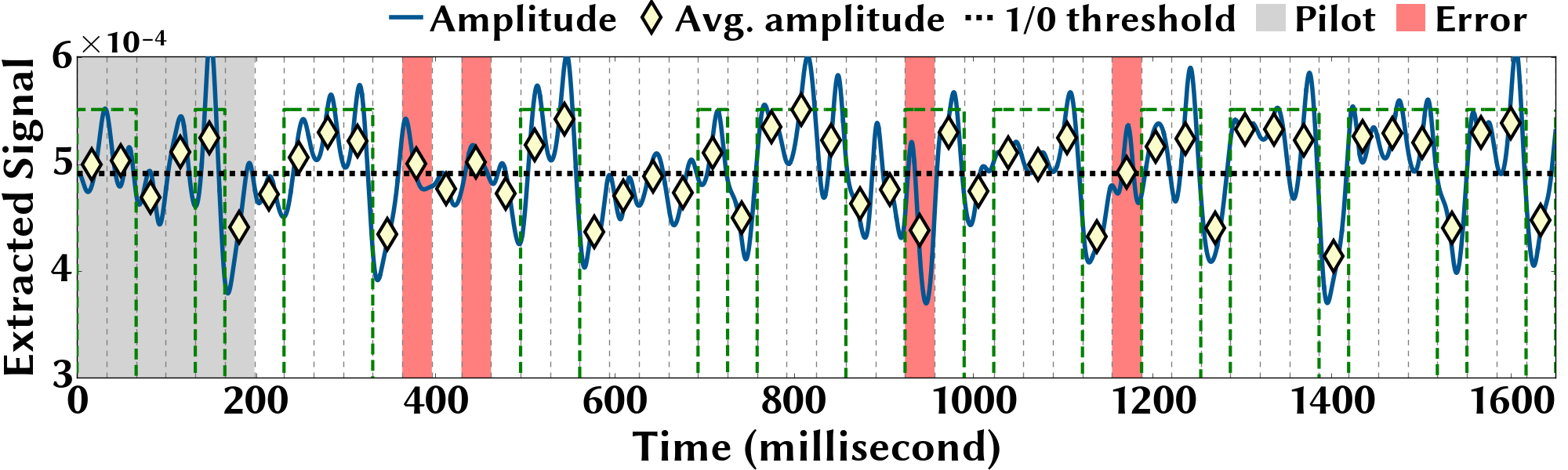}}
	\vspace{-0.2cm}	
	\caption{Data exfiltration from different computers.}\label{fig:detection_other}
	\vspace{-0.0cm}	
\end{figure*}

Fig.~\ref{fig:detection_other}
shows the experiment results for our Acer computer,
our custom built computer \#1,
and our custom built computer \#2.

\subsection{Experiment on iMac}\label{sec:appendix_imac}

We conduct our experiment on an Apple iMac 27-inch computer in Lab \#1. The transmitter and the receiver are placed 55 feet away from each other.
The iMac has an all-in-one compact design with monitor and other components assembled together and powered by a single power supply unit.

\begin{figure}[!t]
	\centering	
	\subfigure[Bit length = 100ms]{\includegraphics[trim=0cm 0cm 0cm 0cm,clip, width=0.46\textwidth]{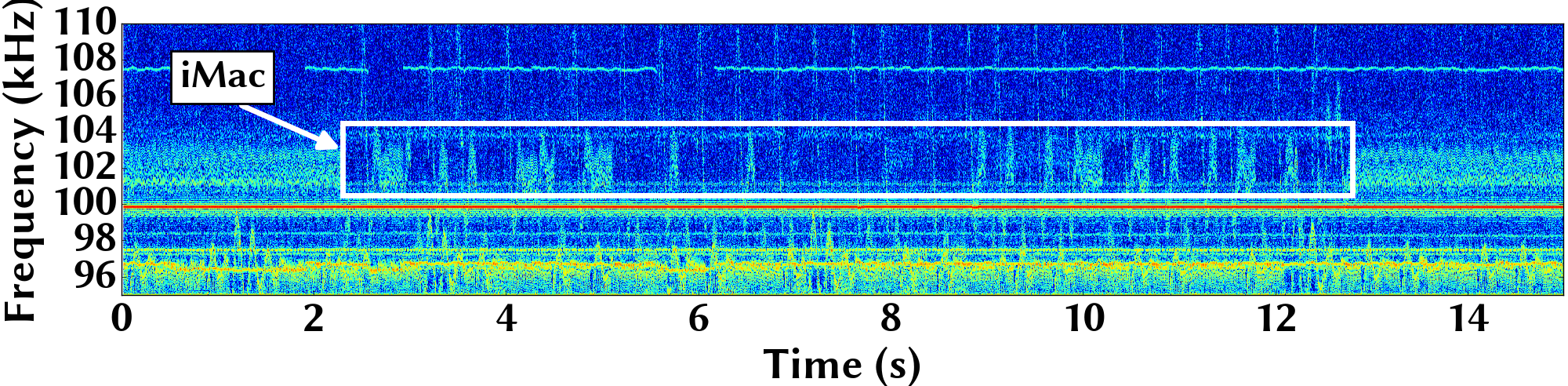}\label{fig:psd_spectrum_iMac_100ms}}
	\subfigure[Bit length = 50ms]{\includegraphics[trim=0cm 0cm 0cm 0cm,clip, width=0.46\textwidth]{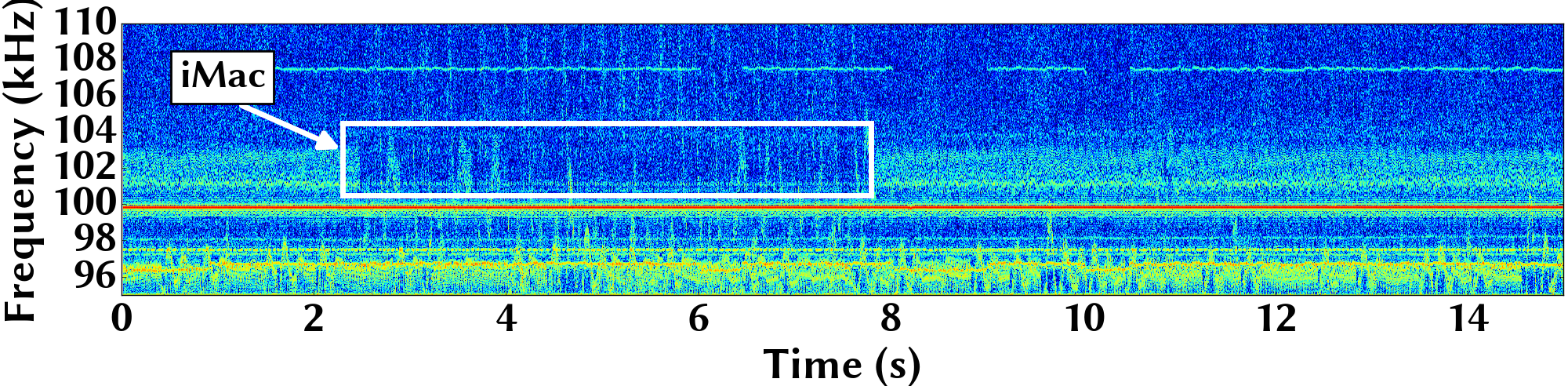}\label{fig:psd_spectrum_iMac_50ms}}
	\caption{Frequency spectrum showing a data frame transmitted by the iMac.}\label{fig:psd_spectrum_iMac}
\end{figure}

We first show the frequency spectrum for a data frame transmitted from the iMac with a symbol length of 100ms in Fig.~\ref{fig:psd_spectrum_iMac_100ms}, which reveals that the PFC-induced
frequency spike of the iMac is around 101kHz.
It also shows that frequency band that carries the transmitted data has a much wider bandwidth of $\sim$1kHz compared to the Dell computers' frequency bands in Fig.~\ref{fig:spectrum_voltage_multi}.
Further, while not explicitly indicated in Fig.~\ref{fig:psd_spectrum_iMac_100ms}, the frequency amplitudes in the transmitted band is reversed, where the ``1''s has lower amplitude and ``0''s has
a higher amplitude. We also show the frequency spectrum for a symbol length of 50ms in Fig.~\ref{fig:psd_spectrum_iMac_50ms}. The figures show the detection accuracy significantly deteriorates as we increase the symbol rate. In fact, we can not have successful data transmission for our default symbol length of 33ms. This is mainly due to
the iMac's power supply unit has a slower response to the power demand compared to
other computers in our experiments.

\begin{figure*}[!h]
	\centering
	\subfigure[Symbol length = 50ms, bit error rate = 16\%, and bit per second = 15.79]{\includegraphics[width=0.46\textwidth]{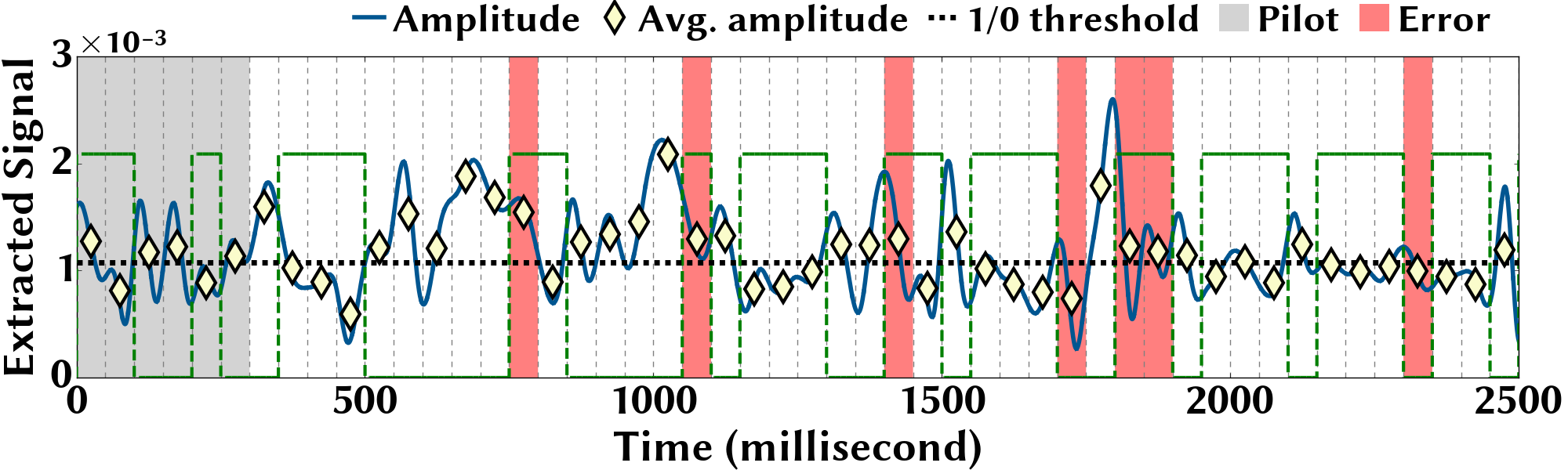}}\hspace{0.1cm}
	\subfigure[Symbol length = 66ms, bit error rate = 8\%, and bit per second = 13.1]{\includegraphics[width=0.46\textwidth]{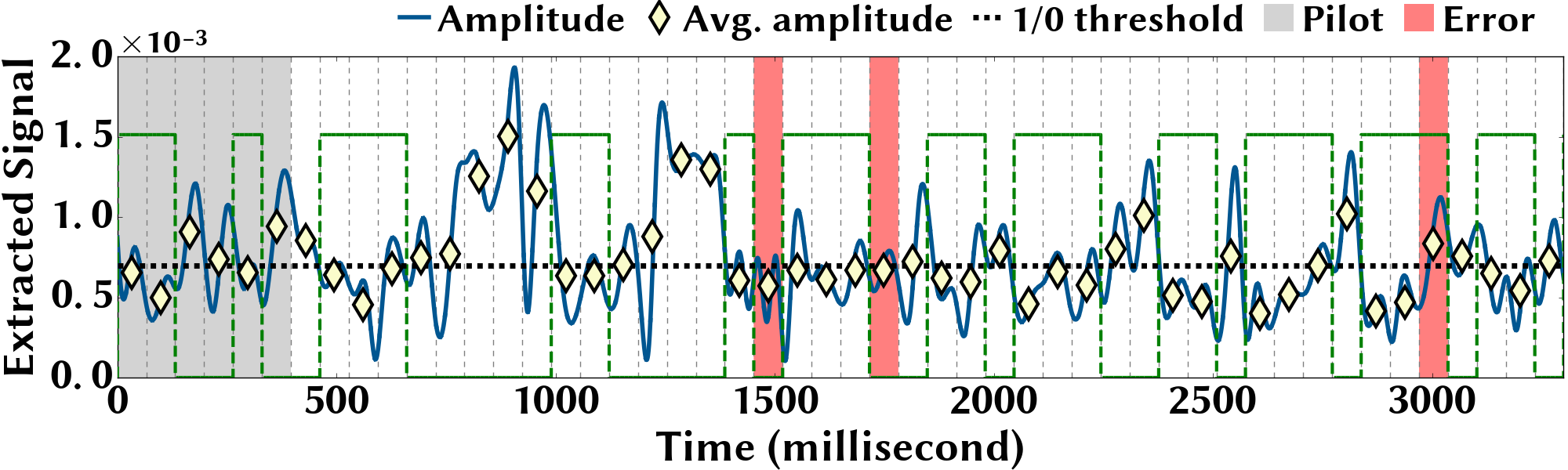}}
	\subfigure[Symbol length = 100ms, bit error rate = 2\%, and bit per second = 9.2]{\includegraphics[width=0.46\textwidth]{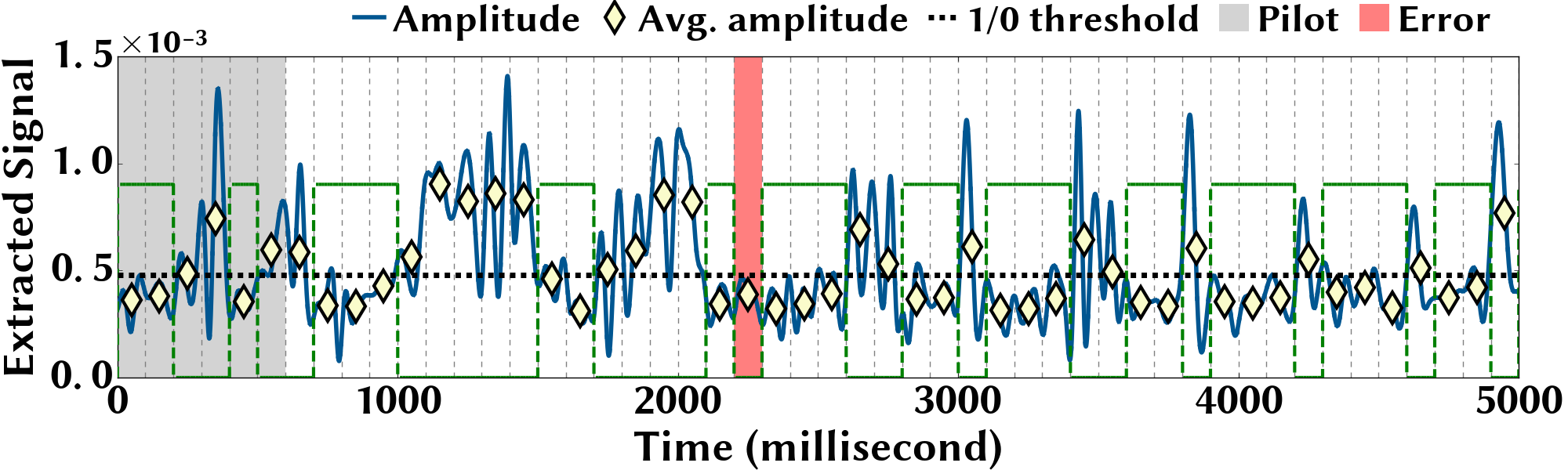}}\hspace{0.1cm}
	\subfigure[Symbol length = 125ms, bit error rate = 0\%, and bit per second = 7.52]{\includegraphics[width=0.46\textwidth]{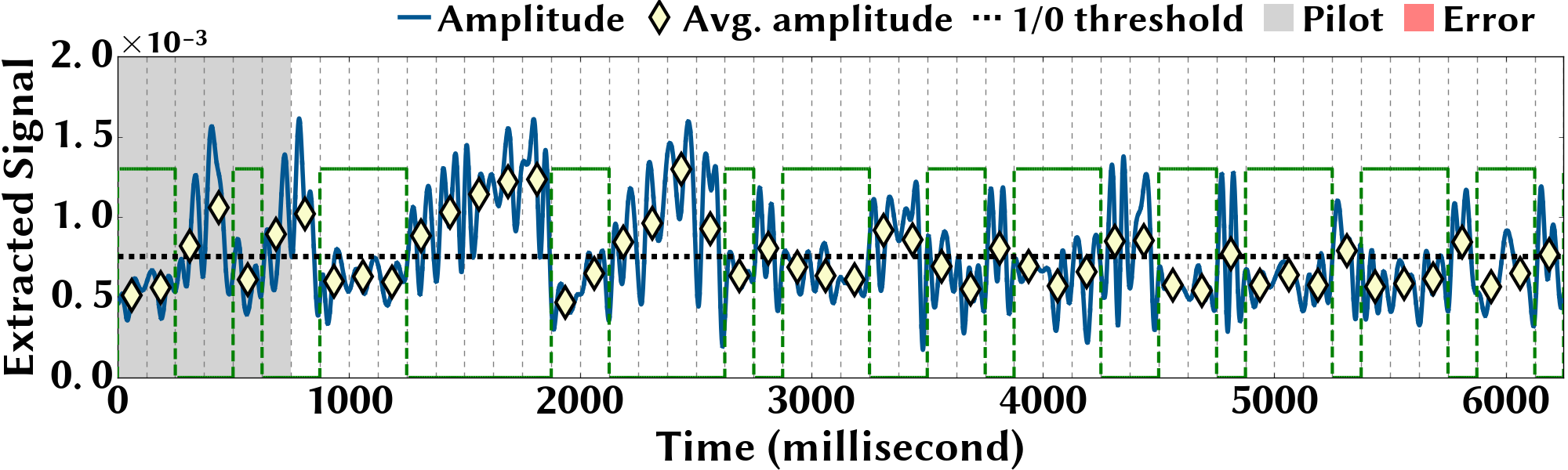}}
	\vspace{-0.2cm}	
	\caption{Snapshots of data exfiltration with iMac computer for different bit durations.}\label{fig:detection_iMac}
	\vspace{-0.0cm}	
\end{figure*}

Next, we show a snapshot of data exfiltration from the iMac for four different symbol lengths ranging from 50ms to 125ms in Fig.~\ref{fig:detection_iMac}. As discussed before, the bit error rate decreases as we increase the symbol length. However, since a higher symbol length means a lower maximum bits/second, the overall bit transfer rate decreases. 


\section{PMF of Noise in Extracted Signal Amplitudes}

\begin{figure}[!h]
	\centering
	\subfigure[Dell Optiplex computer]{\label{fig:detection_histogram_noise_lab_demo}\includegraphics[width=0.23\textwidth]{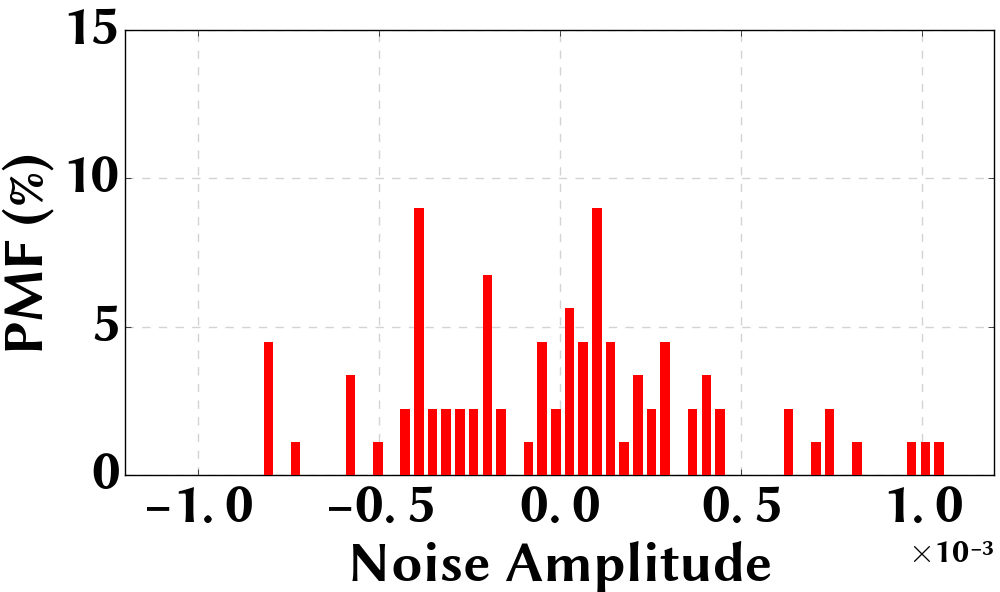}}\hspace{0.1cm}
	\subfigure[Dell PowerEdge computer]{\label{fig:detection_histogram_noise_office_other_room}\includegraphics[width=0.23\textwidth]{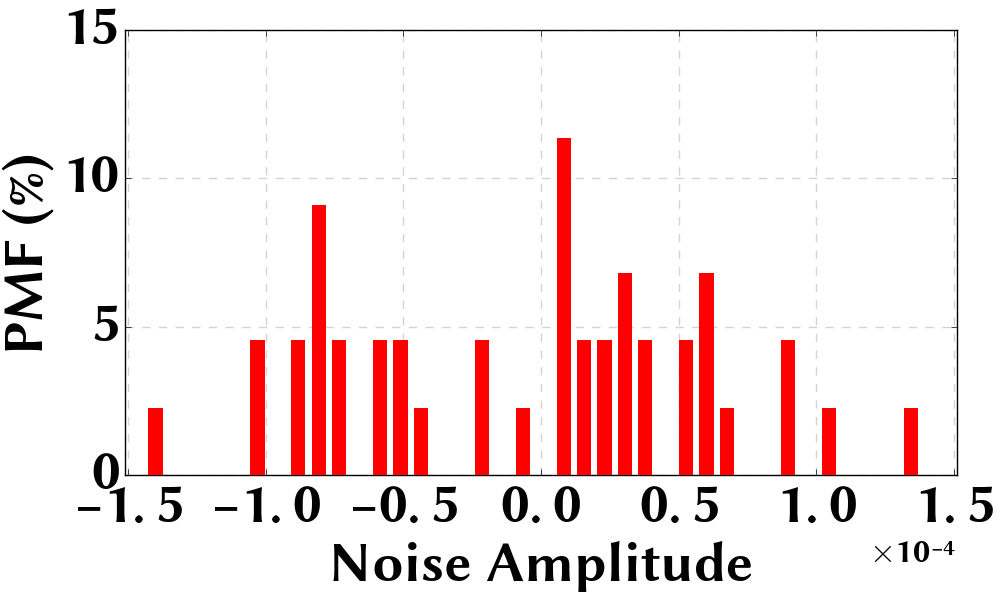}}
	\caption{PMF of noise in extracted signal amplitudes.}
	\label{fig:detection_histogram_noise_both}
\end{figure}

We set the average signal amplitude of filtered voltage
signals when transmitted bits are 1 as
the reference signal amplitude for bit 1.
When the transmitted bit is 1, any deviation of
an actually received bit-wise signal amplitude from the reference value is considered as noise.
Similarly, we also obtain the reference signal amplitude for bit 0, and obtain the noise.
Next, we show the PMF of noise for both Dell Optiplex and Dell PowerEdge computers in Fig.~\ref{fig:detection_histogram_noise_both}. The noise amplitude
distribution does not seem to be Gaussian. In other words, our covert channel
is likely corrupted by non-Gaussian noises.


\section{Defense Mechanism Experiments}

\subsection{Experiment on a UPS-Powered Computer}\label{sec:appendix_ups_computer}

\begin{figure*}[!h]
	\centering
	\includegraphics[width=0.44\textwidth]{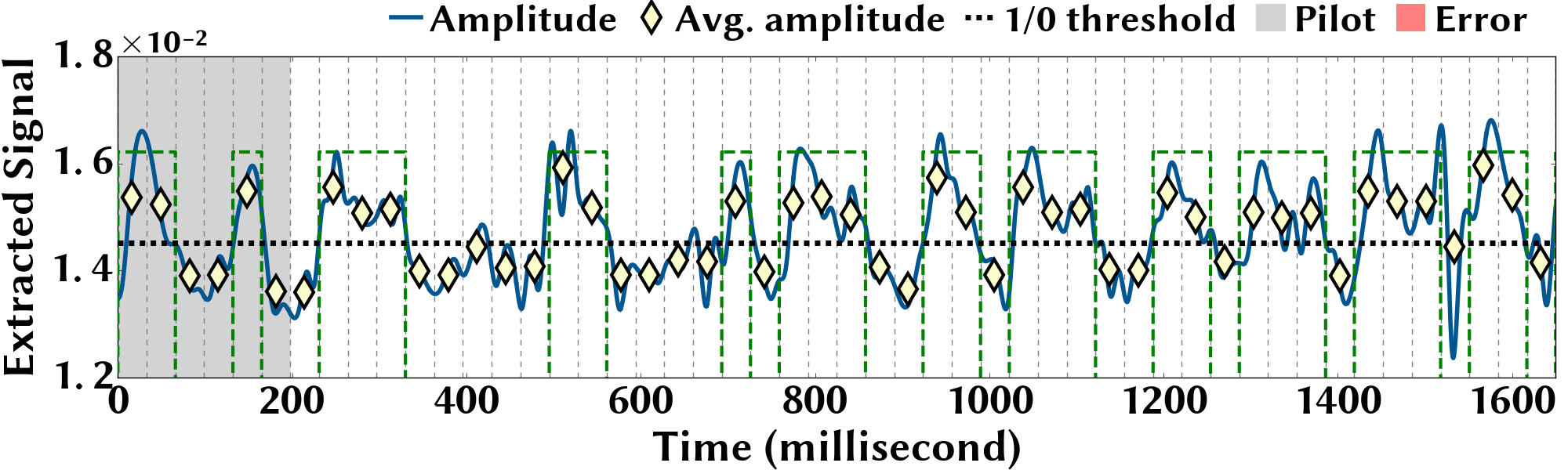}
	\vspace{-0.2cm}
	\caption{Dell PowerEdge computer powered by a CyberPower UPS. The voltage signal is filtered
		with a passband of $<$65.77kHz, 65.83kHz$>$. The resulting bit error rate is 0.0\% and the effective bit rate is 28.48 bits/second.}\label{fig:detection_server_ups}\label{fig:detection_result_ups_all}
	\vspace{-0.0cm}	
\end{figure*}

We power our Dell PowerEdge computer through
a CyberPower UPS, which is plugged into a power outlet in
the transmitter's room in Building B.
In Fig.~\ref{fig:detection_server_ups},
we show that the receiver can still extract information without errors.
Thus, an UPS-powered computer does not
necessarily mean it is immune to the threat of \ouralg, let alone
its added UPS cost.

\subsection{Experiment in the Presence of Noise Filters}\label{sec:appendix_powerline_noise_Filter}

\begin{figure*}[!h]
	\centering
	\includegraphics[width=0.44\textwidth]{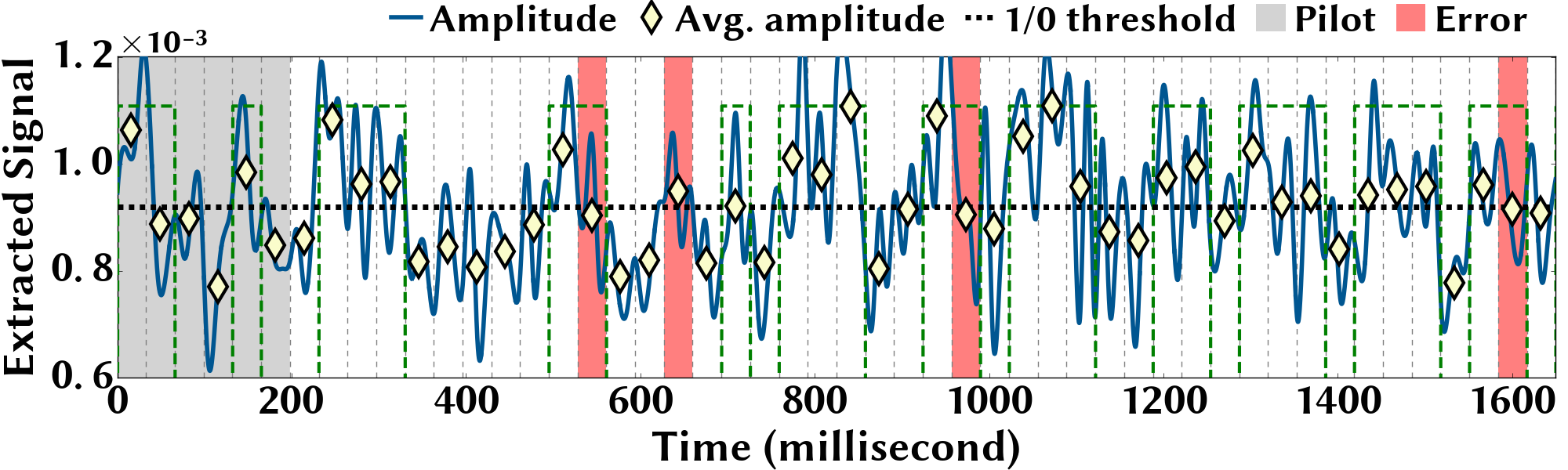}
	\vspace{-0.2cm}
	\caption{Dell PowerEdge computer with a power line noise filter plugged into
		the power outlet. The voltage signal is filtered
		with a passband of $<$65.78kHz, 65.84kHz$>$. The resulting bit error rate is 10.2\% and the effective bit rate is 25.57 bits/second.}\label{fig:detection_server_plc_filter}\label{fig:detection_result_plc_filter_all}
	\vspace{-0.0cm}	
\end{figure*}

We plug in noise filter (X10 XPPF) \cite{PFC_power_line_filter}
into
a power outlet in the transmitter's room in Building B,
and then plug in the power cord of our PowerEdge computer
into the noise filter.
Although the intensity of high-frequency PSD spikes
in the receiver's voltage signal is reduced, it is still much higher
than the power line background noise and detectable. In Fig.~\ref{fig:detection_server_plc_filter},
we show that the receiver can have an effective rate of 25.57 bits/second. Although
the transmission distance is reduced compared to the no-filter case, the receiver
can still be located in another room without being restricted
to the line of sight of the transmitter.

\subsection{Random Power Load Defense}\label{sec:appendix_random_power}

\begin{figure}[!h]
	\centering
	\subfigure[]{\label{fig:power_different_noise_main}\includegraphics[width=0.23\textwidth]{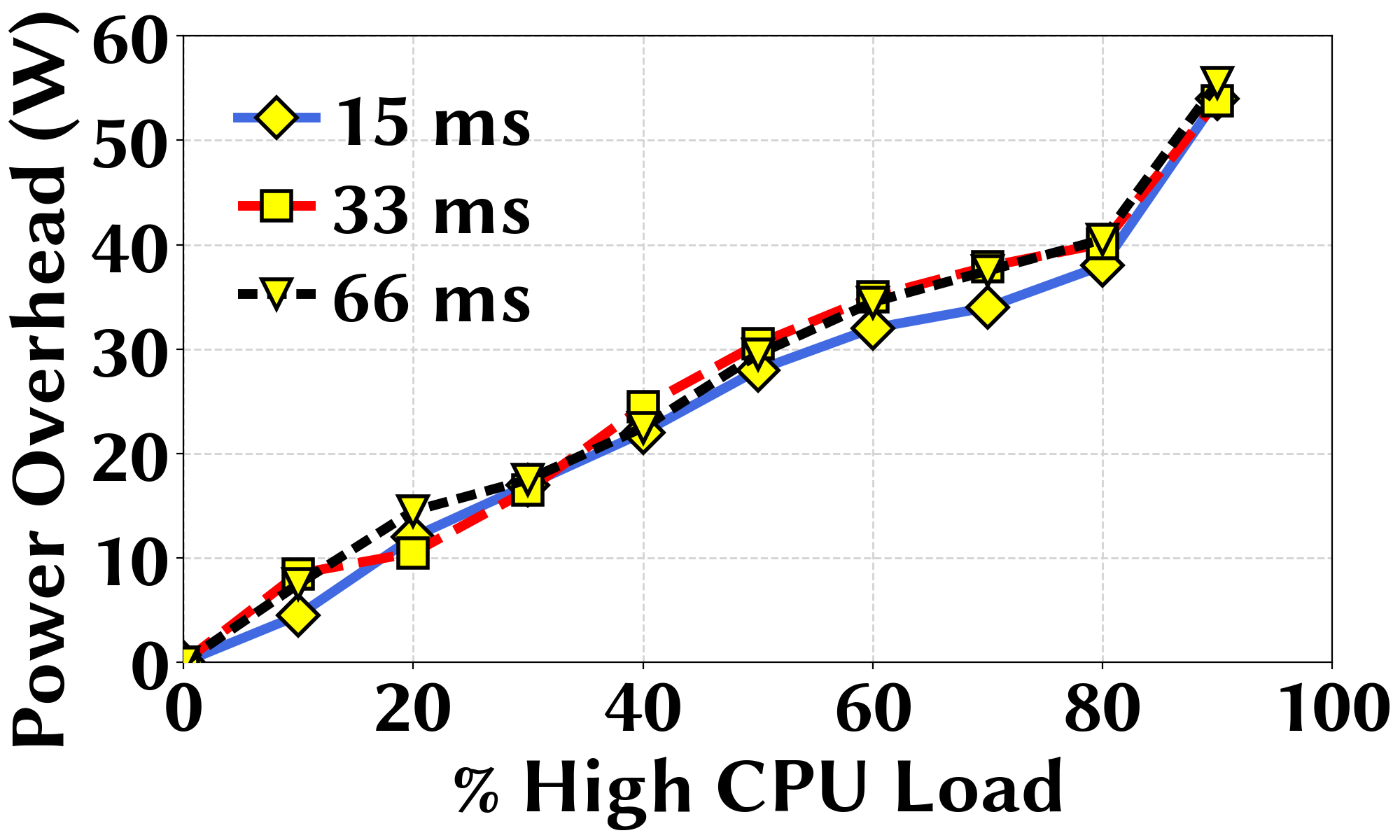}}
	\subfigure[]{\label{fig:power_different_core_main}\includegraphics[width=0.23\textwidth]{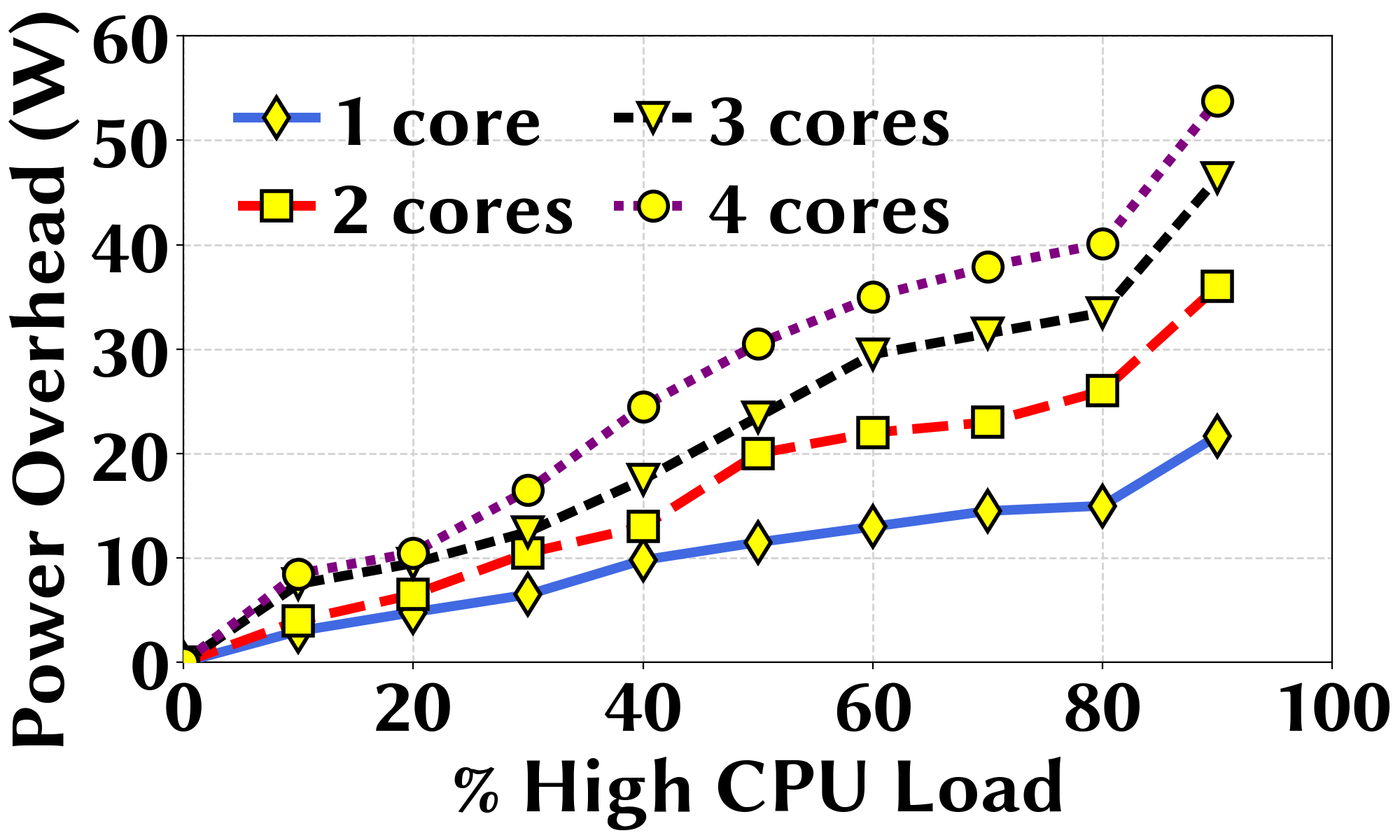}}
	\caption{Power overhead for the defense program. (a) Power overhead remains same for different CPU load intervals. (b) Impact of the number of cores utilized by the defense program.}\label{fig:defense_noise_power_overheads}
\end{figure}

\begin{figure}[!h]
	\centering
	\subfigure[20\% high CPU load]{\label{fig:revision_simulTX_current_noise_33ms_20p_NoTX_apx}\includegraphics[width=0.46\textwidth]{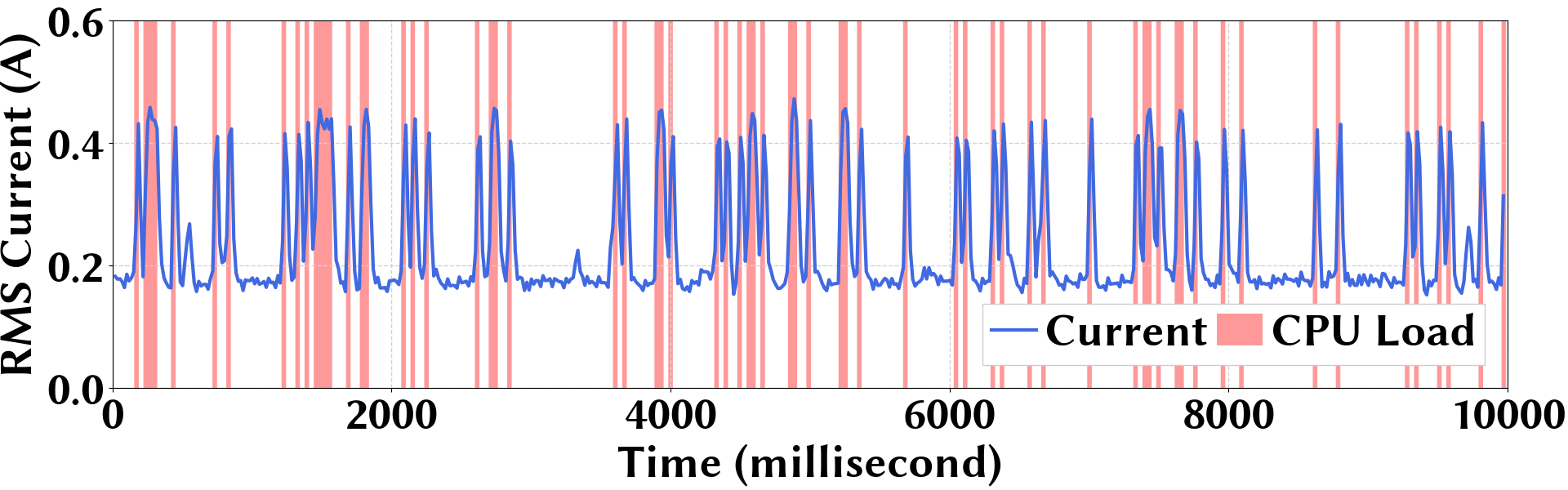}}
	\subfigure[60\% high CPU load]{\label{fig:revision_simulTX_current_noise_33ms_60p_NoTX_apx}\includegraphics[width=0.46\textwidth]{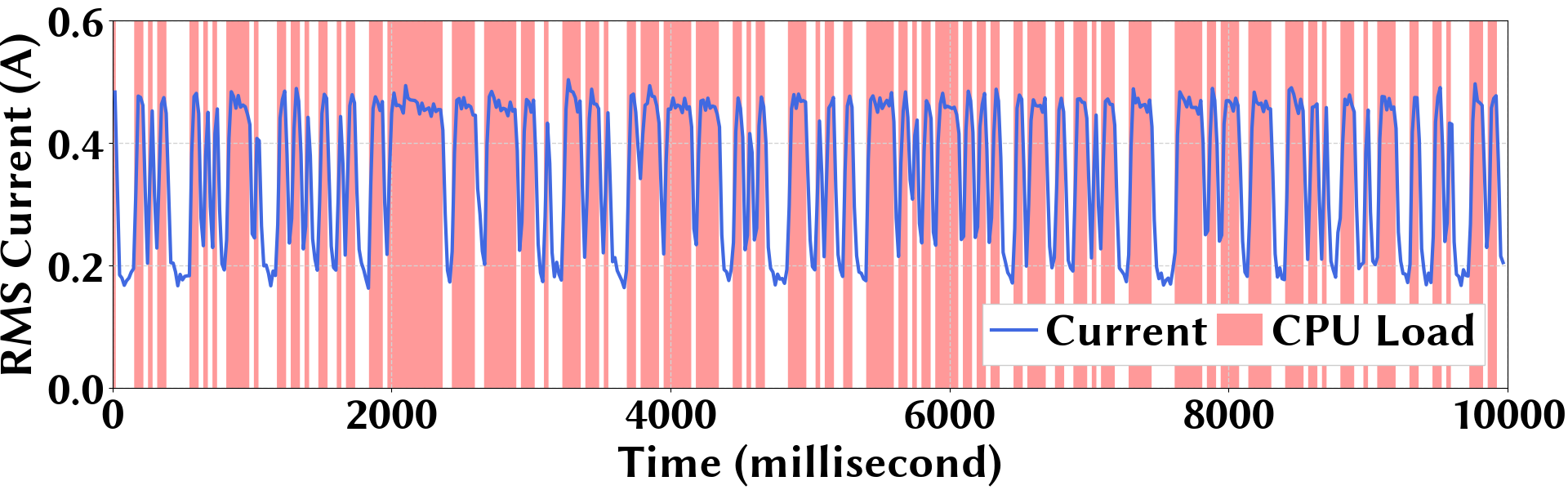}}
	\caption{Current draw of the defense program running with random high CPU loads. (a) 20\% high CPU load. (b) 60\% high CPU load.}\label{fig:revision_simulTX_current_noise_33ms_NoTX_apx}
\end{figure}

\begin{figure*}[!h]
	\centering
	\subfigure[Symbol length = 15 ms, high CPU load = 20\% times, bit error rate = 0\%]{\includegraphics[width=0.46\textwidth]{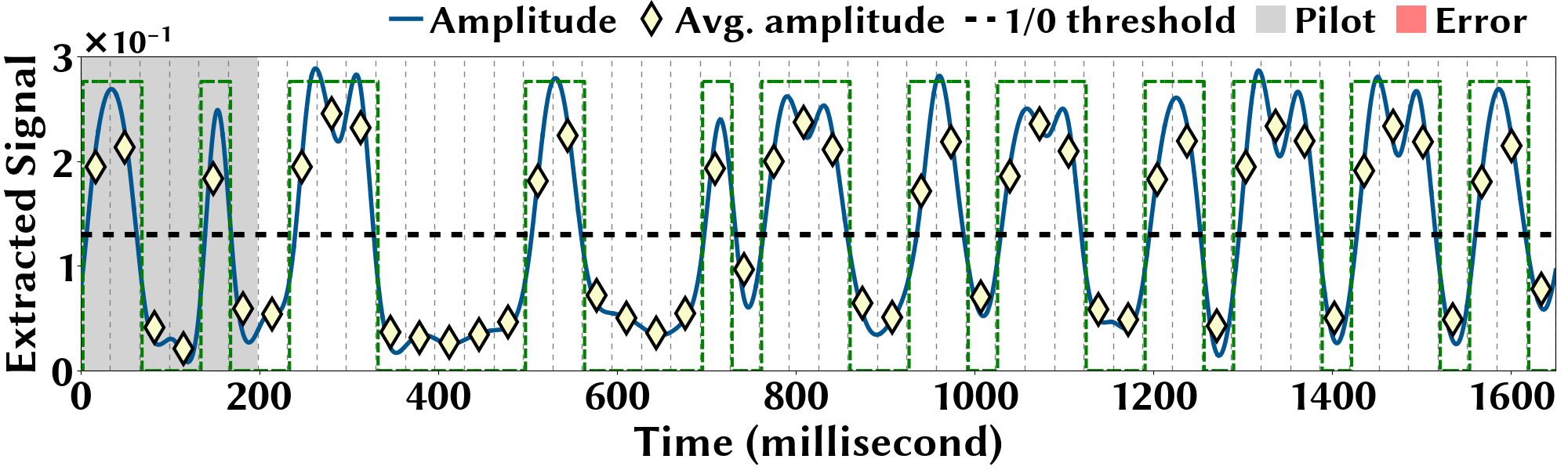}}\hspace{0.1cm}
	\subfigure[Symbol length = 33 ms, high CPU load = 20\% times, bit error rate = 0\%]{\includegraphics[width=0.46\textwidth]{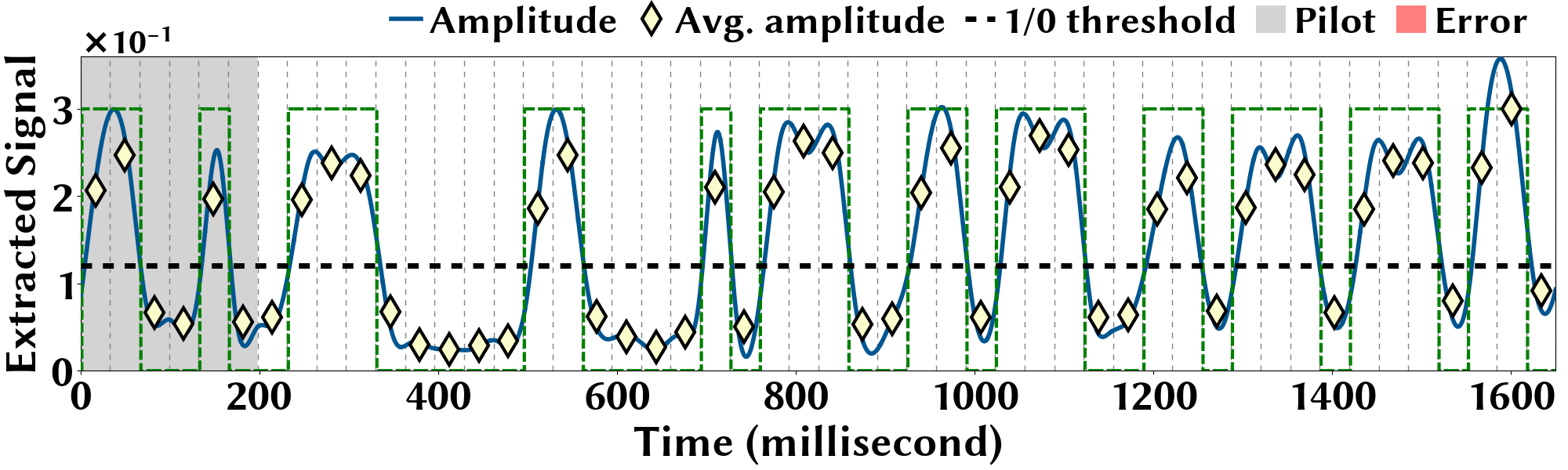}}
	\subfigure[Symbol length = 66 ms, high CPU load = 20\% times, bit error rate = 0\%]{\includegraphics[width=0.46\textwidth]{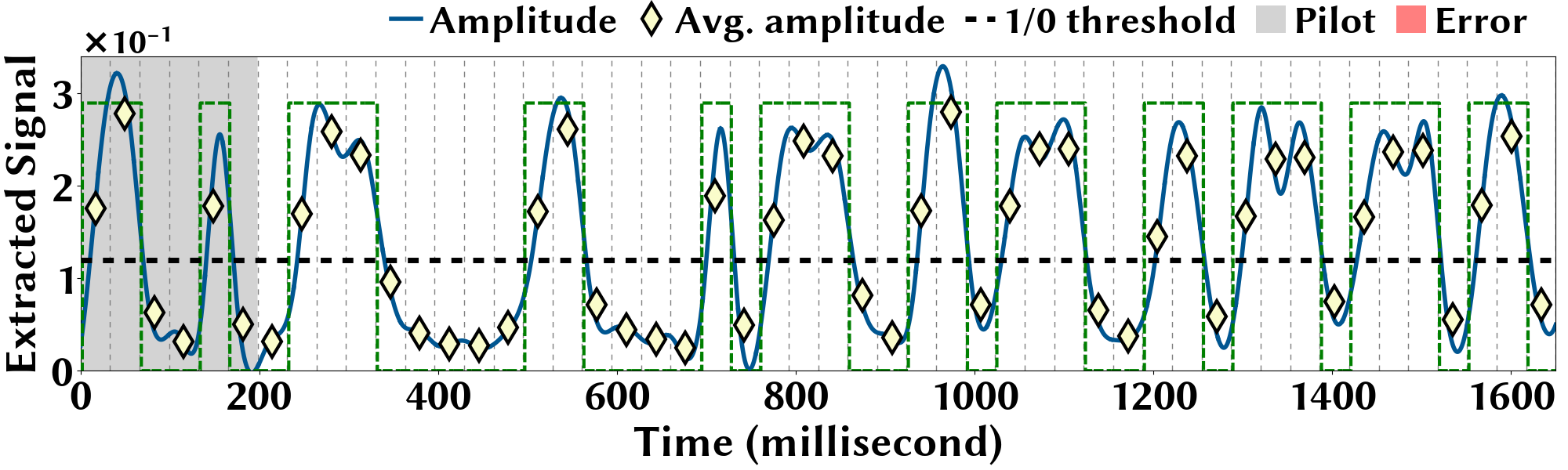}}
	\vspace{-0.2cm}	
	\caption{Impact CPU loading intervals with CPU loads 20\% of the times..}\label{fig:detection_noise_interval_20p}
	\vspace{-0.0cm}	
\end{figure*}

\begin{figure*}[!h]
	\centering
	\subfigure[Symbol length = 15 ms, high CPU load = 60\% times, bit error rate = 20\%]{\includegraphics[width=0.46\textwidth]{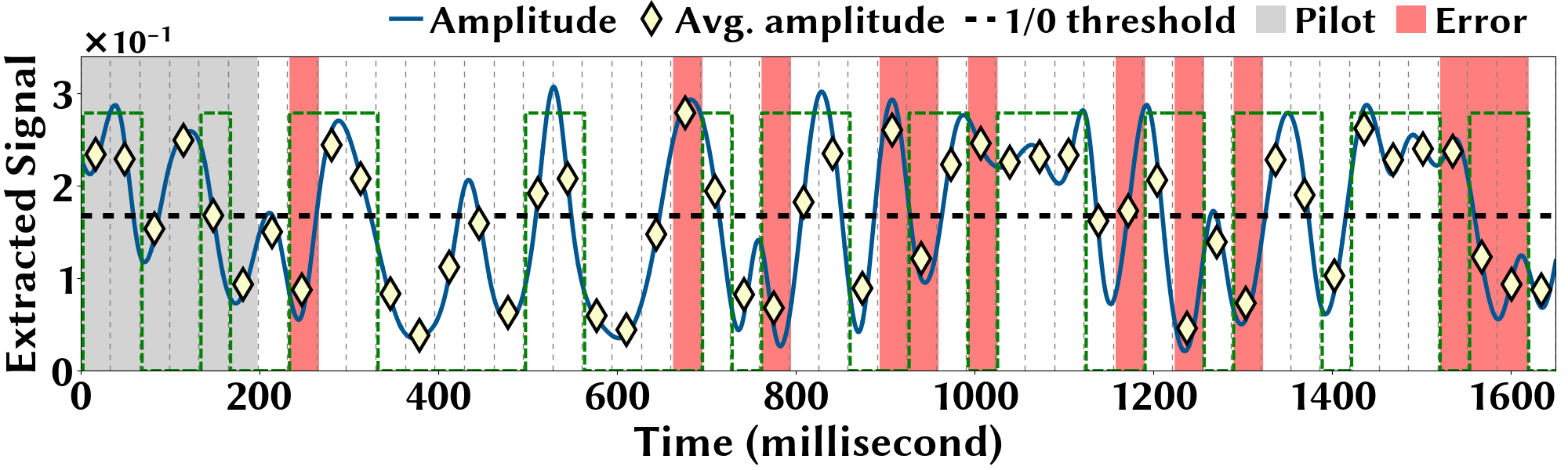}}\hspace{0.1cm}
	\subfigure[Symbol length = 33ms, high CPU load = 60\% times, bit error rate = 30\%]{\includegraphics[width=0.46\textwidth]{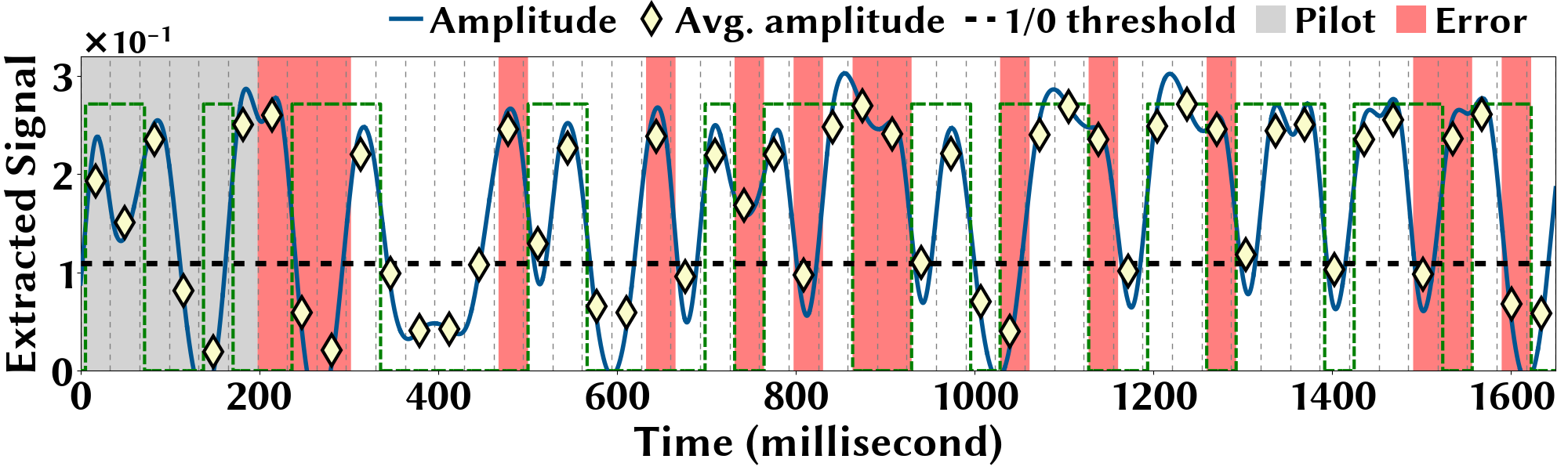}}
	\subfigure[Symbol length = 66 ms, high CPU load = 60\% times, bit error rate = 30\%]{\includegraphics[width=0.46\textwidth]{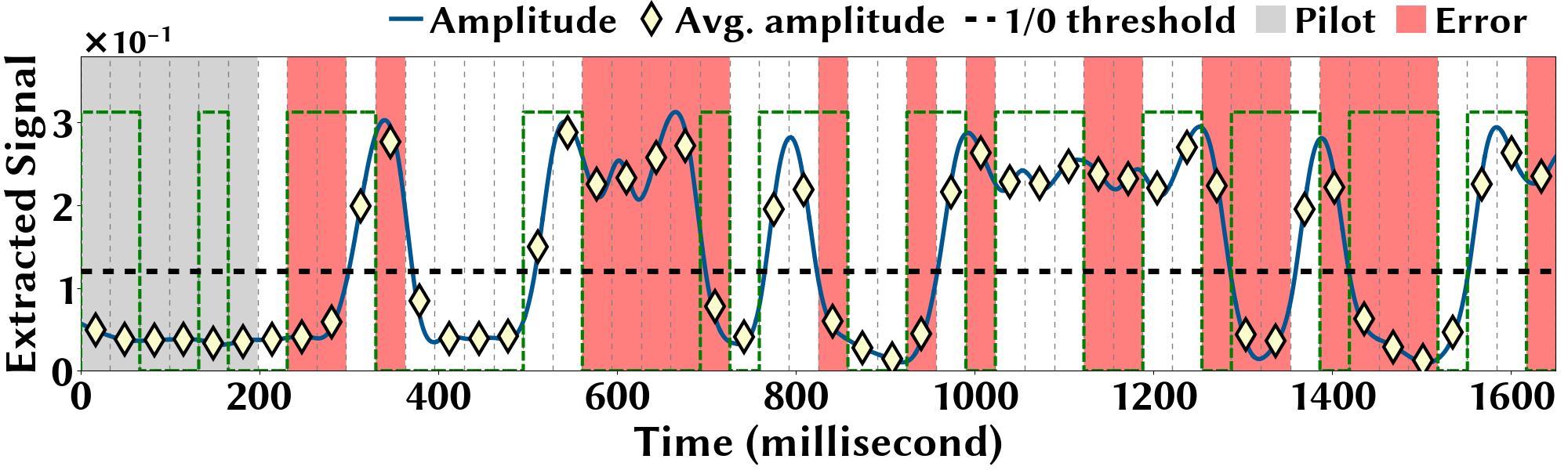}}
	\vspace{-0.2cm}	
	\caption{Impact CPU loading intervals with CPU loads 60\% of the times..}\label{fig:detection_noise_interval_60p}
	\vspace{-0.0cm}	
\end{figure*}

\begin{figure*}[!h]
	\centering
	\subfigure[CPU core = 1, high CPU load = 20\% times, bit error rate = 0\%]{\includegraphics[width=0.46\textwidth]{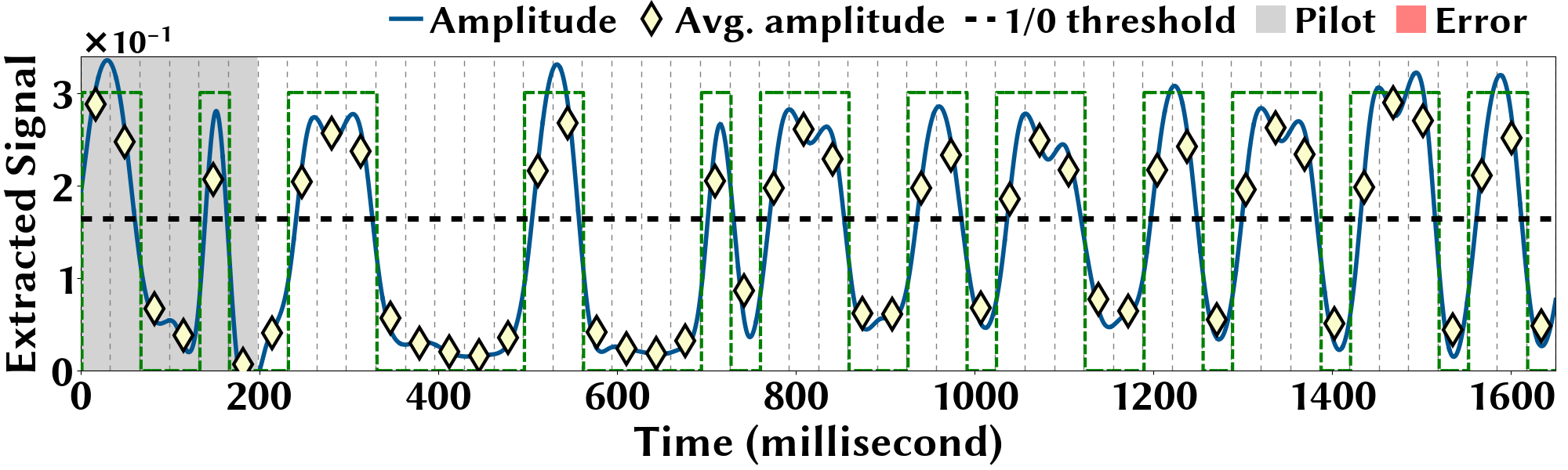}}\hspace{0.1cm}
	\subfigure[CPU core = 2, high CPU load = 20\% times, bit error rate = 0\%]{\includegraphics[width=0.46\textwidth]{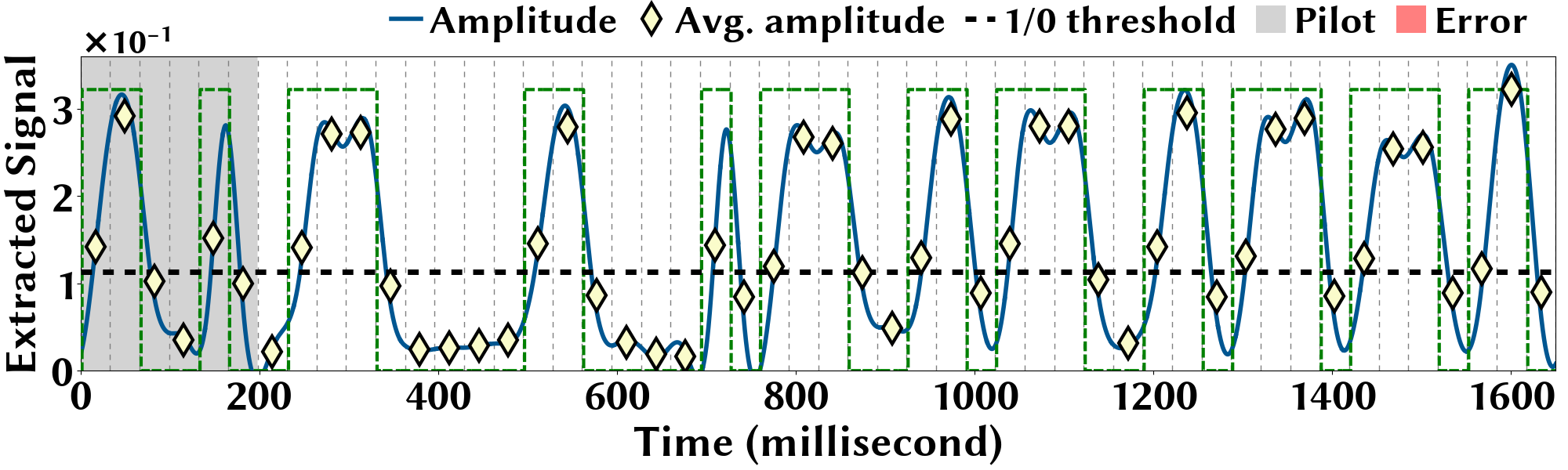}}
	\subfigure[CPU core = 3, high CPU load = 20\% times, bit error rate = 0\%]{\includegraphics[width=0.46\textwidth]{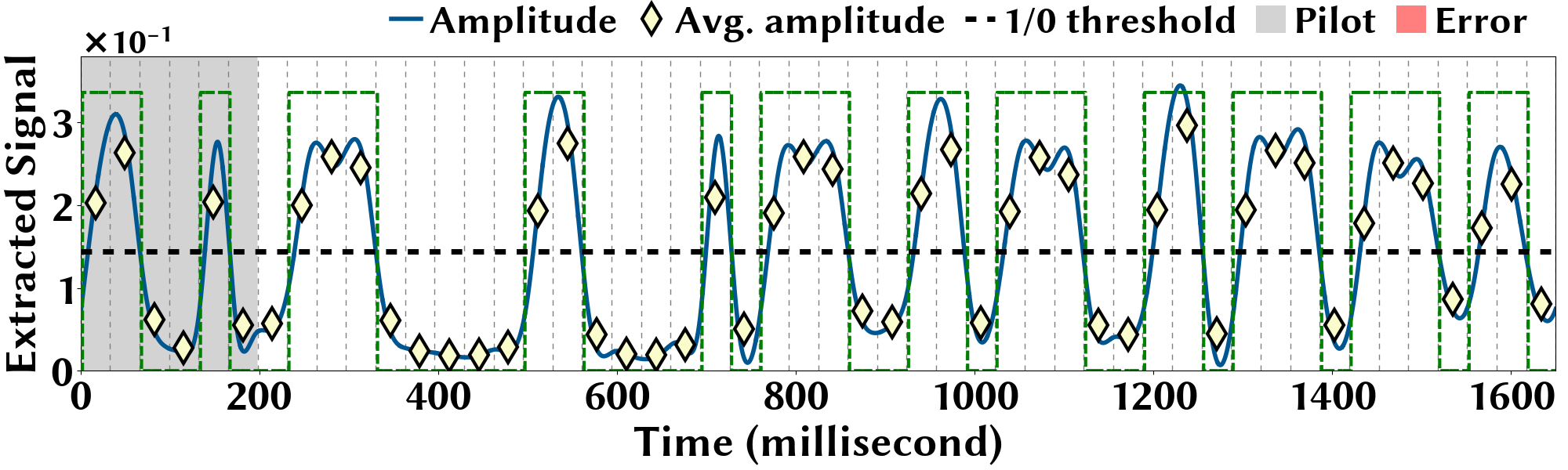}}\hspace{0.1cm}
	\subfigure[CPU core = 4, high CPU load = 20\% times, bit error rate = 0\%]{\includegraphics[width=0.46\textwidth]{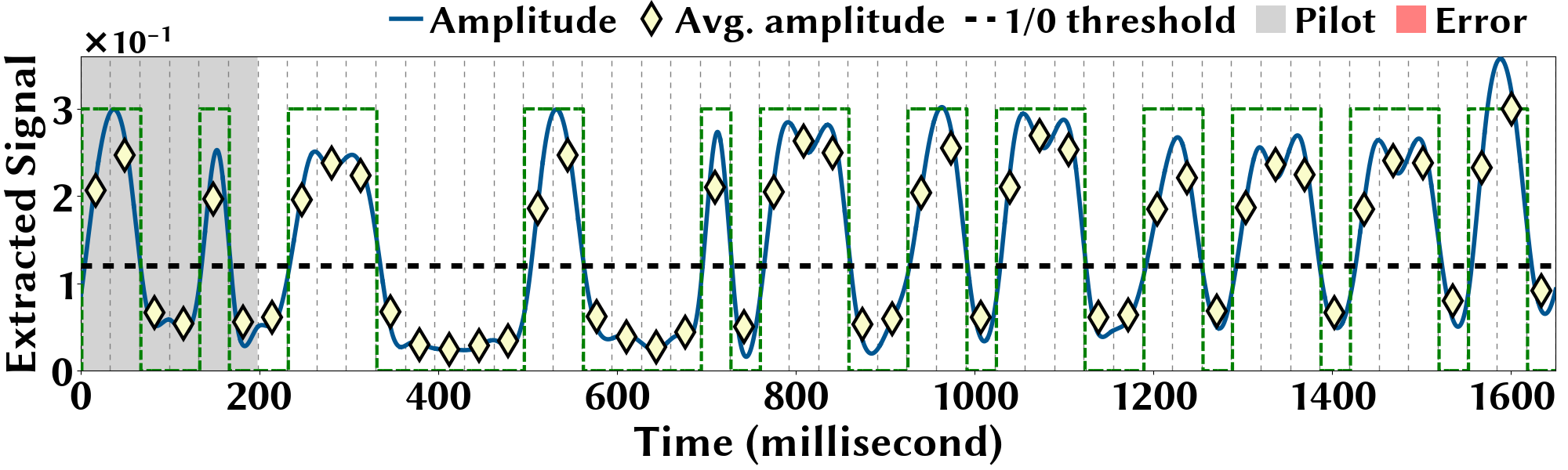}}
	\vspace{-0.2cm}	
	\caption{Impact of number of cores used by the defense program with CPU loads 20\% of the times.}\label{fig:detection_20pnoise_core}
	\vspace{-0.0cm}	
\end{figure*}

\begin{figure*}[!h]
	\centering
	\subfigure[CPU core = 1, high CPU load = 60\% times, bit error rate = 6\%]{\includegraphics[width=0.46\textwidth]{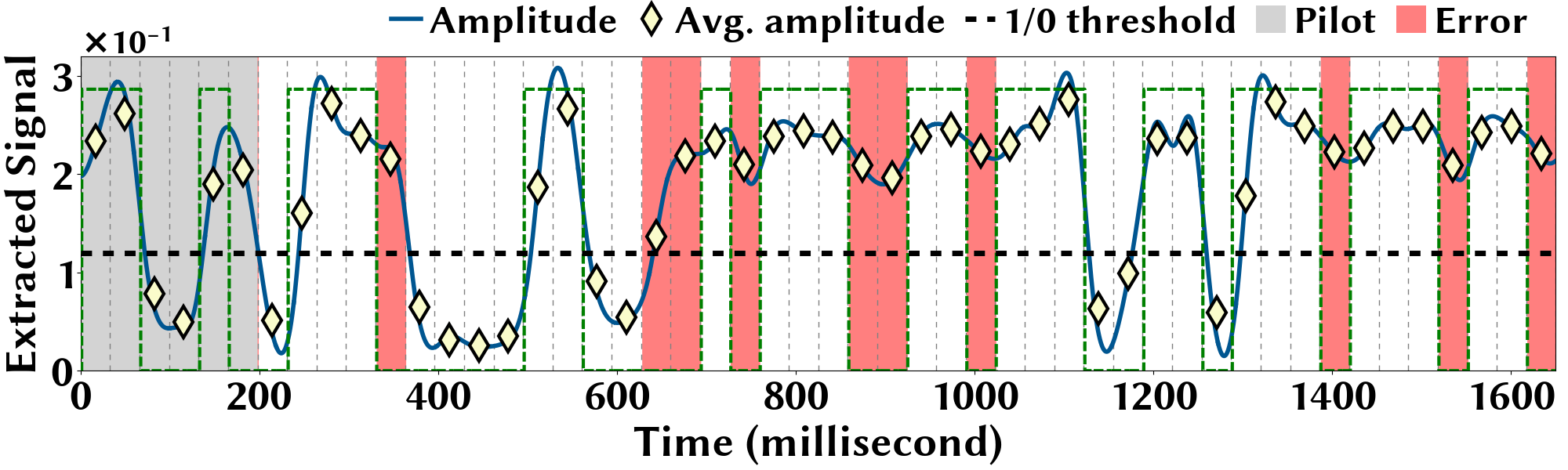}}\hspace{0.1cm}
	\subfigure[CPU core = 2, high CPU load = 60\% times, bit error rate = 30\%]{\includegraphics[width=0.46\textwidth]{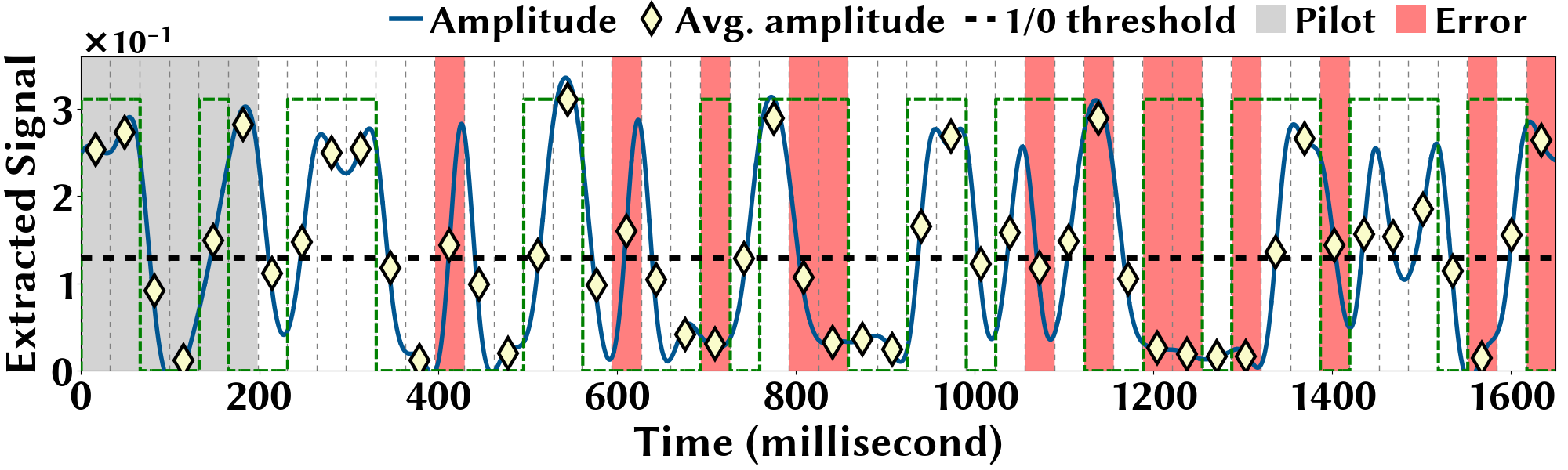}}
	\subfigure[CPU core = 3, high CPU load = 60\% times, bit error rate = 44\%]{\includegraphics[width=0.46\textwidth]{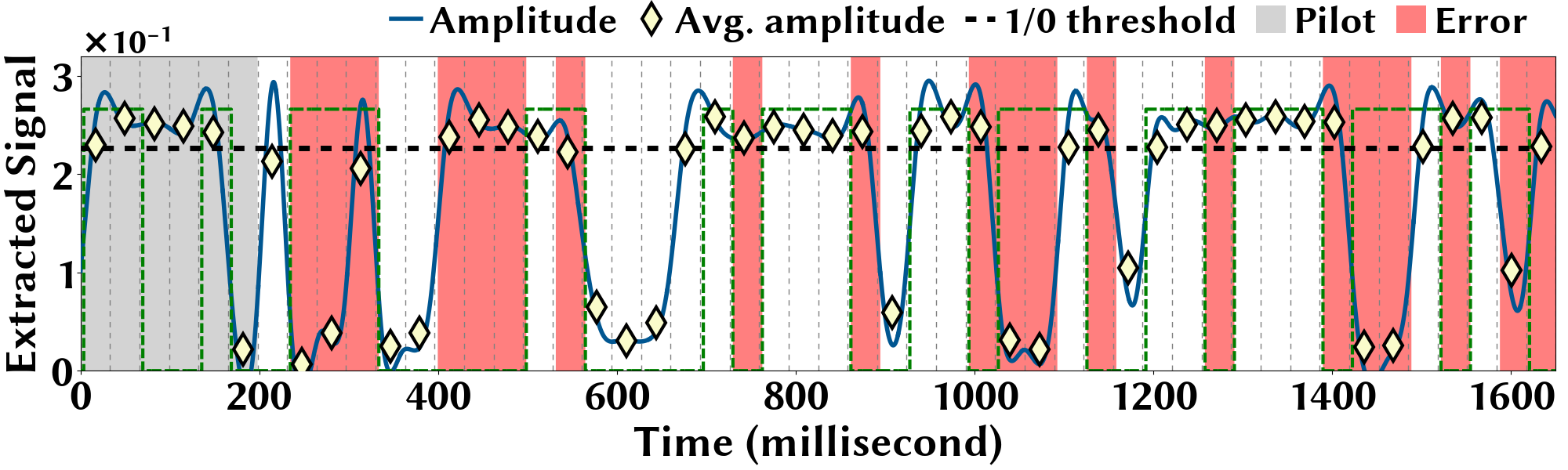}}\hspace{0.1cm}
	\subfigure[CPU core = 4, high CPU load = 60\% times, bit error rate = 44\%]{\includegraphics[width=0.46\textwidth]{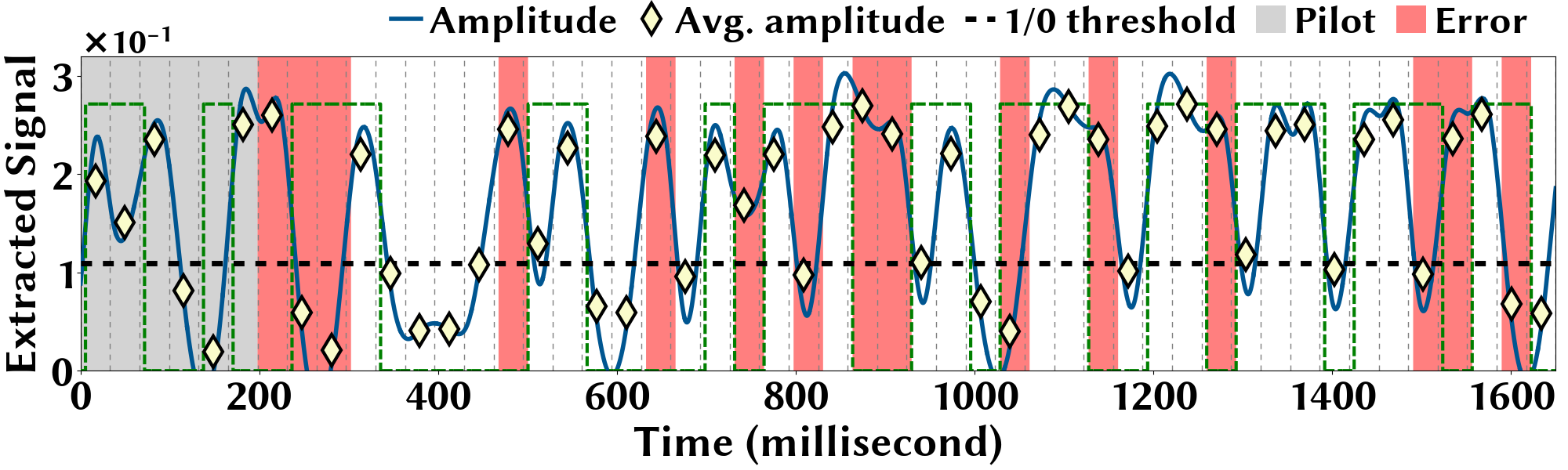}}
	\vspace{-0.2cm}	
	\caption{Impact of number of cores used by the defense program with CPU loads 60\% of the times}\label{fig:detection_60p_noise_core}
	\vspace{-0.0cm}	
\end{figure*}

We test the performance of a defense program that that injects
random power loads by randomly deciding to either run a CPU-intensive computation or remain idle.
We test \OurAlg's performance under varying settings of the defense program's time interval of the CPU loads, percent time of high CPU load, and number of CPU cores used by the defense program.
Figs.~\ref{fig:detection_noise_interval_20p}, \ref{fig:detection_noise_interval_60p}, \ref{fig:detection_20pnoise_core}, and \ref{fig:detection_60p_noise_core} shows the snapshot of detection results for some selected cases of interest, while Fig.~\ref{fig:defense_noise_power_overheads} shows the power overhead for different settings. Fig.~\ref{fig:revision_simulTX_current_noise_33ms_NoTX_apx} shows the current drawn by the defense program with the CPU load 20\% and 60\% of the times.

\newpage
\section{Snapshots of All Detection Results}\label{sec:appendix_all_snapshots}


\end{document}